\documentclass{scrartcl}

\makeatletter
\DeclareOldFontCommand{\rm}{\normalfont\rmfamily}{\mathrm}
\DeclareOldFontCommand{\sf}{\normalfont\sffamily}{\mathsf}
\DeclareOldFontCommand{\tt}{\normalfont\ttfamily}{\mathtt}
\DeclareOldFontCommand{\bf}{\normalfont\bfseries}{\mathbf}
\DeclareOldFontCommand{\it}{\normalfont\itshape}{\mathit}
\DeclareOldFontCommand{\sl}{\normalfont\slshape}{\@nomath\sl}
\DeclareOldFontCommand{\sc}{\normalfont\scshape}{\@nomath\sc}
\makeatother

\usepackage{xcolor}
\usepackage{color}
\definecolor{blue1}{RGB}{0,102,189}
\definecolor{blue2}{RGB}{98,160,214}
\definecolor{my_orange}{RGB}{243,98,33}

\colorlet{tumblue}  {blue1}
\colorlet{tumorange}{my_orange}

\usepackage{arxiv}
\usepackage{authblk}

\usepackage{url}
\usepackage{graphicx}
\setkeys{Gin}{keepaspectratio} 

\usepackage{tikz}
\usetikzlibrary{mindmap,trees,shadows}
\usetikzlibrary{arrows,intersections}
\usetikzlibrary{angles}
\usetikzlibrary{shapes.misc} \pgfdeclarelayer{background}
\pgfsetlayers{background, main}
\usetikzlibrary{decorations.markings}

\colorlet{incolor}{tumorange}
\colorlet{outcolor}{tumblue}

\usepackage{tikz-dimline}

\tikzset{inNode/.style={draw=incolor, fill=incolor, circle, inner sep=1.5pt}}
\tikzset{outNode/.style={draw=outcolor,  fill=outcolor, circle, inner sep=1pt}}
\tikzset{projNode/.style={draw=outcolor, fill=outcolor, circle, inner sep=1pt}}

\usetikzlibrary{shapes.arrows, fadings, calc, positioning}
\tikzfading[name=fade right,
  left color=transparent!0, right color=transparent!100]
\tikzfading[name=fade left,
  left color=transparent!100, right color=transparent!0]
\tikzfading[name=fade top,
  bottom color=transparent!0, top color=transparent!100]
  
\pgfdeclarelayer{bg}    \pgfdeclarelayer{fg}    \pgfsetlayers{bg,main,fg}  

\definecolor{participantAcolor}{RGB}{243,98,33}
\definecolor{participantBcolor}{RGB}{0,102,189}
\definecolor{participantCcolor}{RGB}{66,70,50}
\definecolor{cplColor}{RGB}{0,0,0}
\definecolor{preciceColor}{RGB}{ 161, 177, 25}

\definecolor{ci_pantone300}{RGB}{ 0, 101, 189}

\definecolor{ci_pantone301}{RGB}{ 0, 82, 147}
\definecolor{ci_pantone540}{RGB}{ 0, 51, 89}

\definecolor{ci_pantone283}{RGB}{ 152, 198, 234}
\definecolor{ci_pantone542}{RGB}{ 100, 160, 200}
\definecolor{ci_ivory}{RGB}{ 218, 215, 203}
\definecolor{ci_orange}{RGB}{ 227, 114, 34}
\definecolor{ci_green}{RGB}{ 162, 173, 0}  
 
\usepackage[frozencache]{minted}
\usepackage{listings}
  \usepackage{amsmath}
\usepackage{amssymb} \usepackage{graphicx}
\usepackage[hidelinks]{hyperref}
\pgfplotsset{compat=1.16}

\usepackage{multirow}
\usepackage{booktabs}
\usepackage[export]{adjustbox}
\usepackage{gensymb}
\usepackage{subcaption}
\usepackage{xspace} \usepackage{algorithm}
\usepackage{algpseudocode}

\usepackage{geometry}
\newcommand{\incode}[1]{\mintinline{text}{#1}}
\newcommand{\incpp}[1]{\mintinline{c++}{#1}}
\newcommand{\inpython}[1]{\mintinline{python}{#1}}

\newcommand{\inbash}[1]{\mintinline{bash}{#1}}

\usepackage{siunitx}

\usepackage{framed}

\usepackage{enumitem}

\newcommand\github[1]{\href{https://github.com/#1}{\texttt{#1}}}

\title{preCICE v2: A Sustainable and User-Friendly Coupling Library}

\author[1,*]{Gerasimos Chourdakis}
\author[2,*]{Kyle Davis}
\author[1,*]{Benjamin Rodenberg}
\author[2,*]{Miriam Schulte}
\author[1,*]{Frédéric Simonis}
\author[3,*,$\dagger$]{Benjamin Uekermann}
\author[2]{Georg Abrams}
\author[1]{Hans-Joachim Bungartz}
\author[1]{Lucia Cheung Yau}
\author[3]{Ishaan Desai}
\author[1]{Konrad Eder}
\author[1]{Richard Hertrich}
\author[2]{Florian Lindner}
\author[1]{Alexander Rusch}
\author[1]{Dmytro Sashko}
\author[3]{David Schneider}
\author[2]{Amin Totounferoush}
\author[1]{Dominik Volland}
\author[2]{Peter Vollmer}
\author[1]{Oguz Ziya Koseomur}

\affil[1]{Scientific Computing in Computer Science, Department of Informatics, Technical University of Munich}
\affil[2]{Simulation of Large Systems, Institute for Parallel and Distributed Systems, University of Stuttgart}
\affil[3]{Usability and Sustainability of Simulation Software, Institute for Parallel and Distributed Systems, University of Stuttgart}
\affil[*]{Equal contributors}
\affil[$\dagger$]{Corresponding author. \textit{Email address:} \url{benjamin.uekermann@ipvs.uni-stuttgart.de} (Benjamin Uekermann)}

\begin{document}

\maketitle

\begin{abstract}
preCICE is a free/open-source coupling library. It enables creating partitioned multi-physics simulations by gluing together separate software packages.
This paper summarizes the development efforts in preCICE of the past five years. During this time span, we have turned the software from a working prototype -- sophisticated numerical coupling methods and scalability on ten thousands of compute cores -- to a sustainable and user-friendly software project with a steadily-growing community. Today, we know through forum discussions, conferences, workshops, and publications of more than 100 research groups using preCICE. We cover the fundamentals of the software alongside a performance and accuracy analysis of different data mapping methods. Afterwards, we describe ready-to-use integration with widely-used external simulation software packages, tests and continuous integration from unit to system level, and community building measures, drawing an overview of the current preCICE ecosystem.

\end{abstract}


\begin{framed}
    \def\roleconceptualization{Conceptualization}
    \def\rolemethodology{Methodology}
    \def\rolesoftware{Software}
    \def\rolewriting{Writing}
    \def\rolevisualization{Visualization}
    \def\roleadministration{Project administration}
    \def\rolesupervision{Supervision}
    \def\rolefunding{Funding acquisition}
    
    \textbf{Author roles:}\vspace{-2ex}
    \begin{flushleft}
        \textbf{Gerasimos Chourdakis:} \rolemethodology, \rolesoftware, \rolewriting, \rolevisualization, \rolesupervision \ \  
        \textbf{Kyle Davis:} \rolesoftware, \rolewriting \  
        \textbf{Benjamin Rodenberg:} \rolemethodology, \rolesoftware, \rolewriting, \rolevisualization, \rolesupervision \  
        \textbf{Miriam Schulte:} \roleconceptualization, \rolewriting, \roleadministration, \rolesupervision, \rolefunding \  
        \textbf{Frédéric Simonis:} \rolemethodology, \rolesoftware, \rolewriting, \rolevisualization, \rolesupervision \  
        \textbf{Benjamin Uekermann:} \roleconceptualization, \rolemethodology, \rolesoftware, \rolewriting, \rolevisualization, \roleadministration, \rolesupervision, \rolefunding \  
        \textbf{Georg Abrams:} \rolesoftware \ 
        \textbf{Hans-Joachim Bungartz:} \rolesupervision, \rolefunding \ 
        \textbf{Lucia Cheung Yau:} \rolesoftware \ 
        \textbf{Ishaan Desai:} \rolesoftware \  
        \textbf{Konrad Eder:} \rolesoftware \ 
        \textbf{Richard Hertrich:} \rolesoftware \ 
        \textbf{Florian Lindner:} \rolemethodology, \rolesoftware, \rolesupervision \ 
        \textbf{Alexander Rusch:} \rolesoftware \ 
        \textbf{Dmytro Sashko:} \rolesoftware \ 
        \textbf{David Schneider:} \rolesoftware \  
        \textbf{Amin Totounferoush:} \rolesoftware \ 
        \textbf{Dominik Volland:} \rolesoftware \ 
        \textbf{Peter Vollmer:} \rolesoftware \ 
        \textbf{Oguz Ziya Koseomur:} \rolesoftware
    \end{flushleft}
\end{framed}

\section{Introduction} 

Flexible, modular simulation environments are key to many important application fields such as aerospace engineering~\cite{Slotnick2014}, biomedical engineering~\cite{HellersteibBiomed2019}, climate and environmental research~\cite{Climate2020} and many others.
The need to provide smart mathematical and software solutions to combine different aspects of such
simulations in a modular way has been showcased in many publications, e.g.,~\cite{Keyes2013}. With increasing complexity of the respective software environments, the usability and maintainability of the involved software components become a critical issue, which is addressed by a growing research software engineering community (see, e.g., a recent position paper of the German community~\cite{Anzt2021_RSE}).

We present the software package preCICE, which enables black-box coupling of separate solvers for different types of numerical models. It has originally been developed for modular, so-called partitioned, simulations of fluid-structure interactions, i.e., the combination of a flow solver with a structural mechanics solver via a common surface at which forces and displacements are exchanged. Over the past ten years, preCICE has developed into a far more general tool for partitioned simulations, which can handle different types of coupling (weak/strong, explicit/implicit, surface/volume) and any type of equations. Examples range from fluid-structure-acoustics interactions~\cite{Lindner2020_ExaFSA}, over blood flow simulation in the human body~\cite{naseri2020scalable}, free-flow porous media coupling~\cite{Jaust2020}, conjugate heat transfer~\cite{fan2020study}, muscle-tendon system simulations~\cite{Maier2021_Diss}, flow-particle coupling~\cite{besseron2021eulerian}, to coupling between subsurface flow and planning tools for geothermal energy infrastructure~\cite{geoKW}. The coupling is not restricted to a pair of solvers, but has been extended to enable multi-component coupling of arbitrarily many solvers~\cite{Bungartz2014_MIQN_CompuMech}.
preCICE offers comprehensive functionality far beyond simple data exchange: It provides (i) a variety of mapping methods for data transfer between non-matching meshes of different solvers, (ii) quasi-Newton acceleration methods for iterative implicit coupling, and (iii) bottleneck-free point-to-point communication between processes of parallel solvers. preCICE was originally designed with surface coupling in mind, but most features can and have been used for volume coupling as well. All coupling numerics and communication are implemented in a library approach and are fully parallelized. The library can be used via a high-level application programming interface (API) in a minimally-invasive way (from the perspective of the coupled solvers).

The first version of preCICE, presented in~\cite{Gatzhammer2015}, used a server process per coupled solver and was, thus, not very efficient for the coupling of parallel solvers. The authors of~\cite{Uekermann2016, Lindner2019, Bungartz2016_ExaFSA_LNCSE} developed a fully parallel version based on point-to-point communication, which shows good scalability on 10,000s of compute cores. A first overview paper of preCICE was published in 2016~\cite{Bungartz2016_preCICE}, summarizing basic functionality, the API, and the user-specific configuration as well as showing example applications and validation cases. Semantic versioning of preCICE was introduced in 2017.

In this paper, we summarize new developments from 2016 to 2021, i.e., from~\cite{Bungartz2016_preCICE} to the release v2.2, which is part of the first preCICE distribution v2104.0~\cite{Chourdakis2021_Distribution} -- a complete ecosystem of preCICE components. We present an overview of the functionality of preCICE in \autoref{sec:library}, in particular the numerical coupling methods, i.e., quasi-Newton iterations and data mapping. We complement the description of methods for data mapping with a performance and accuracy study using realistic 3D turbine blade meshes. Section~\ref{sec:library}, in addition, gives details on the installation process of preCICE. Beyond the core of preCICE, we describe the newly-developed, ready-to-use adapters~\cite{Uekermann2017_Adapters} for many widely-used simulation software projects in \autoref{sec:adapters}. Section~\ref{sec:tutorials} introduces a simple conjugate heat transfer (CHT) scenario and a simple fluid-structure interaction (FSI) scenario as illustrative examples on how to use preCICE with any of these simulation software projects, followed by the presentation of a systematic multi-level testing infrastructure in \autoref{sec:tests}. In the past years, preCICE has become a widely used software ecosystem, for which we have built up a community of users, as we show in \autoref{sec:community}.

Our description focuses on (i) usability (by providing robust numerical choices, the multitude of ready-to-use adapter codes, well-structured documentation, and easily-accessible illustrative examples), (ii) reliability (by the systematic multi-level testing concept), and (iii) sustainability (by continuous integration, well-defined development and release cycles and a concept to involve the community in the software development). For a more classical description of preCICE including classical validation with benchmarks, we refer the reader to~\cite{Bungartz2016_preCICE}. For an analysis of recent performance and scalability improvements, we refer to~\cite{Bungartz2016_ExaFSA_LNCSE, AminPerf2021}. The contributions of this paper enable new scientific insights in the research fields of our users, but also provide new experiences in scientific software engineering. Results for various applications run with preCICE have been published in many other papers such as~\cite{naseri2020scalable, Jaust2020, fan2020study, andrun2020simulating, Cocco2020} (see also \autoref{sec:community}).

Naturally, preCICE is not the only general-purpose coupling software, which has been developed during the past decades. In the following short summary of related tools, we focus particularly on user-focused and open-source software (i.e., we do not focus on in-house or commercial coupling software, e.g., MpCCI~\cite{Wolf2017}).  
There is a number of more multi-scale oriented tools, such as Amuse~\cite{Amuse}, MuMMI~\cite{MuMMI}, MUSCLE 3~\cite{veen2020easing}, MaMiCo~\cite{neumann2016mamico}, or MUI~\cite{tang2015multiscale}. Often, the categories of use cases are not strict. MUI, for example, has recently also been used for fluid-structure interaction~\cite{liu2021parallel}. At the same time, current work on preCICE aims towards certain multi-scale coupling patterns (cf.~\autoref{sec:conclusion}). A good review on multi-scale coupling software is provided in~\cite{groen2019mastering}.
For climate simulations, a number of specialized tools are available, for example OASIS3-MCT~\cite{Craig2017_OASIS3_MCT}, YAC~\cite{Hanke2016_YAC}, and C-Coupler2~\cite{Liu2018_C-Coupler2}.
In principle, the term \textit{coupling} is not well-defined. For example, software such as pyiron~\cite{Janssen2019_Pyiron} or the Kepler Project~\cite{Ludascher2006_Kepler} are referred to as \textit{coupling software} as well, whereas they refer to workflow coupling and not a strong coupling between different simultaneously running simulations.

The two software projects most similar to preCICE are presumably DTK~\cite{Slattery2013_DTK} and OpenPALM~\cite{Duchaine2015_OpenPALM}.
DTK offers an API that targets lower-level operations compared to preCICE. Its main job is to map and communicate data between different meshes in parallel. The implementation of the actual coupling logic is left to the user, which leads to greater flexibility, but also to more development effort for the user. OpenPALM employs a similar API, provided by CWIPI. The \textit{front end} of OpenPALM, however, provides a higher-level API that includes coupling logic. On top, a graphical user interface is available to configure and steer coupled simulations.
The largest difference of preCICE compared to DTK and OpenPALM is presumably the large number of ready-to-use adapters to widely-used simulation software packages (e.g., OpenFOAM, SU2, FEniCS, deal.II, or CalculiX) (cf.~\autoref{sec:adapters}).

 \section{The preCICE library}
\label{sec:library}

\begin{figure}[h!]
\centering
\includegraphics[width=\textwidth]{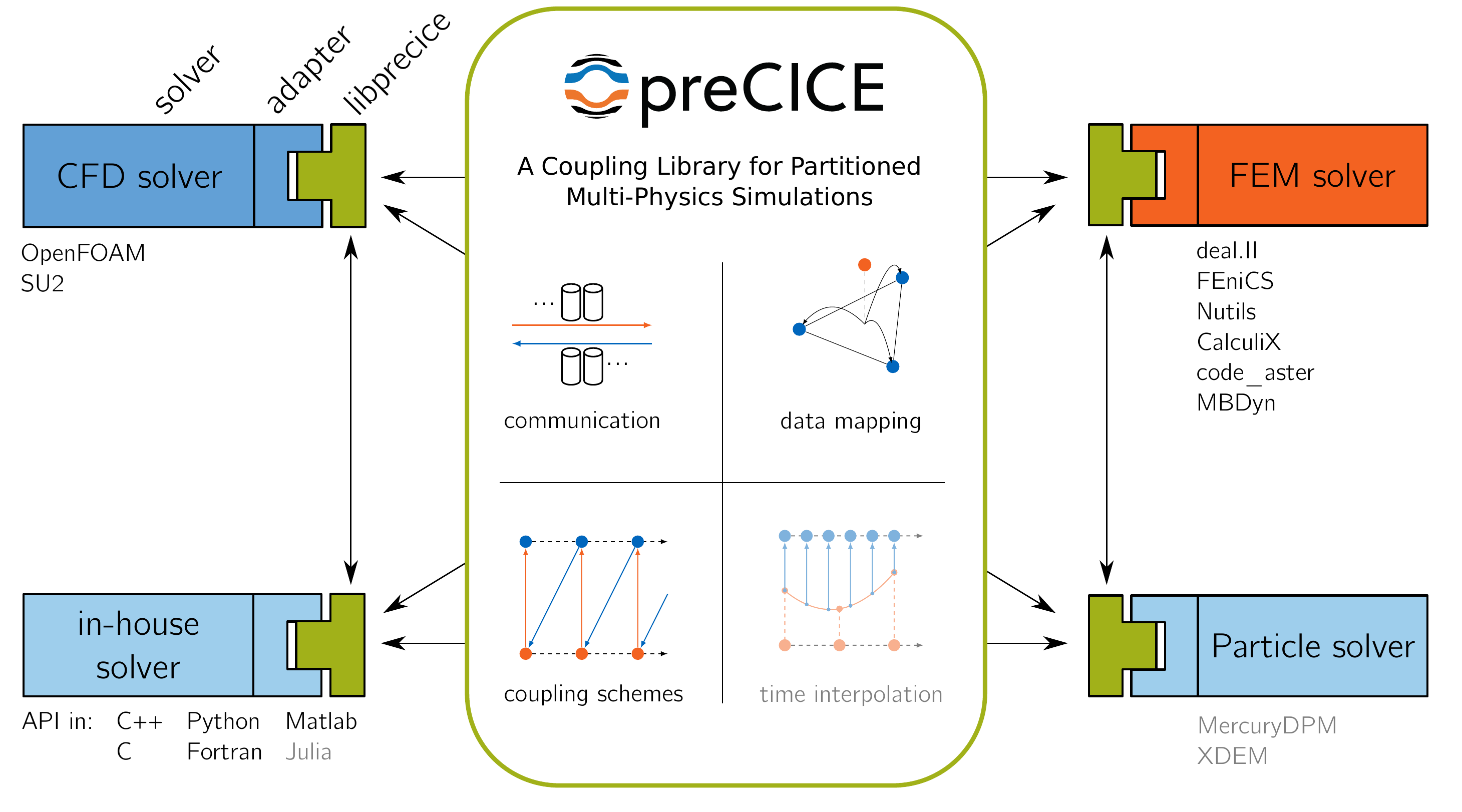}
\caption{
\label{fig:overview} Concept and basic functional components of preCICE. Components that are work in progress and not yet released are shown grayed out.}
\end{figure}

In this section, we present the core library of preCICE in a nutshell. 
\autoref{fig:overview} visualizes the concept and basic functional components of preCICE.
In the figure, several types of simulation codes are coupled: computational fluid dynamics (CFD) solvers, finite-element method (FEM) solvers, in-house solvers, and particle solvers. Please note that we use these types as examples to introduce the overall concept, not as strict non-overlapping categories.
In the following, we refer to coupled simulation codes as \textit{participants} of a coupled simulation.
The glue code between a participant's code and the preCICE library is called adapter.
Depending on the participant, an adapter can be a module or a class of the participant's code or a complete stand-alone software, which uses some callback interface of the participant. 
In some cases, an adapter can also be a sophisticated script that calls the participant as well as preCICE, but such a software design contradicts the main idea behind the library approach of preCICE to some extent.
preCICE comes with several ready-to-use adapters, which are listed in the picture.
Not all adapters, though, feature the same level of maturity.
Users of preCICE can develop adapters for their own (in-house) codes by using the preCICE API, which is available in the most important languages used in Scientific Computing.
An adapter is responsible for what we call \textit{coupling physics}, meaning how to translate nodal coupling values from preCICE into boundary conditions or forcing terms and, reciprocally, how to extract nodal coupling values from internal fields to provide them to preCICE.  
If we say that an adapter can handle a certain type of coupling physics, for example conjugate-heat transfer, it basically means which type of variables, e.g., temperature or heat flux, the adapter is able to read and write.

preCICE itself has no notion of physics.
Instead, preCICE itself is responsible for the technical aspects of coupling and the coupling numerics, depicted in the middle of \autoref{fig:overview}. We now give a first brief overview of these components going from top left to bottom right:
(i) Coupled participants are separate executables, potentially running on different nodes in a heterogeneous compute cluster with independent MPI communicators.
preCICE handles the communication between these executables. 
The communication is asynchronous and completely parallel. 
Only those ranks of the participants that need to exchange coupling data communicate with each other.  
Technically, the communication is based on either MPI Ports or TCP/IP, configurable at runtime.
(ii) preCICE implements coupling schemes. Coupling schemes, on the one hand, define the logical coupling flow, i.e., which participant sends which data to which other participant and how the execution of time steps is synchronized between the participants.
On the other hand, coupling schemes comprise acceleration methods for implicit coupling such as Aitken under-relaxation or quasi-Newton methods. 
(iii) Moreover, preCICE allows to map coupling data between non-matching and non-conforming coupling meshes. 
To this end, the user can choose between projection-based methods (nearest neighbor or nearest projection) or radial-basis function interpolation.
(iv) Finally, preCICE also handles interpolation in time.
Currently, only plain sub-cycling is supported, but higher-order interpolation is under development~\cite{QNWI} and will be available in future releases.
 
Please note that, even though \autoref{fig:overview} depicts a significant green box in the middle, there is no central server-like instance running, even for parallel simulations. 
preCICE uses a pure peer-to-peer library approach.
The only executables that are started are the participants, which all call preCICE.

The current section describes the main concepts of the core library and is structured as follows. 
In \autoref{ssec:cplscheme}, \autoref{ssec:mapping}, and \autoref{ssec:com}, we describe the methods preCICE uses for coupling schemes, data mapping, and communication, respectively. Different options to get and if necessary build preCICE are listed in \autoref{ssec:building}. Finally, \autoref{ssec:api} explains the API and the runtime configuration of preCICE.

\subsection{Coupling schemes and acceleration}
\label{ssec:cplscheme}

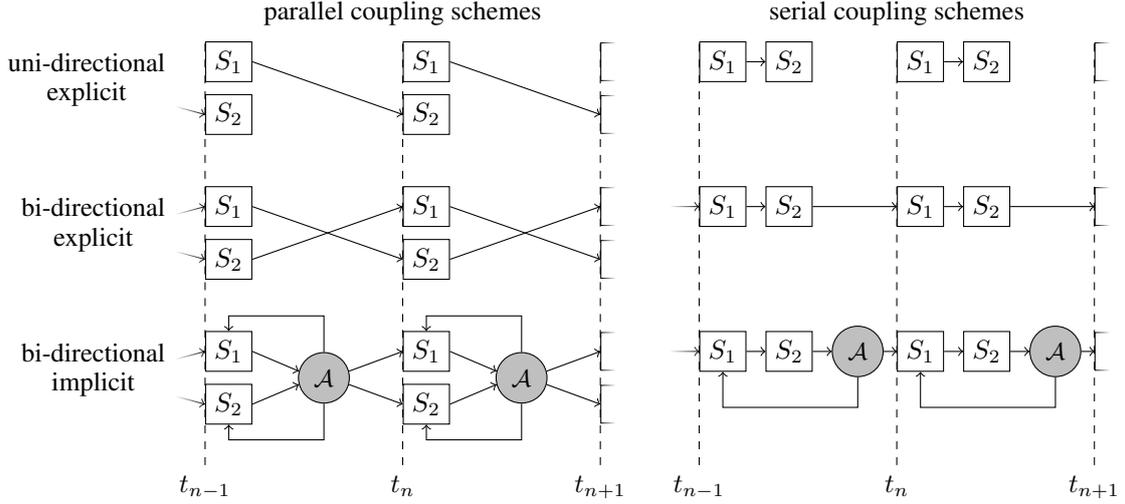
\begin{figure}[h!]
\centering
\begin{tikzpicture}[
scale=.8, 
every node/.style={minimum width=1cm, 
                   minimum height=0.5cm,
                   minimum size=0.5cm,
                   node distance=0.7cm and 0.25cm}]

\coordinate(origin1) at (0,0);
\coordinate(sizex) at ($(1,0)$);
\coordinate(sizey) at ($(0,0.6)$);
\coordinate(size) at ($(sizex)+(sizey)$);
\coordinate(step) at (3.25,0);
\coordinate(maxy) at (0,-7);

\coordinate(origin3) at ($(origin1)-4*(sizey)$);
\coordinate(origin5) at ($(origin3)-4*(sizey)$);

\coordinate(origin2) at ($(origin1)+2.5*(step)$);
\coordinate(origin4) at ($(origin2)-4*(sizey)$);
\coordinate(origin6) at ($(origin4)-4*(sizey)$);

\draw[dashed] ($(origin1)+.5*(sizey)$) -- ++ ($(maxy)$) node[below]{$t_{n-1}$};

\draw[dashed] ($(origin1)+.5*(sizey)+(step)$) -- ++ ($(maxy)$) node[below]{$t_{n}$};

\draw[dashed] ($(origin1)+.5*(sizey)+2*(step)$) -- ++ ($(maxy)$) node[below]{$t_{n+1}$};

\draw[dashed] ($(origin2)+.5*(sizey)$) -- ++ ($(maxy)$) node[below]{$t_{n-1}$};

\draw[dashed] ($(origin2)+.5*(sizey)+(step)$) -- ++ ($(maxy)$) node[below]{$t_{n}$};

\draw[dashed] ($(origin2)+.5*(sizey)+2*(step)$) -- ++ ($(maxy)$) node[below]{$t_{n+1}$};

\node[anchor=east,align=center] at ($(origin1)-0.5*(sizex)-0.5*(sizey)$) {uni-directional\\ explicit};

\node[anchor=east,align=center] at ($(origin3)-0.5*(sizex)-0.5*(sizey)$) {bi-directional\\ explicit};

\node[anchor=east,align=center] at ($(origin5)-0.5*(sizex)-0.5*(sizey)$) {bi-directional\\ implicit};

\node[anchor=south] at ($(origin1)+(step)+0.75*(sizey)$) {parallel coupling schemes};
\node[anchor=south] at ($(origin2)+(step)+0.75*(sizey)$) {serial coupling schemes};

\node[draw=black, anchor=west] (Suep11) at (origin1) {$S_1$};
\node[below of = Suep11, draw=black] (Suep21) {$S_2$};

\begin{scope}[opacity=1,transparency group] 
\path [scope fading=west] ($(origin1)+(sizey)$) rectangle ++ ($-.5*(sizex)-2*(sizey)$); 
\clip ($(origin1)+(sizey)$) rectangle ($(origin1)-.5*(sizex)+(maxy)$);  \draw[<-] (Suep21.west) -- ($(origin1)-(step)$);
\end{scope}

\node[draw=black, anchor=west] (Suep12) at ($(origin1)+(step)$) {$S_1$};
\node[below of = Suep12, draw=black] (Suep22) {$S_2$};

\draw[->] (Suep11.east) -- (Suep22.west);

\begin{scope}[opacity=1,transparency group] 
\path [scope fading=east] ($(origin1)+(sizey)+2*(step)$) rectangle ++ ($.25*(sizex)-2*(sizey)$); 
\clip ($(origin1)+2*(step)+.24*(sizex)+(sizey)$) rectangle ($(origin1)+2*(step)+(maxy)$);  \node[draw=black, anchor=west] (Suep13) at ($(origin1)+2*(step)$) {};
\node[below of = Suep13, draw=black] (Suep23) {};
\end{scope}

\draw[->] (Suep12.east) -- (Suep23.west);

\node[draw=black, anchor=west] (Sbep11) at (origin3) {$S_1$};
\node[below of = Sbep11, draw=black] (Sbep21) {$S_2$};

\begin{scope}[opacity=1,transparency group] 
\path [scope fading=west] ($(origin3)+(sizey)$) rectangle ++ ($-.5*(sizex)-2*(sizey)$); 
\clip ($(origin1)+(sizey)$) rectangle ($(origin1)-.5*(sizex)+(maxy)$);  \draw[<-] (Sbep21.west) -- ($(Sbep11.west)-(step)$);
\draw[<-] (Sbep11.west) -- ($(Sbep21.west)-(step)$);
\end{scope}

\node[draw=black, anchor=west] (Sbep12) at ($(origin3)+(step)$) {$S_1$};
\node[below of = Sbep12, draw=black] (Sbep22) {$S_2$};

\draw[->] (Sbep11.east) -- (Sbep22.west);
\draw[->] (Sbep21.east) -- (Sbep12.west);

\begin{scope}[opacity=1,transparency group] 
\path [scope fading=east] ($(origin3)+(sizey)+2*(step)$) rectangle ++ ($.25*(sizex)-2*(sizey)$); 
\clip ($(origin1)+2*(step)+.24*(sizex)$) rectangle ($(origin1)+2*(step)+(maxy)$);  \node[draw=black, anchor=west] (Sbep13) at ($(origin3)+2*(step)$) {};
\node[below of = Sbep13, draw=black] (Sbep23) {};
\end{scope}

\draw[->] (Sbep12.east) -- (Sbep23.west);
\draw[->] (Sbep22.east) -- (Sbep13.west);

\node[draw=black, anchor=west] (Sbip11) at (origin5) {$S_1$};
\node[below of = Sbip11, draw=black] (Sbip21) {$S_2$};

\begin{scope}[opacity=1,transparency group] 
\path [scope fading=west] ($(origin5)+(sizey)$) rectangle ++ ($-.5*(sizex)-2*(sizey)$); 
\clip ($(origin1)+(sizey)$) rectangle ($(origin1)-.5*(sizex)+(maxy)$);  \draw[<-] (Sbip21.west) -- ($(Sbip11.west)-(step)$);
\draw[<-] (Sbip11.west) -- ($(Sbip21.west)-(step)$);
\end{scope}

\coordinate(Sbip1mid) at ($.5*(Sbip11.east)+.5*(Sbip21.east)$);

\node[draw=black, fill=gray!50, circle, right = .6cm of Sbip1mid, font=\footnotesize](SbipAcc1){$\mathcal{A}$};

\draw[->] (Sbip11.east) -- (SbipAcc1.165);
\draw[->] (Sbip21.east) -- (SbipAcc1.195);

\coordinate(SbipAcc1topcorner) at ($(SbipAcc1.north)+(0,.6)$);
\coordinate(SbipAcc1bottomcorner) at ($(SbipAcc1.south)+(0,-.6)$);

\draw[->] (SbipAcc1.north) -- (SbipAcc1topcorner) -- (Sbip11.north|-SbipAcc1topcorner) -- (Sbip11.north);
\draw[->] (SbipAcc1.south) -- (SbipAcc1bottomcorner) -- (Sbip21.south|-SbipAcc1bottomcorner) -- (Sbip21.south);

\node[draw=black, anchor=west] (Sbip12) at ($(origin5)+(step)$) {$S_1$};
\node[below of = Sbip12, draw=black] (Sbip22) {$S_2$};

\draw[->] (SbipAcc1.15) -- (Sbip12.west);
\draw[->] (SbipAcc1.345) -- (Sbip22.west);

\coordinate(Sbip2mid) at ($.5*(Sbip12.east)+.5*(Sbip22.east)$);

\node[draw=black, fill=gray!50, circle, right = .6cm of Sbip2mid, font=\footnotesize](SbipAcc2){$\mathcal{A}$};

\draw[->] (Sbip12.east) -- (SbipAcc2.165);
\draw[->] (Sbip22.east) -- (SbipAcc2.195);

\coordinate(SbipAcc2topcorner) at ($(SbipAcc2.north)+(0,.6)$);
\coordinate(SbipAcc2bottomcorner) at ($(SbipAcc2.south)+(0,-.6)$);

\draw[->] (SbipAcc2.north) -- (SbipAcc2topcorner) -- (Sbip12.north|-SbipAcc2topcorner) -- (Sbip12.north);
\draw[->] (SbipAcc2.south) -- (SbipAcc2bottomcorner) -- (Sbip22.south|-SbipAcc2bottomcorner) -- (Sbip22.south);

\begin{scope}[opacity=1,transparency group] 
\path [scope fading=east] ($(origin5)+(sizey)+2*(step)$) rectangle ++ ($.25*(sizex)-2*(sizey)$); 
\clip ($(origin1)+2*(step)+.24*(sizex)$) rectangle ($(origin1)+2*(step)+(maxy)$);  \node[draw=black, anchor=west] (Sbip13) at ($(origin5)+2*(step)$) {};
\node[below of = Sbip13, draw=black] (Sbip23) {};
\end{scope}

\draw[->] (SbipAcc2.15) -- (Sbip13.west);
\draw[->] (SbipAcc2.345) -- (Sbip23.west);

\node[draw=black, anchor=west] (Sues11) at (origin2) {$S_1$};

\node[right = of Sues11, draw=black] (Sues21) {$S_2$};

\draw[->] (Sues11) -- (Sues21);

\node[draw=black, anchor=west] (Sues12) at ($(origin2)+(step)$) {$S_1$};

\node[right = of Sues12, draw=black] (Sues22) {$S_2$};

\draw[->] (Sues12) -- (Sues22);

\begin{scope}[opacity=1,transparency group] 
\path [scope fading=east] ($(origin2)+(sizey)+2*(step)$) rectangle ++ ($.25*(sizex)-2*(sizey)$); 
\clip ($(origin2)+2*(step)+.24*(sizex)+(sizey)$) rectangle ($(origin2)+2*(step)+(maxy)$);  \node[draw=black, anchor=west] (Sues13) at ($(origin2)+2*(step)$) {};
\end{scope}

\node[draw=black, anchor=west] (Sbes11) at (origin4) {$S_1$};

\begin{scope}[opacity=1,transparency group] 
\path [scope fading=west] ($(origin4)+.5*(sizey)$) rectangle ++ ($-.5*(sizex)-(sizey)$); 
\clip ($(origin2)+(sizey)$) rectangle ($(origin2)-.5*(sizex)+(maxy)$);  \draw[<-] (Sbes11.west) -- ($(origin4)-(step)$);
\end{scope}

\node[right = of Sbes11, draw=black] (Sbes21) {$S_2$};

\draw[->] (Sbes11) -- (Sbes21);

\node[draw=black, anchor=west] (Sbes12) at ($(origin4)+(step)$) {$S_1$};

\draw[->] (Sbes21) -- (Sbes12);

\node[right = of Sbes12, draw=black] (Sbes22) {$S_2$};

\draw[->] (Sbes12) -- (Sbes22);

\begin{scope}[opacity=1,transparency group] 
\path [scope fading=east] ($(origin4)+(sizey)+2*(step)$) rectangle ++ ($.25*(sizex)-2*(sizey)$); 
\clip ($(origin2)+2*(step)+.24*(sizex)$) rectangle ($(origin2)+2*(step)+(maxy)$);  \node[draw=black, anchor=west] (Sbes13) at ($(origin4)+2*(step)$) {};
\end{scope}

\draw[->] (Sbes22) -- (Sbes13);

\node[draw=black, anchor=west] (Sbis11) at (origin6) {$S_1$};

\begin{scope}[opacity=1,transparency group] 
\path [scope fading=west] ($(origin6)+.5*(sizey)$) rectangle ++ ($-.5*(sizex)-(sizey)$); 
\clip ($(origin2)+(sizey)$) rectangle ($(origin2)-.5*(sizex)+(maxy)$);  \draw[<-] (Sbis11.west) -- ($(origin6)-(step)$);
\end{scope}

\node[right = of Sbis11, draw=black] (Sbis21) {$S_2$};

\draw[->] (Sbis11) -- (Sbis21);

\node[draw=black, fill=gray!50, circle, right = of Sbis21, font=\footnotesize](SbisAcc1){$\mathcal{A}$};

\draw[->] (Sbis21) -- (SbisAcc1);

\coordinate(SbisAcc1bottomcorner) at ($(SbisAcc1.south)+(0,-0.5)$);
\draw[->] (SbisAcc1) -- (SbisAcc1bottomcorner) -- (SbisAcc1bottomcorner-|Sbis11) -- (Sbis11);

\node[draw=black, anchor=west] (Sbis12) at ($(origin6)+(step)$) {$S_1$};

\draw[->] (SbisAcc1) -- (Sbis12);

\node[right = of Sbis12, draw=black] (Sbis22) {$S_2$};

\draw[->] (Sbis12) -- (Sbis22);

\node[draw=black, fill=gray!50, circle, right = of Sbis22, font=\footnotesize](SbisAcc2){$\mathcal{A}$};

\draw[->] (Sbis22) -- (SbisAcc2);

\coordinate(SbisAcc2bottomcorner) at ($(SbisAcc2.south)+(0,-0.5)$);
\draw[->] (SbisAcc2) -- (SbisAcc2bottomcorner) -- (SbisAcc2bottomcorner-|Sbis12) -- (Sbis12);

\begin{scope}[opacity=1,transparency group] 
\path [scope fading=east] ($(origin6)+(sizey)+2*(step)$) rectangle ++ ($.25*(sizex)-2*(sizey)$); 
\clip ($(origin2)+2*(step)+.24*(sizex)$) rectangle ($(origin2)+2*(step)+(maxy)$);  \node[draw=black, anchor=west] (Sbis13) at ($(origin6)+2*(step)$) {};
\end{scope}

\draw[->] (SbisAcc2) -- (Sbis13);

\end{tikzpicture} \caption{
\label{fig:cpl_simple}
Different coupling options in preCICE for two participants $S_1$ and $S_2$ defined by combinations of (i) uni-directional or bi-directional (data transfer between two participants only in one or in both directions); (ii) explicit or implicit (execution of both participants once per time step or iterative solution of a fixed-point equation); (iii) parallel or sequential (simultaneous or one-after-the-other execution of two participants). $\mathcal{A}$ symbolizes a convergence acceleration method.}
\end{figure}

Coupling schemes and acceleration methods are at the very center of the preCICE core and define the coupling flow. As they have been studied in numerous publications (e.g.,~\cite{Mehl2016, Lindner2015_MVQN, Scheuf:QN}), we restrict the description to 
a short summary showcasing which combinations of coupling options and acceleration
schemes lead to robust and efficient partitioned simulations.
The coupling options can be configured at runtime. preCICE distinguishes: (i) \textit{uni-directional} or \textit{bi-directional} coupling, i.e., data dependencies between the participants in one direction only (example: full flow simulation coupled to an acoustic far field, where the acoustic far fields receives background velocity and pressure values as well as acoustic perturbations at the coupling interface, but we do not observe acoustic waves traveling back into the flow region) or data dependencies in both directions (e.g., fluid-structure interaction); (ii) \textit{explicit} or \textit{implicit} coupling, i.e., execution of each participant once per time step or execution of multiple iterations per time step, such that the values at the end of the time step fulfill all coupling conditions; and (iii) \textit{parallel} or \textit{serial} coupling, i.e., simultaneous
or one-after-the-other execution of participants.
Uni-directional coupling requires data transfer only from one participant to the other. Thus, only an explicit coupling makes sense in this case, which can, however, be both serial or parallel. For bi-directional coupling, we have four different coupling scheme options: (i) parallel-explicit, (ii) serial-explicit, (iii) parallel-implicit, (iv) serial-implicit. We show the six different resulting coupling schemes in
 \autoref{fig:cpl_simple}.
In the following, we focus on two non-trivial aspects of coupling: the coupling of more than two participants (multi-code coupling) and the choice of suitable convergence acceleration schemes for implicit coupling.

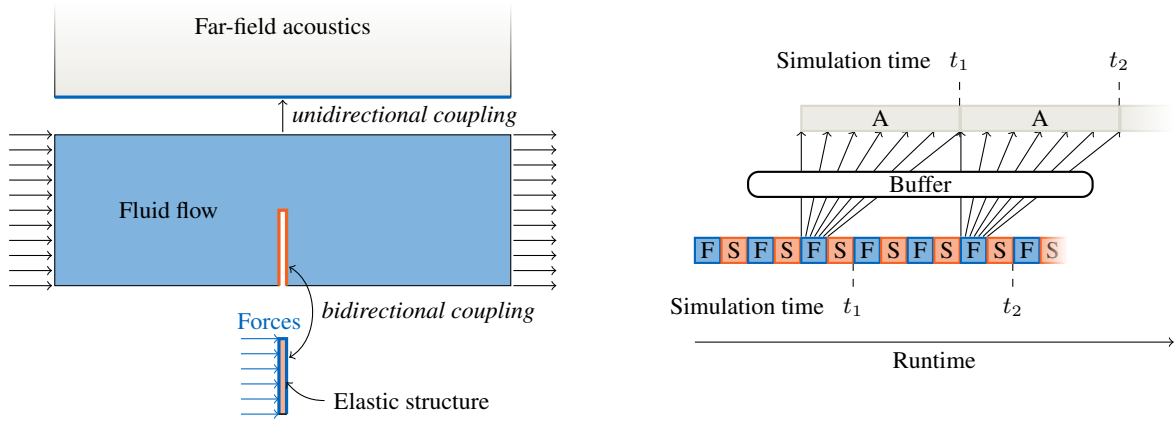
\begin{figure}[h!]
  \centering
  \begin{minipage}{0.48\textwidth}
  \begin{tikzpicture}[scale=1, every node/.style={font=\small}]

\coordinate(origin) at (0,0);

\coordinate(lengthFlap) at (0,1);
\coordinate(thicknessFlap) at (0.1,0);

\coordinate(leftTopFlap) at ($(origin) + (0,-2em)$);
\coordinate(rightTopFlap) at ($(leftTopFlap)+(thicknessFlap)$);
\coordinate(leftBottomFlap) at ($(leftTopFlap)-(lengthFlap)$);
\coordinate(rightBottomFlap) at ($(rightTopFlap)-(lengthFlap)$);

\coordinate(distance) at (0,.5);
\coordinate(thicknessFarField) at (0,1.25);

\coordinate(channelWidth) at (0,2);
\coordinate(channelPreFlap) at (-2.95,0);
\coordinate(channelPostFlap) at (2.95,0);

\coordinate(leftBottomInflow) at ($(origin)+(channelPreFlap)$);
\coordinate(leftTopInflow) at ($(leftBottomInflow)+(channelWidth)$);

\coordinate(rightBottomOutflow) at ($(origin)+(thicknessFlap)+(channelPostFlap)$);
\coordinate(rightTopOutflow) at ($(rightBottomOutflow)+(channelWidth)$);

\coordinate(leftBottomFarField) at ($(leftTopInflow)+(distance)$);
\coordinate(rightBottomFarField) at ($(rightTopOutflow)+(distance)$);
\coordinate(leftTopFarField) at ($(leftBottomFarField)+(thicknessFarField)$);
\coordinate(rightTopFarField) at ($(rightBottomFarField)+(thicknessFarField)$);

\draw[fill=participantBcolor!50](leftBottomInflow) -- (origin) -- ($(origin)+(lengthFlap)$) -- ($(origin)+(thicknessFlap)+(lengthFlap)$) -- ($(origin)+(thicknessFlap)$) -- (rightBottomOutflow) -- (rightTopOutflow) -- (leftTopInflow) -- cycle;
\node[] at ($.375*(leftBottomInflow)+.375*(leftTopInflow)+.125*(rightBottomOutflow)+.125*(rightTopOutflow)$){Fluid flow};
\draw[fill=participantAcolor!50](leftTopFlap) -- (rightTopFlap) -- (rightBottomFlap) -- (leftBottomFlap) -- cycle;
\node[anchor=west, align=left](FEniCSLabel) at ($.25*(rightTopFlap)+.75*(rightBottomFlap)+(.5,-.2em)$){Elastic structure};

\draw[participantBcolor,very thick](leftBottomFlap) -- (leftTopFlap) -- (rightTopFlap) -- (rightBottomFlap);  \coordinate(midStructure) at ($.25*(leftBottomFlap)+.25*(leftTopFlap)+.25*(rightTopFlap)+.25*(rightBottomFlap)$);

\begin{scope}[opacity=1,transparency group] 
\path [scope fading=fade top] (leftBottomFarField) rectangle (rightTopFarField); 
\draw[fill=ci_ivory!50, path fading=fade top, draw=none] (leftBottomFarField) rectangle (rightTopFarField);
\draw[black] (leftTopFarField) -- (leftBottomFarField) -- (rightBottomFarField) -- (rightTopFarField);
\draw[black, dotted] (leftTopFarField) -- (rightTopFarField);
\end{scope}

\draw[participantBcolor, very thick] (leftBottomFarField) -- (rightBottomFarField);

\node[] at ($.125*(leftBottomFarField)+.125*(rightBottomFarField)+.375*(rightTopFarField)+.375*(leftTopFarField)$){Far-field acoustics};

\path [name path=A--B](FEniCSLabel.west) to[in=-60, out=180] (midStructure);
\path [name path=C--D](rightTopFlap) -- (rightBottomFlap);
\path [name intersections={of=A--B and C--D,by=E}];
\draw [->](FEniCSLabel.west) to[in=-50, out=180] (E);

\draw[participantAcolor,very thick](origin) -- ($(origin)+(lengthFlap)$) -- ($(origin)+(lengthFlap)+(thicknessFlap)$) -- ($(origin)+(thicknessFlap)$);  \draw(leftBottomFlap) -- (rightBottomFlap);  

\foreach \i in {0,...,5}{\draw[->,participantBcolor] ([yshift=\i * 0.2cm, xshift = -.5cm]leftBottomFlap) -- ([yshift=\i * .2cm]leftBottomFlap);}
      \path([xshift = -.25cm]leftTopFlap) -- node[above, participantBcolor]{Forces} (leftTopFlap);
      
\foreach \i in {0,...,10}{\draw[->] ([yshift=\i * 0.2cm, xshift = -.6cm]leftBottomInflow) -- ([yshift=\i * .2cm, xshift = -.1em]leftBottomInflow);}
\foreach \i in {0,...,10}{\draw[->] ([yshift=\i * 0.2cm, xshift = +.1em]rightBottomOutflow) -- ([yshift=\i * .2cm, xshift = +.6cm]rightBottomOutflow);}      
      
\draw[<->] ($(origin)+(thicknessFlap)+.25*(lengthFlap)+(.1em,0)$) to[in=30, out=-30] node[right]{\textit{bidirectional coupling}}($.25*(rightBottomFlap)+.75*(rightTopFlap)+(.1em,0)$);

\draw[->] ($.5*(rightTopOutflow)+.5*(leftTopInflow)+(0,.1em)$) -- node[right]{\textit{unidirectional coupling}}($.5*(leftBottomFarField)+.5*(rightBottomFarField)-(0,.1em)$);

\end{tikzpicture}     \end{minipage}
    \hfill
  \begin{minipage}{0.48\textwidth}
    \begin{tikzpicture}[scale=.35, every node/.style={font=\small}]

\coordinate(origin) at (0,0);
\coordinate(sizex) at ($(1,0)-(.05, 0)$);
\coordinate(sizey) at ($(0,1)-(0, .05)$);
\coordinate(size) at ($(sizex)+(sizey)$);
\coordinate(step) at (1,0);
\coordinate(distance) at (0,5);

\draw[draw=participantBcolor, thick, fill=participantBcolor!50] ($(origin)$) rectangle node{F} ++(size);
\draw[draw=participantAcolor, thick, fill=participantAcolor!50] ($(origin)+(step)$) rectangle node{S} ++(size);
\draw[draw=participantBcolor, thick, fill=participantBcolor!50] ($(origin)+2*(step)$) rectangle node{F} ++(size);
\draw[draw=participantAcolor, thick, fill=participantAcolor!50] ($(origin)+3*(step)$) rectangle node{S} ++(size);

\draw[->] ($(origin)-(0,3)$) -- node[below]{Runtime} ++($18*(step)$);

\draw[dashed] ($(origin)+(sizex)+5*(step)$) -- ++ ($(0,-1)$) node[below, label={west:Simulation time\vphantom{$t_1$}}]{$t_1$};

\foreach \i in {0,...,6}{\draw[->] ([xshift = \i * 0.166cm]$(origin)+4*(step)+(0,1)$) -- ([yshift=4cm, xshift = \i * 1cm]$(origin)+4*(step)+(0,1)$);}
      
\draw[draw=ci_ivory, thick,fill=ci_ivory!50] ($(origin)+(distance)+4*(step)$) rectangle node{A} ++($(size)+5*(step)$);

\draw[dashed] ($(origin)+(distance)+9*(step)+(size)$) -- ++ ($(0,1)$)  node[above, label={west:Simulation time\vphantom{$t_1$}}]{$t_1$};

\draw[draw=participantBcolor, thick,fill=participantBcolor!50] ($(origin)+4*(step)$) rectangle node{F} ++(size);
\draw[draw=participantAcolor, thick,fill=participantAcolor!50] ($(origin)++5*(step)$) rectangle node{S} ++(size);
\draw[draw=participantBcolor, thick,fill=participantBcolor!50] ($(origin)+6*(step)$) rectangle node{F} ++(size);
\draw[draw=participantAcolor, thick, fill=participantAcolor!50] ($(origin)+7*(step)$) rectangle node{S} ++(size);
\draw[draw=participantBcolor, thick,fill=participantBcolor!50] ($(origin)+8*(step)$) rectangle node{F} ++(size);
\draw[draw=participantAcolor, thick, fill=participantAcolor!50] ($(origin)+9*(step)$) rectangle node{S} ++(size);

\draw[dashed] ($(origin)+11*(step)+(sizex)$) -- ++ ($(0,-1)$) node[below]{$t_2$};

\foreach \i in {0,...,6}{\draw[->] ([xshift = \i * 0.166cm]$(origin)+10*(step)+(0,1)$) -- ([yshift=4cm, xshift = \i * 1cm]$(origin)+10*(step)+(0,1)$);}
      
\draw[draw=ci_ivory, thick,fill=ci_ivory!50] ($(origin)+(distance)+10*(step)$) rectangle node{A} ++($(size)+5*(step)$);

\draw[dashed] ($(origin)+(distance)+15*(step)+(size)$) -- ++ ($(0,1)$) node[above]{$t_2$};

\draw[draw=participantBcolor, thick,fill=participantBcolor!50] ($(origin)+10*(step)$) rectangle node{F} ++(size);
\draw[draw=participantAcolor, thick, fill=participantAcolor!50] ($(origin)+11*(step)$) rectangle node{S} ++(size);
\draw[draw=participantBcolor, thick, fill=participantBcolor!50] ($(origin)+12*(step)$) rectangle node{F} ++(size);
\draw[draw=participantAcolor, thick, fill=participantAcolor!50, path fading=fade right] ($(origin)+13*(step)$) rectangle node{S} ++(size);

\draw[draw=ci_ivory, thick,fill=ci_ivory!50, path fading=fade right] ($(origin)+(distance)+16*(step)$) rectangle ++($(size)+1*(step)$);

\draw[draw=black, fill=white, rounded corners,  thick] ($(origin)+2*(step)+.5*(distance)$) rectangle node{Buffer} ++($(size)+12*(step)$);

\end{tikzpicture}     \end{minipage}
  \caption{
    \label{fig:exafsa} Example for a pairwise multi-code coupling: bi-directional implicit sequential coupling between a structure solver and a flow solver (assuming three iterations per time step) and uni-directional explicit parallel coupling between the flow solver and an acoustics solver.~\cite{Lindner2020_ExaFSA} presents more details on the required data buffering allowing to achieve parallel efficiency by overlapping the acoustic far field solver with the fluid-structure iterations of the next time step.}
\end{figure}

\paragraph{Multi-code coupling} 
For multi-code coupling, coupling schemes can be configured in two different ways: (i) for each pair of participants separately or (ii) as an overall multi-coupling scheme. For (i), theoretically, all combinations of coupling options are possible. However, combinations of several pairwise implicit coupling schemes have been shown to be numerically unstable~\cite{Bungartz2014_MIQN_CompuMech}. Still, pairwise coupling can be the best option for some combinations of explicit and implicit coupling. One such example is the extension of the fluid-acoustics example mentioned above shown in~\cite{Lindner2020_ExaFSA}: a bi-directional, implicit, and serial coupling of a structure solver with a flow solver, combined with a uni-directional, explicit, and parallel coupling of the flow solver to the acoustics solver (see \autoref{fig:exafsa}). To equally balance computational load, this overall scheme requires buffering of the data to be communicated to the acoustics solver, which is another feature provided by preCICE (introduced in~\cite{Lindner2020_ExaFSA}). In contrast to pairwise coupling, for multi-coupling schemes, the only reasonable realization is parallel coupling, i.e., their combined input and output can be used in the coupling acceleration for implicit coupling as described below.

\paragraph{Acceleration of implicit coupling iterations} 
Implicit parts of the coupling schemes described above always require solving a fixed-point equation
\begin{equation*}
H(x) = x.
\end{equation*}
To give some example, consider serial coupling of two participants $S_1: x_1 \mapsto x_2$ and $S_2: x_2 \mapsto x_1$ or the parallel multi-coupling of three participants $S_1: (x_2, x_3) \mapsto (y_2, y_3)$, $S_2: y_2 \mapsto x_3$, and $S_3: y_3 \mapsto x_3$. The corresponding fixed-point equations read
\begin{eqnarray*}
  S_2 \circ S_1(x_1) = x_1 \; \mbox{ and } \;
  \left\{ \begin{array}{ccc}
    S_1(x_2,x_3) & = & (y_2, y_3), \\
    S_2(y_2) & = &  x_2, \\
    S_3(y_3) & = & x_3. 
  \end{array} \right\}
\end{eqnarray*}

If multiple coupling data vectors are combined in a single fixed-point equation as in the last example, it is numerically beneficial to bring all data to the same scale by an automatic weighting (called \textit{preconditioner} in preCICE, see~\cite{Uekermann2016} for details). 

The pure fixed-point iterations can be enhanced with an accelerator $\mathcal{A}$, which uses all input and output information of the operator $H$ collected in previous iterations:
\begin{equation*}
x^{k+1}  =  \mathcal{A}(x^0,\ldots, x^k, H(x^0), \ldots, H(x^k)).
\end{equation*}

The simplest acceleration scheme is Aitken's under-relaxation as presented, e.g., in~\cite{Kuettler2008}.
It reuses only information from the last iteration. For most applications, quasi-Newton schemes  
\begin{equation}
\mathcal{A}(x^0,\ldots, x^k, H(x^0), \ldots, H(x^k)) = \tilde{x}^k + \left( W_k - J^{\text{prev}} V_k\right) \alpha^k - J^{\text{prev}} R(x^k) \label{eq:QN}
\end{equation}
are significantly more efficient and robust (e.g.,~\cite{Uekermann2016}). Here, $\alpha^k$ is a coefficient vector, $\tilde{x}^k := H(x^k)$, $R(x^k) := H(x^k)-x^k$.
We use the matrix $J^{\text{prev}}$ to include knowledge about the inverse Jacobian already achieved in previous time steps (similar to~\cite{Bogaers2014_MVQN}).
In the classical quasi-Newton methods for fluid-structure interaction as introduced in~\cite{Degroote2009}, $J^{\text{prev}}$ is zero. 
The matrices $V_k$ and $W_k$ collect residual and value differences throughout previous iterations:
\begin{eqnarray*}
  ~~~~~ W_k = \left[\Delta \tilde{x}^k_0, \Delta
  \tilde{x}^k_{1}, \cdots , \Delta \tilde{x}^k_{k-1}\right],   &\mbox{with}
                                                               ~~  &\Delta \tilde{x}^k_i = \tilde{x}^k - \tilde{x}^i \;,\\
  ~~~~~ V_k = \left[\Delta r^k_0, \Delta r^k_{1}, \cdots
  , \Delta r^k_{k-1}\right],   &\mbox{with} ~~  &\Delta r^k_i = R(x^k) -
                                                  R(x^i)\;
\end{eqnarray*} 
with the number $k$ of iterations done so far\footnote{$k\geq 1$. A plain or underrelaxed fixed-point iteration can be used to start the procedure.}. 
In practice, the leftmost columns of $V_k$ and $W_k$ can always be dropped in cases where several iterations ($k$) are required for convergence.

Equation \eqref{eq:QN} is an approximation of the modified Newton iteration
\[ x^{k+1} = \tilde{x}^k - J_{\tilde{R}}^{-1} R(x^k), \]
where $J^{-1}_{\tilde{R}}$ is the inverse Jacobian of $\tilde{R}: \tilde{x}^i \mapsto R(x^i)$. To derive Equ.~\eqref{eq:QN}, $J^{-1}_{\tilde{R}}$ is approximated by the solution $J_k^{-1} = (W_k - J^{\text{prev}} V_k) V_k^T (V_k V_k^T)^{-1} + J^{\text{prev}}$ of the multi-secant equation\footnote{corresponding to the coefficient vector $\alpha_k = - V_k^T (V_k V_k^T)^{-1} R(x^k)$.}
\begin{equation}
J_k^{-1} V^k = W^k \; \mbox{ under the norm minimization } \; \| J_k^{-1} - J^{\text{prev}} \|_F \leftarrow \text{min}, \label{eq:multi_secant}
\end{equation}
where $\| \cdot \|_F$ denotes the Frobenius norm. In the following, we shortly present the main two quasi-Newton classes, IQN-ILS and IQN-IMVJ, provided by preCICE and introduce the so-called filtering that can improve the robustness of both. 

\paragraph{IQN-ILS} For $J^{\text{prev}}=0$, we get the classical \textit{interface quasi-Newton inverse least squares} method as introduced in~\cite{Degroote2009}. For this approach, re-using information from previous time steps by adding further columns to $V_k$ and $W_k$ can help speed up the coupling iterations significantly. However, the optimal number of reused time steps strongly depends on the involved equations, on the discretization of the respective fields and even their mesh resolution~\cite{Scheuf:QN}.

\paragraph{IQN-IMVJ} For $J^{\text{prev}}$ chosen as the last inverse Jacobian approximation of the previous time step, an idea adopted from~\cite{Bogaers2014_MVQN} in~\cite{Lindner2015_MVQN,Scheufele2015}, we get the method called \textit{interface quasi-Newton inverse multi-vector Jacobian}. A variety of re-start mechanism allows us to implement this method with linear complexity in the number of coupling unknowns by avoiding storing the full matrix $J^{\text{prev}}$. Only low-rank additive components are stored and the respective sum is re-set after a chunk of time steps. The size of these chunks is configurable by the user, but its influence on the convergence behavior is not very significant~\cite{Scheuf:QN,Scheufele2019_Diss}.

\paragraph{Filtering} Since linear dependencies of columns in $V_k$ can not be avoided in both IQN-ILS and IQN-IMVJ, we implemented various filtering algorithms, which automatically delete columns that cause (near) linear dependencies. This eliminates both contradicting and outdated information~\cite{Haelterman2015_Filtering}.

All quasi-Newton variants are implemented in a fully parallel way based on parallel $QR$-solvers for the calculation of components of $J_k^{-1}$ and the matrix-vector product $J_k^{-1} R(x^k)$. We do not present numerical results in this section, but refer to~\cite{Mehl2016, Scheuf:QN,Uekermann2016} for detailed comparisons of quasi-Newton variants and examples showcasing their efficiency and robustness. 

\subsection{Data mapping} 
\label{ssec:mapping}

In a coupled simulation, participants exchange data via coupling meshes. 
The coupling meshes of each pair of participants discretize either the common coupling interface (for surface coupling) or a common volume (for volume coupling).
However, the discretization approaches of the two participants are usually different, leading to non-matching meshes.
In order to transfer physical variables between these non-matching meshes, we use data mappings.

Let the dimension of the scenario $d$ be two or three and let us consider a data mapping from the coupling mesh of participant $S_1$ to the coupling mesh of participant $S_2$. 
The minimum mesh information required (during initialization) is the vertex coordinates of both meshes, which we define as:
\[
\mathcal{M}_{S_1} = \lbrace \vec{x}_{1}^{S_1}, \ldots, \vec{x}_{n_{1}}^{S_1} \rbrace \quad \mbox{with} \quad \vec{x}_{i}^{S_1} \in \mathbb{R}^{d}, \quad 
\mathcal{M}_{S_2} = \lbrace \vec{x}_{1}^{S_2}, \ldots, \vec{x}_{n_2}^{S_2} \rbrace \quad \mbox{with} \quad \vec{x}_{i}^{S_2} \in \mathbb{R}^{d}.
\]
Data mapping aims to map the vector $\vec{v}^{S_1} = (v_1^{S_1}, v_2^{S_1}, \ldots, v_{n_1}^{S_1})^T$ of values at the vertices in $\mathcal{M}_{S_1}$ to the vector $\vec{v}^{S_2} = ( v_1^{S_2}, v_2^{S_2}, \ldots, v_{n_2}^{S_2})^T$ of values at the vertices in $\mathcal{M}_{S_2}$. 

There are currently three mapping methods available in preCICE: nearest-neighbor mapping, nearest-projection mapping, and data mapping using radial basis functions. Nearest-projection mapping is only available for surface coupling and requires the definition of so-called \textit{mesh connectivity} information: edges that form connections between vertices and in 3D additionally triangles or quads made from edges.
All three mapping methods can be either consistent or conservative. Consistent mapping operations exactly reproduce constant data at $\mathcal{M}_{S_1}$ on $\mathcal{M}_{S_2}$. Conservative mapping methods preserve the sum of all values. Consistent mapping schemes are, thus, used for physical variables such as displacements, velocities, pressure, or stresses, whereas conservative methods have to be used for cumulative variables, such as forces. 
Data mapping can be written as a linear mapping
\[ M \vec{v}^{S_1} = \vec{v}^{S_2} \]
with a matrix $M \in \mathbb{R}^{n_2 \times n_1}$. For a consistent mapping, the sum of entries in each row of the mapping matrix $M$ has to be one, whereas, for a conservative mapping, the requirement is $\sum_i v_i^ {S_1} = \sum_i v_i^{S_2}$, and thus, the sum of entries in each column of $M$ has to be one. Therefore, conservative mapping methods are generated by transposing the 
mapping matrix of a consistent mapping.  
Throughout this section, we restrict our explanation to consistent data mapping.  
For in depth details on both variants, the reader is referred to~\cite{DeBoer2008_ComparisonMapping, Gatzhammer2015, Lindner2019}.

\subsubsection{Projection-based data mapping}
\label{map:projectionMethods}

Two projection-based mappings are available in preCICE: nearest-neighbor and nearest-projection. 
The nearest neighbor mapping establishes an association between each vertex $\vec{x}_i^{S_2}$ of the output mesh $\mathcal{M}_{S_2}$, with the spatially nearest vertex $\vec{x}_{j(i)}^{S_1}$ on the input mesh $\mathcal{M}_{S_1}$.

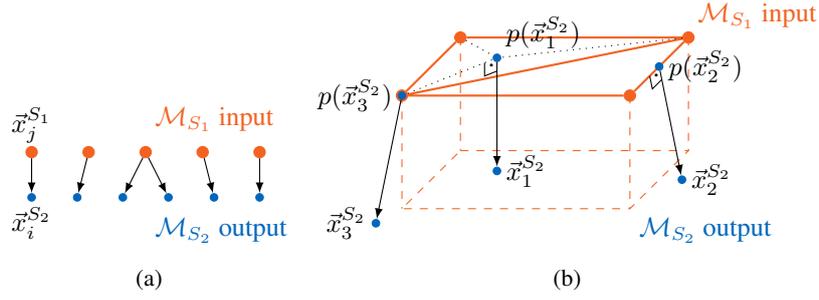
\begin{figure}[h!]
  \centering
  \subcaptionbox{}{\begin{tikzpicture}[x=15mm, y=15mm]
      \foreach \x in {0,1,...,4} {
        \node[inNode] (in\x) at (\x/2, 0) {};
      }
      \foreach \x in {0,1,...,5} {
        \node[outNode] (out\x) at (\x/2.5, -0.4) {};
      }
      \node[anchor=south west] at (1,  0.1) {\textcolor{incolor}{$\mathcal{M}_{S_1}$ input}};
      \node[anchor=north west] at (1, -0.5) {\textcolor{outcolor}{$\mathcal{M}_{S_2}$ output}};

      \node[anchor=south] at (in0)  {$\vec{x}_j^{S_1}$};
      \node[anchor=north] at (out0)  {$\vec{x}_i^{S_2}$};

      \draw[-latex] (in0) -- (out0);
      \draw[-latex] (in1) -- (out1);
      \draw[-latex] (in2) -- (out2);
      \draw[-latex] (in2) -- (out3);
      \draw[-latex] (in3) -- (out4);
      \draw[-latex] (in4) -- (out5);
    \end{tikzpicture}
  }
  \subcaptionbox{}{\begin{tikzpicture}[x=15mm, y=15mm]
\draw[dashed,incolor] (0, -1, 0) -- (2, -1, 0) -- (2, -1, 2) -- (0, -1, 2) -- cycle;
      \foreach \x in {0, 2} \foreach \y in {0, 2} {
        \draw[dashed,incolor] (\x, -0, \y) -- (\x, -1, \y);
      }

\node[inNode] (in0) at (0, 0, 0) {};
      \node[inNode] (in1) at (2, 0, 0) {};
      \node[inNode] (in2) at (2, 0, 2) {};
      \node[inNode] (in3) at (0, 0, 2) {};

\draw[thick, incolor] (in0) -- (in1) -- (in2) -- (in3) -- (in0) (in1) -- (in3);

      \node[anchor=south west] at (2,0,0) {\textcolor{incolor}{$\mathcal{M}_{S_1}$ input}};
      \node[anchor=north west] at (2,-1,2) {\textcolor{outcolor}{$\mathcal{M}_{S_2}$ output}};

\path (0.5, 0, 0.7) 
        node[projNode] (tp) {} 
        node[above right] {$p(\vec{x}_1^{S_2})$}
        +(0,-1,0)
        node[outNode] (to) {} 
        node[right] {$\vec{x}_1^{S_2}$};
      \draw[-latex] (tp) -- (to);
      \draw pic [draw, angle radius=5pt, angle eccentricity=.5,pic text=.] { right angle = in3--tp--to };
      \draw[thin, dotted] (in0) -- (tp);
      \draw[thin, dotted] (in1) -- (tp);
      \draw[thin, dotted] (in3) -- (tp);

\path (2, 0, 1) 
        node[projNode] (ep) {} 
        node[right] {$p(\vec{x}_2^{S_2})$}
        -- +(0.2,-1,0)
        node[outNode] (eo) {} 
        node[right] {$\vec{x}_2^{S_2}$} ;
      \draw[-latex] (ep) -- (eo);
      \draw pic [draw, angle radius=5pt, angle eccentricity=.5,pic text=.] { right angle = in2--ep--eo };

\path (0, 0, 2) 
        node[projNode] (vp) {} 
        node[left] {$p(\vec{x}_3^{S_2})$}
        -- +(-0.1,-1,0.5)
        node[outNode] (vo) {} 
        node[left] {$\vec{x}_3^{S_2}$};
      \draw[-latex] (vp) -- (vo);
    \end{tikzpicture}
  }
  \caption{\label{fig:nn_np}Basic principles of nearest-neighbor and nearest-projection mapping: (a) Transfer of each value $v_{j(i)}^{S_1}$ at the nearest neighbor $\vec{x}_{j(i)}^{S_1}$ in the coupling mesh of $S_1$ to the vertex $\vec{x}_i^{S_2}$ in the coupling mesh of $S_2$. (b) Projection of points $\vec{x}_1^{S_2}$, $\vec{x}_2^{S_2}$, and $\vec{x}_3^{S_2}$ of the coupling mesh of $S_2$ to a triangle, an edge, and a vertex, respectively, of the coupling mesh of $S_1$.}
\end{figure}

The mapping is then simply defined as
\[ v_{i}^{S_2} = v_{j(i)}^{S_1}. \] 
The nearest projection mapping uses connectivity information between multiple vertices on the input mesh to interpolate to a vertex on the output mesh. To calculate the value at an output vertex $\vec{x}_j^{S_2}$, we calculate a projection point $p(\vec{x}_j^{S_2})$ on the entities of the input mesh, interpolate a value to this projection point and copy this to the output vertex $\vec{x}_j^{S_2}$. The projection point $p(\vec{x}_j^{S_2})$ is a projection on a triangle of the input mesh defined by participant ${S_1}$. If such a triangle does not exist, we determine $p(\vec{x}_j^{S_2})$ via orthogonal projection to an edge in the coupling mesh of participant ${S_1}$ or, as the last option, as the closest vertex in the mesh of participant ${S_1}$. This requires a search operation over triangles and potentially edges and vertices of the mesh of participant ${S_1}$ for each output vertex $\vec{x}_j^{S_2}$. See \autoref{fig:nn_np} for the relation between mesh entities of the two meshes. In the second step of the mapping, we use barycentric interpolation (if $p(\vec{x}_j^{S_2})$ is in a triangle), linear interpolation (if $p(\vec{x}_j^{S_2})$ is on an edge) or the respective vertex value (if $p(\vec{x}_j^{S_2})$ is a vertex) to determine a value at $p(\vec{x}_j^{S_2})$ and then use this value as $v_j^{S_2}$ at the output point. In other words, the interpolation is a combination of a second-order accurate interpolation inside a triangle in $S_1$ and a first-order accurate extrapolation in normal direction, i.e., the error is $\mathcal{O}(h^2) + \mathcal{O}(\delta)$ with the mesh width $h$ of the mesh of $S_1$ and the normal distance $\delta$ between the two coupling meshes.  
In cases where the mesh of participant ${S_1}$ does not provide suitable mesh connectivity information, a simple projection onto the closest vertex is performed, i.e., the nearest-projection mapping falls back to nearest-neighbor.

Both nearest-neighbor and nearest-projection mapping require neighbor search between mesh entities of both participants. 
To implement this search efficiently, we generate r-start index-trees for vertices, edges, and triangles of meshes using the Geometry package of Boost\footnote{Boost C++ libraries: \url{https://boost.org/}.}.
The complexity of generating the index tree is $O(n \log(n))$ and the complexity of each nearest neighbor query is $O(\log(n))$ if $n$ is the number of entities in the involved meshes. 

\subsubsection{Data mapping with radial basis functions}

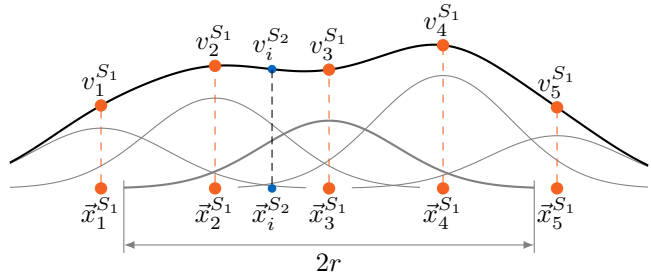
\begin{figure}[h!]
  \centering
\begin{tikzpicture}[x=15mm, y=15mm]
\draw[gray, shift={(1, 0)}, domain=-0.8:1.8,samples=200] plot(\x,{0.8*exp(-pow(\x/0.6, 2)/2)/(0.6*sqrt(2*pi))});
    \draw[gray, shift={(2, 0)}, domain=-1.8:1.8,samples=200] plot(\x,{1.2*exp(-pow(\x/0.6, 2)/2)/(0.6*sqrt(2*pi))});
    \draw[gray, thick, shift={(3, 0)}, domain=-1.8:1.8,samples=200] plot(\x,{0.9*exp(-pow(\x/0.6, 2)/2)/(0.6*sqrt(2*pi))});
    \draw[gray, shift={(4, 0)}, domain=-1.8:1.8,samples=200] plot(\x,{1.5*exp(-pow(\x/0.6, 2)/2)/(0.6*sqrt(2*pi))});
    \draw[gray, shift={(5, 0)}, domain=-1.8:0.8,samples=200] plot(\x,{0.7*exp(-pow(\x/0.6, 2)/2)/(0.6*sqrt(2*pi))});

\coordinate (mgl) at (1.2,0);
    \coordinate (mgr) at (4.8,0);
    \coordinate (sr) at (0,-0.5);

    \draw[gray, thin] (mgl) +(0,3pt) -- (mgl|-sr) -- +(0,-3pt);
    \draw[gray, thin] (mgr) +(0,3pt) -- (mgr|-sr) -- +(0,-3pt);
    \draw[gray, thin, latex-latex] (mgl|-sr) -- node[black,midway, below] {$2r$} (mgr|-sr);

\draw[thick, name path=in, shift={(0, 0)}, domain=0.2:5.8,samples=200] plot(\x,{0.8*exp(-pow((\x-1)/0.6, 2)/2)/(0.6*sqrt(2*pi)) +
        1.2*exp(-pow((\x-2)/0.6, 2)/2)/(0.6*sqrt(2*pi)) +
        0.9*exp(-pow((\x-3)/0.6, 2)/2)/(0.6*sqrt(2*pi)) +
        1.5*exp(-pow((\x-4)/0.6, 2)/2)/(0.6*sqrt(2*pi)) +
        0.7*exp(-pow((\x-5)/0.6, 2)/2)/(0.6*sqrt(2*pi))});

\foreach \i in {1,2,3,4,5} {
      \node[inNode] (in\i) at (\i, 0) {};
      \node[anchor=north] at (in\i) {$\vec{x}_\i^{S_1}$};

      \path[name path=h\i] (\i,0) -- (\i,2);
      \path[name intersections={of=in and h\i, by=p\i}];
      \node[inNode] at (p\i) {};
      \node[anchor=south] at (p\i) {$v_\i^{S_1}$};
      \draw[dashed, incolor] (in\i) -- (p\i);
    }

    \node[outNode] (inI) at (2.5, 0) {};
    \node[anchor=north] at (inI) {$\vec{x}_i^{S_2}$};

    \path[name path=hI] (inI) -- +(0,2);
    \path[name intersections={of=in and hI, by=pI}];

    \node[projNode] at (pI) {};
    \draw[densely dashed] (inI) -- (pI);
    \node[anchor=south] at (pI) {$v_i^{S_2}$};
  \end{tikzpicture}
  \caption{\label{fig:rbf}Simplified one-dimensional view of the generation of the interpolant and the evaluation at the target point $\vec{x}_i^{S_2}$ based on a linear combination of Gaussian basis functions with support radius $r$, neglecting the global polynomial.}
\end{figure}

Radial basis function (RBF) mapping uses a linear combination of radially symmetric basis functions centered at vertices $\vec{x}^{S_1}_{i}$ of the input mesh to create a global interpolation function, which is afterwards sampled at the vertices of the output mesh $\vec{x}^{S_1}_{i}$. In order to ensure that constant and linear functions are interpolated exactly, an additional global first-order polynomial term $q(\vec{x})$ is added to the interpolant $s : \mathbb{R}^{d} \rightarrow \mathbb{R}$:

\begin{equation*}
s \left( \vec{x} \right) = \sum\limits_{i=1}^{n_1} \lambda_{i} \cdot \phi \left( \| \vec{x} - \vec{x}^{S_1}_{i} \| _{2}\right) + q \left( \vec{x} \right) ,
\end{equation*}

where the radial basis function is given by $\phi$, and the polynomial term $q \left( \vec{x} \right) = \beta_{0} + \beta_{1} x_{1} + \ldots + \beta_{d} x_{d}$.
Several basis functions available in preCICE are listed in \autoref{BasisFunc}.
See \autoref{fig:rbf} for a schematic view of the relation between the vertices of both coupling meshes and the construction and evaluation of the interpolant.

\begin{table}[h!]
\centering
\caption{Radial basis functions available in preCICE (excerpt). Local basis functions have a support radius $r$, i.e., $\phi \left(\Vert \vec{x} \Vert_{2}\right) = 0$ for $\Vert \vec{x} \Vert_{2} > r$. C-TPS use normalized variables $\xi = \Vert \vec{x} \Vert_{2}/r$ and are set to zero for $\xi>1$. We enforce a finite support radius for Gaussians by setting the basis function to zero when falling below a threshold of $10^{-9}$~\cite{Lindner2017}. For a given support $r$, we can, thus, compute the necessary shape parameter $\zeta$.}
\label{BasisFunc}
\begin{tabular}{lcl}\toprule
& Basis Function & Support \\ \midrule
Gaussians    & $\exp^{-(\zeta \cdot \Vert \vec{x} \Vert)^{2}}$  & Local  \\
Global Thin Plate Splines (G-TPS) & $\Vert \vec{x} \Vert ^{2} \log(\Vert \vec{x} \Vert)$ & Global \\
Compact Thin Plate Splines C2 (C-TPS)  & $1 - 30\xi ^{2} - 10\xi ^{3} + 45\xi ^{4} - 6\xi ^{5} - 60\xi ^{3}\log\xi$   & Local  \\
\bottomrule
\end{tabular}
\end{table}

The set of coefficients $\lambda_{i} \in \mathbb{R}, i=1,...,n_{S_1}$ is determined such that the interpolation conditions
\[
s \left( \vec{x}_{i}^{S_1} \right) = v_{i}^{S_1} \quad \forall i = 1,...,n_1 
\]
are fulfilled. The addition of the polynomial term leads to an under-determined system which is regularized by the polynomial conditions
\[
\sum\limits_{i=1}^{n_{S_1}} \lambda_{i} \cdot \vec{x}_{i}^{S_1} = 0 \; \mbox{ and } \; \sum\limits_{i=1}^{n_{S_1}} \lambda_{i} = 0.
\]
In matrix notation, this leads to the linear system
\[ \left( \begin{array}{cc} C & Q \\ Q^T & \mathbf{0} \end{array} \right) 
   \left( \begin{array}{c} \vec{\lambda} \\ \vec{\beta} \end{array} \right) = 
   \left( \begin{array}{c} \vec{v}^{S_1} \\ \vec{0} \end{array} \right), \]
where $\vec{\lambda} = \left( \lambda_1, \lambda_2, \ldots, \lambda_{n_1} \right)^T$, $\beta = \left(\beta_0, \beta_1, \ldots, \beta_d\right)^T$, $C = \left( \phi \left( \Vert \vec{x}_{i}^{S_1} - \vec{x}_{j}^{S_1} \Vert_{2} \right) \right)_{i,j = 1, \ldots, n_1} \in \mathbb{R}^{n_1 \times n_1}$, $Q = \left( 1,\ x_{1,i}^{S_1},\ \ldots,\ x_{d,i}^{S_1} \right)_{i=1,\ldots,n_1} \in \mathbb{R}^{n_1 \times (d+1)}$
and, finally, the mapping reads
\[ \vec{v}^{S_2}  = \left( \begin{array}{cc} \tilde{C} & \tilde{Q} \end{array} \right) \left( \begin{array}{cc} C & Q \\ Q^T & \mathbf{0} \end{array} \right)^{-1} \left( \begin{array}{c} \vec{v}^{S_1} \\ \vec{0} \end{array} \right) \] 
with $\tilde{C} = \left( \phi \left( \Vert \vec{x}_{i}^{S_2} - \vec{x}_{j}^{S_1} \Vert_{2} \right)\right)_{\substack{i=1,\ldots, n_2 \\ j=1,\ldots,n_1}} \in \mathbb{R}^{n_2 \times n_1}$, 
$\tilde{Q} = \left( 1,\ x_{1,i}^{S_2},\ x_{2,i}^{S_2},\ x_{3,i}^{S_2} \right)_{i=1,\ldots,n_2} \in \mathbb{R}^{n_2 \times 4}$.

Local basis functions result in a sparse matrix $C$. 
However, the polynomial term matrix $Q$ is always densely populated, which hampers the favorable properties of the sparse matrix. Solving the polynomial term separately in a least squares approach via QR-decomposition of the matrix $Q$ capitalizes on the sparsity of the matrix $C$. For a full description of this separated polynomial approach, the reader is referred to~\cite{Lindner2017}.

As radial basis functions are radially symmetric in all spatial dimensions, distances between the two involved coupling meshes normal to a coupling surface do not have to be explicitly tackled, in contrast to the nearest-projection method. However, the accuracy of the RBF mapping decreases with an increasing gap or overlap between the two meshes of $S_1$ and $S_2$. In addition, the RBF mapping with local basis functions suffers from a trade-off between high accuracy (achieved for basis functions with wide support) and feasible conditioning of the linear system (only given for moderate support width). We address the latter to some extent by scaling the interpolant with the interpolant of the constant unit function, which allows us to use a smaller support radius without deteriorating accuracy~\cite{deparis2014rescaled, Lindner2017}.

The RBF data mapping is implemented using either (depending on configuration) an iterative GMRES solver from PETSc~\cite{petsc-user-ref} in every mapping step, or an initial dense QR-decomposition from Eigen~\cite{eigenweb} followed by a matrix-vector product and a backward substitution in every mapping step. While the GMRES solver is fully parallelized, the QR-decomposition uses a sequential computation on a single rank.

\subsubsection{Numerical and performance results}

We compare the various data mapping methods in terms of accuracy and computational demand using the Abstract Solver Testing Environment (ASTE)\footnote{ASTE branch used for these tests: \url{https://github.com/precice/aste/tree/mapping-tests}}. ASTE imitates data input and output of participants coupled via preCICE in an artificial setting. In our test setup, two ASTE participants, $S_1$ and $S_2$, are coupled via preCICE. Both define individual surface meshes of the same geometry. We then use an analytical test function,
\[
  f(\vec{x}) = 0.78 \cdot \cos \left(10 \cdot (x_1 + x_2 + x_3)\right)\;,
\]
to set values on $\mathcal{M}_{S_1}$. We compute a single consistent mapping from $\mathcal{M}_{S_1}$ to $\mathcal{M}_{S_2}$ and measure errors on $\mathcal{M}_{S_2}$ with a discrete $l_2$-norm,
\[
\frac{1}{n} \left(\sum_{i=1}^n \left(v_i^{S_1} - f(\vec{x}^{S_1}_i)\right)^2 \right)^{\frac{1}{2}}\;.    
\]
The mapping error procedure is shown in Algorithm~\ref{alg:mapError}.

\begin{algorithm}[H]
  \caption{Data mapping error procedure}
  \label{alg:mapError}
  \begin{algorithmic}[1]
    \State Fix the output mesh $S_2$ while varying the input mesh $S_1$
    \State Apply test function on mesh $S_1$: $v^{S_1}_{i,real} = f(x^{S_1}_i)$ and B: $v^{S_2}_{i,real} = f(x^{S_2}_i)$
    \State Run mapping from $S_1$ to $S_2$ with $v^{S_1}_{i,real}$, resulting in data $v^{S_2}_{i,interpolant}$
    \State Compute per vertex error  $e_i = | v^{S_2}_{i,interpolant} - v^{S_2}_{i,real} |$
    \State Compute relative-l2 error $e' = \frac{1}{n} \left(\sum_{i=1}^n e_{i}^2 \right)^{\frac{1}{2}}\;.$
  \end{algorithmic}
\end{algorithm}

\begin{figure}[h!]
  \centering
  \begin{tikzpicture}[x=.2\textwidth]
    \node at (0,0) {\includegraphics[width=.2\textwidth,keepaspectratio]{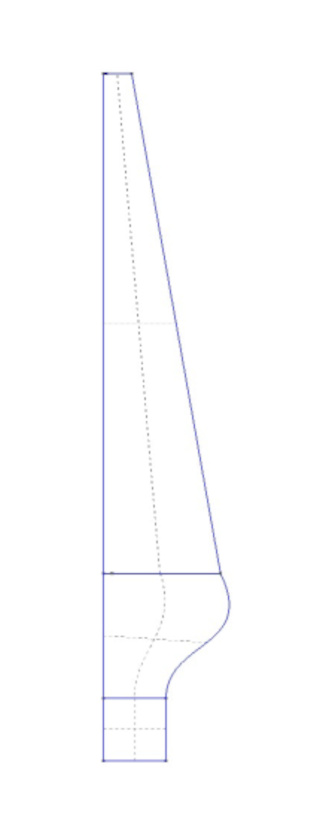}};
    \node at (1,0) {\includegraphics[width=.2\textwidth,keepaspectratio]{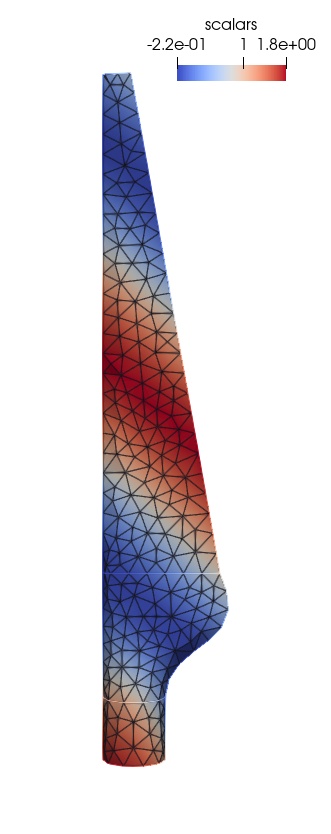}};
    \node at (2,0) {\includegraphics[width=.2\textwidth,keepaspectratio]{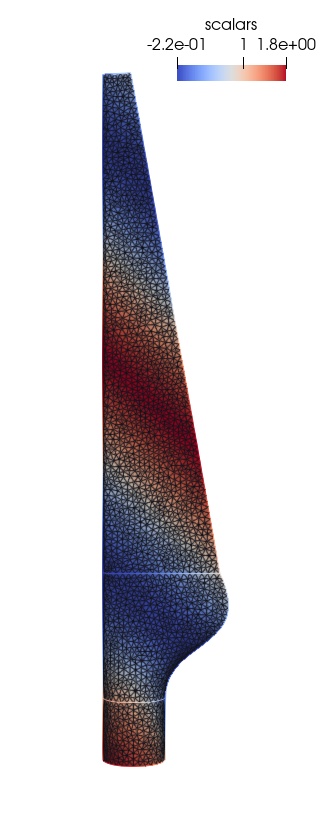}};
    \node at (3,0) {\includegraphics[width=.2\textwidth,keepaspectratio]{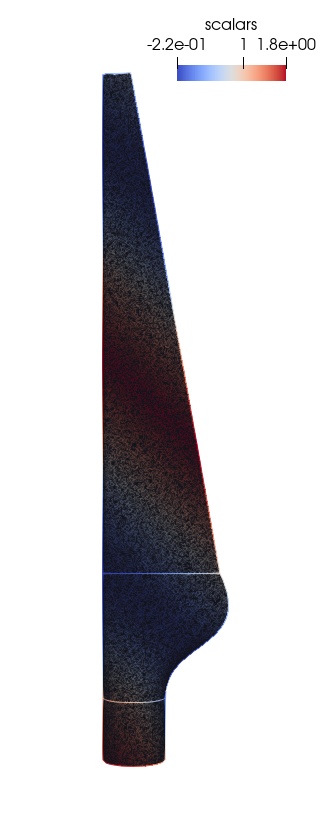}};
    \node at (4,0) {\includegraphics[width=.2\textwidth,keepaspectratio]{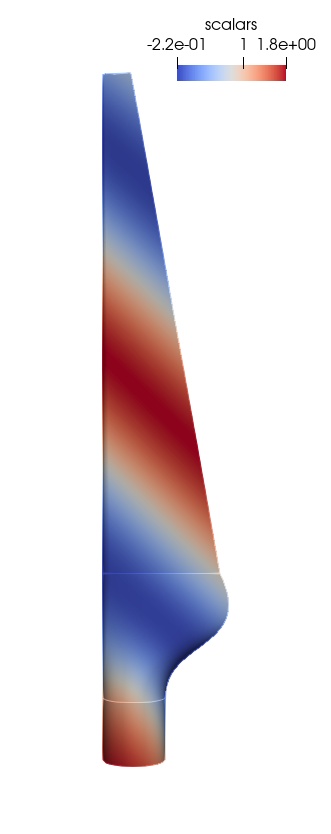}};
  \end{tikzpicture}
  \caption{Mapping test case: Different meshes of the turbine-blade test geometry.
    From left to right: the geometry used to generate the meshes, $h=0.03$, $h=0.009$, $h=0.004$, $h=0.0005$ without edges.
    The mesh surface color indicated the test function, edges are drawn in black.
  }
  \label{fig:geometry}
\end{figure}

As test geometry, we use a turbine blade\footnote{Wind Turbine Blade created by Ivan Zerpa \url{https://grabcad.com/library/wind-turbine-blade--4}}. We use GMSH\cite{geuzaine2009gmsh} to generate almost uniform surface meshes with different resolutions as listed in \autoref{tab:meshes}. In \autoref{fig:geometry}, we visualize the geometry, different meshes, and the test function. 

\begin{table}[h!]
  \centering
  \caption{The meshes used for the mapping tests sorted from fine to coarse. Bold typesetting indicates the output meshes (associated to participant $S_2$). All other meshes are used as input meshes (associated to participant $S_1$).}
  \label{tab:meshes}
  \begin{tabular}{lrrc}
    \toprule
    $h$    & Vertices & Triangles & Series          \\
    \midrule
    0.03   & 438      & 1007      & coarse          \\
    0.02   & 924      & 2027      & coarse          \\
    0.01   & 3458     & 7246      & coarse          \\
    0.009  & 4302     & 8970      & \textbf{coarse} \\
    0.008  & 5310     & 11025     & coarse          \\
    0.006  & 9588     & 19712     & coarse          \\
    0.004  & 21283    & 43352     & fine, coarse    \\
    0.003  & 38112    & 77271     & fine            \\
    0.002  & 84882    & 171319    & fine            \\
    0.0014 & 172803   & 347815    & \textbf{fine}   \\
    0.001  & 338992   & 681069    & fine            \\
    0.0007 & 691426   & 1387249   & fine            \\
    0.0005 & 1354274  & 2714699   & fine            \\
    \bottomrule
  \end{tabular}
\end{table}

\begin{figure}[h!]
  \centering
  \includegraphics{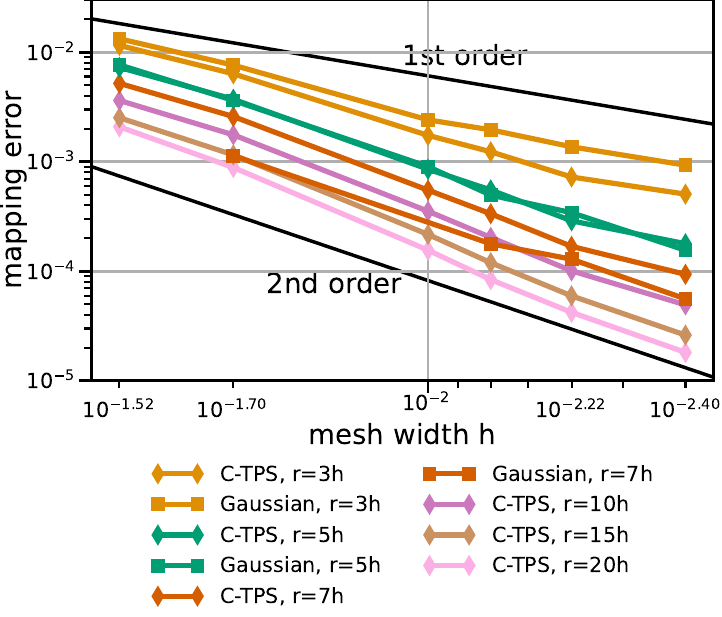}
  \includegraphics{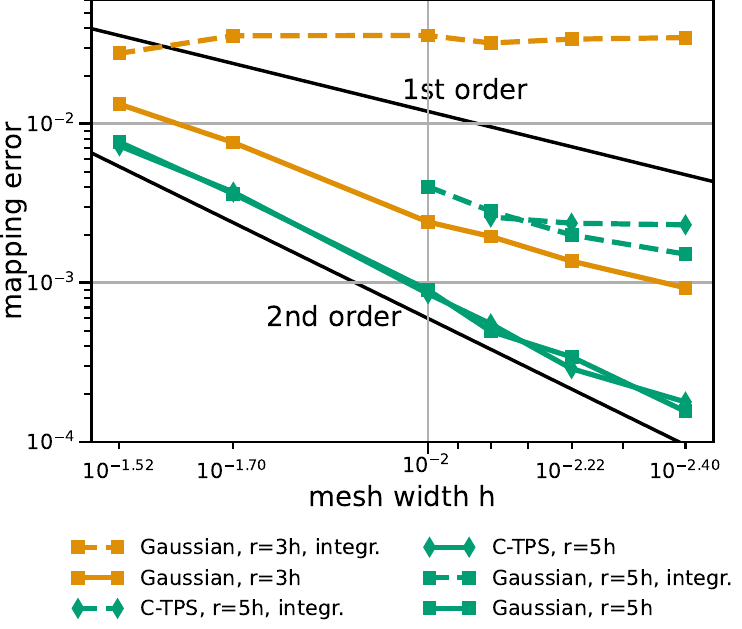}
  \caption{RBF data mapping with local basis functions for coarse series. Comparison of various support radii (left) and integrated and separated handling of the global polynomial (right).  
    Participant $S_2$ uses $h=0.009$. Missing data points mark diverging cases.
    Gaussians with $r=7h$ or $r=10h$ diverge for some or all cases, respectively (left).
    With integrated handling of the polynomial, both basis functions diverge for $r=5h$ and coarser meshes (right).}
  \label{fig:mapping_rbf_small}
\end{figure}

We run the mapping tests on the thin-nodes partition of SuperMUC-NG, hosted at the Leibniz Supercomputing Centre.  Each thin-node contains two 3.1GHz Intel Xeon Platinum 8174 (SkyLake) processors with a total of 48 cores and 96GB of system memory per node.  The tests used the Intel Omni-Path interconnect as primary network connection.
We run participant $S_1$ on a single node (48 MPI ranks) and participant $S_2$ on two nodes (96 MPI ranks) as participant $S_2$ computes the actual mapping. Runtime is the maximum of a given event over all ranks of $S_2$\footnote{preCICE measures timings for a wide range of events across MPI ranks using the \github{precice/EventTimings} framework.} and memory is the sum of the peak memory usage of all ranks of $S_2$. These results are in addition averaged over five runs.
For the RBF data mapping variants, we use GMRES as linear solver with a relative convergence threshold of $10^{-6}$, except for G-TPS, where we use the sequential QR-decomposition.

We compare the data mapping methods in two series of computations. The first, \textit{coarse} series is based on meshes with $h \geq 0.004$, where participant $S_2$ always uses $h=0.009$ and participant $S_1$ varies the value of $h$ through all other values of the series listed in \autoref{tab:meshes}. The second, \textit{fine} series is based on meshes with $h \leq 0.004$. Participant $S_2$ uses $h=0.00014$ and participant $S_1$ the rest. While we can compare all mapping variants for the coarse series, RBF data mapping with G-TPS is too expensive in terms of computation and memory for the fine series.
For reproduction of our results, all data used as well as all steps are available in an archive\footnote{Test Setup: \url{https://gitlab.lrz.de/precice/precice2-ref-paper-setup}}.

\autoref{fig:mapping_rbf_small} gives results for RBF data mapping with local basis functions -- C-TPS and Gaussians -- for varying support radii and for integrated and separated handling of the global polynomial.
Separated handling of the polynomial clearly outperforms integrated handling in terms of accuracy and robustness. For C-TPS, an increase in accuracy is observed for increasing support radius. This is even true for rather large radii ($r=20h$), where we get more than quadratic convergence. Gaussians, however, show robustness issues for larger support radii. We assume that this behavior is caused by the increasing ratio of large to small matrix entries. In a further test (not shown here), we observed that increasing the (hard-coded) threshold value from $10^{-9}$ to $10^{-5}$ seems to improve robustness.

\begin{figure}[h!]
  \centering
  \includegraphics{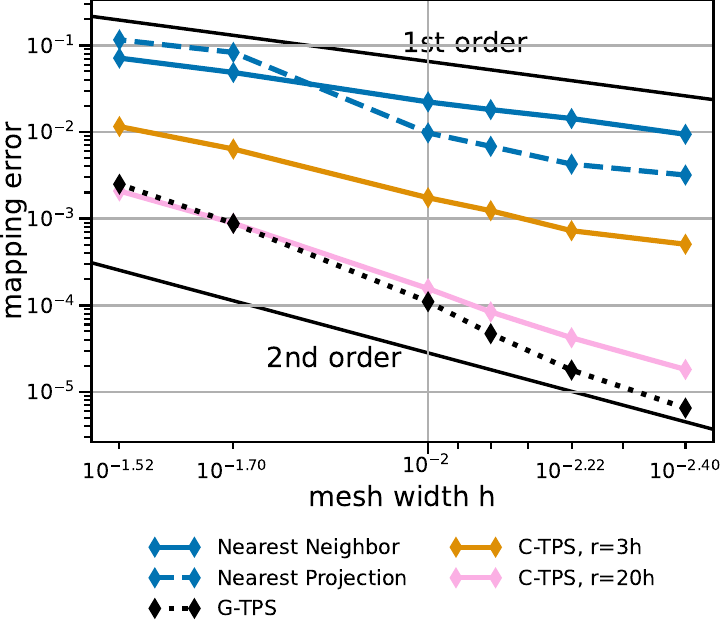}
  \includegraphics{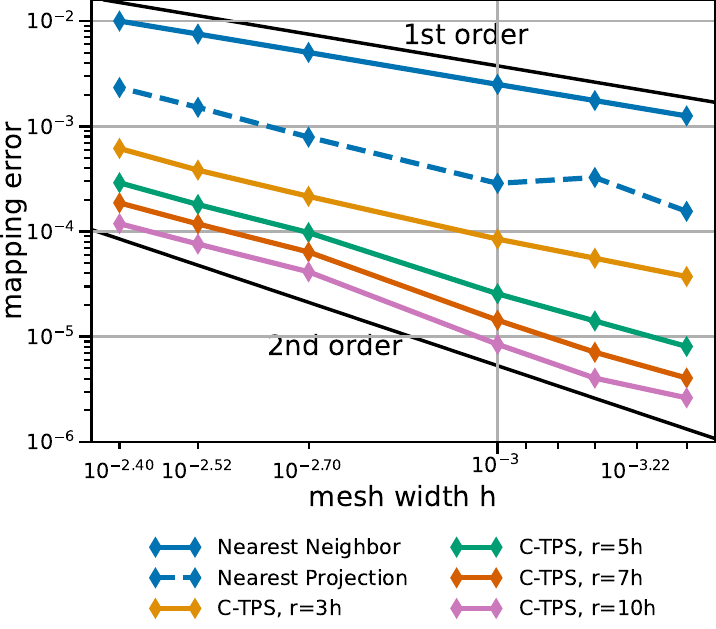}
  \caption{Comparison of nearest-neighbor / nearest-projection mapping and RBF mapping with G-TPS and C-TPS for coarse series (left) and fine series (right) from \autoref{tab:meshes}. RBF mapping with G-TPS is infeasibly expensive for the fine series. All RBF mapping methods use a separated handling of the polynomial.
    Participant $S_2$ uses $h=0.009$ for the coarse series (left) and $h=0.0014$ for the fine series (right).
  }
  \label{fig:mapping_all_small}
\end{figure}

\autoref{fig:mapping_all_small} sets the best local RBF variants in perspective with RBF mapping using G-TPS, nearest-neighbor mapping, and nearest-project mapping. 
RBF mapping using C-TPS with a large support radius is comparable to RBF mapping using G-TPS. The latter only wins for the finest meshes. RBF data mappings clearly outperform nearest-neighbor and nearest-projection mapping -- even for a relatively small support radius of $h=3r$. Nearest-projection mapping, interestingly, does not show a constant second-order convergence for the coarse series, which suggests that the projection error dominates the interpolation error. In fact, similar tests with a cubic geometry (not shown) give constant second-order convergence for nearest-projection mapping as, in this case, all meshes directly lie on the geometry (i.e.~no projection error). For the fine series, nearest-projection mapping gives a rather constant second-order convergence as well.

\begin{figure}[h!]
  \centering
  \includegraphics{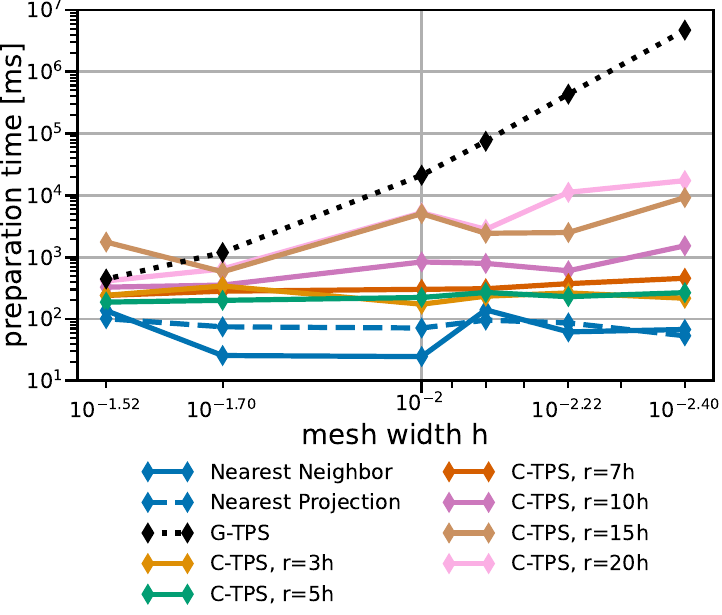}
  \includegraphics{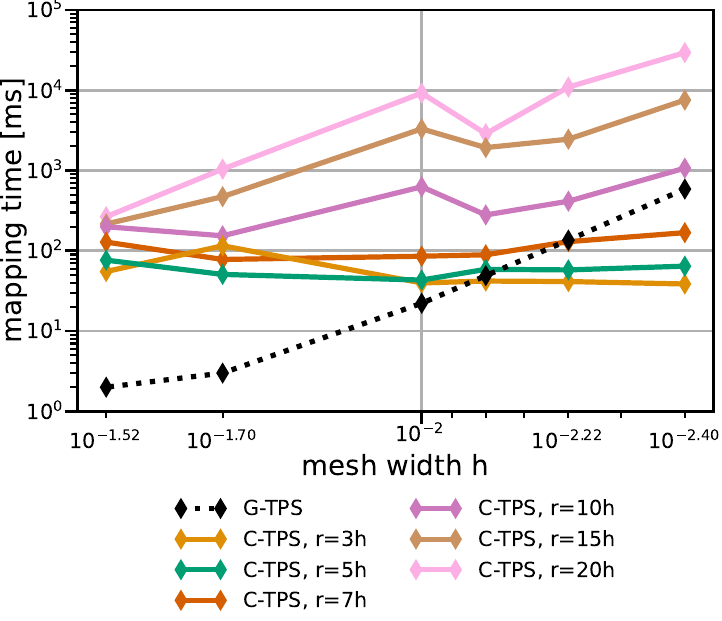}
  \caption{Comparison of one-time preparation time (left) and the recurrent mapping time (right) of various data mapping methods for the coarse series from \autoref{tab:meshes}. All RBF mapping methods use a separated handling of the polynomial.
    The recurrent mapping time of nearest-projection and nearest-neighbor mapping is below the measurement resolution of $1ms$ and hence omitted.
    Participant $S_2$ uses $h=0.009$.
  }
  \label{fig:mapping_time_coarse}
\end{figure}

\begin{figure}[h!]
  \centering
  \includegraphics{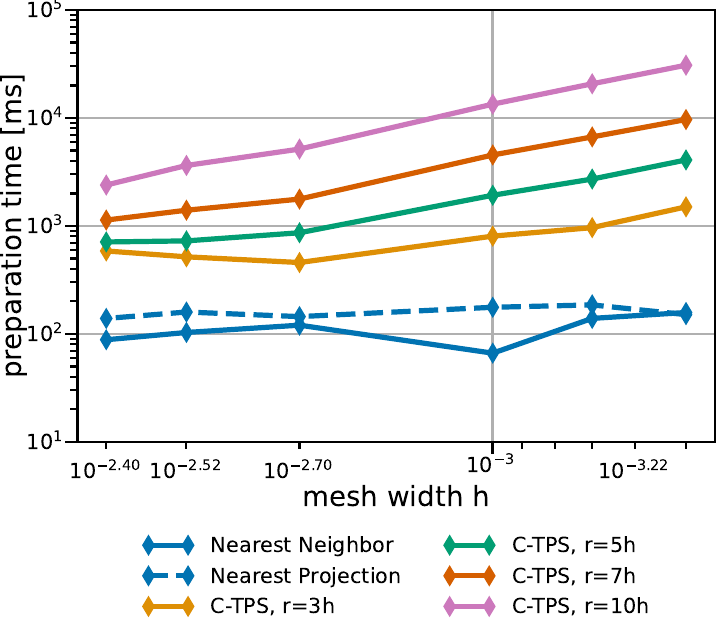}
  \includegraphics{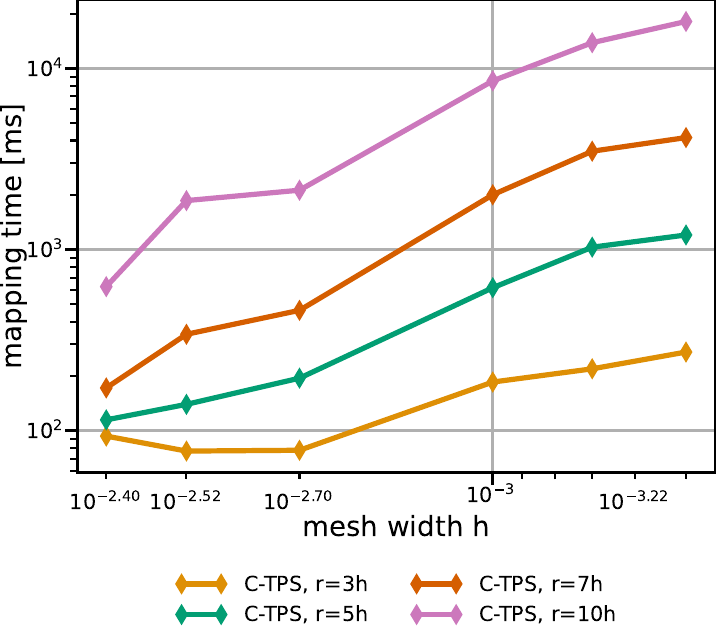}
  \caption{Comparison of one-time preparation time (left) and the recurrent mapping time (right) of various data mapping methods for the fine series. All RBF mapping methods use a separated handling of the polynomial.
    The one-time preparation of the nearest-neighbor mapping is an inexpensive operation and has the tendency to fluctuate, 5 samples are not enough to fully smooth them out resulting in a spike at $0.001$.
    The recurrent mapping time of nearest-projection and nearest-neighbor mapping is below the measurement resolution of $1ms$ and hence omitted.
    Participant $S_2$ uses $h=0.0014$.
  }
  \label{fig:mapping_time_fine}
\end{figure}

\begin{figure}[h!]
  \centering
  \includegraphics{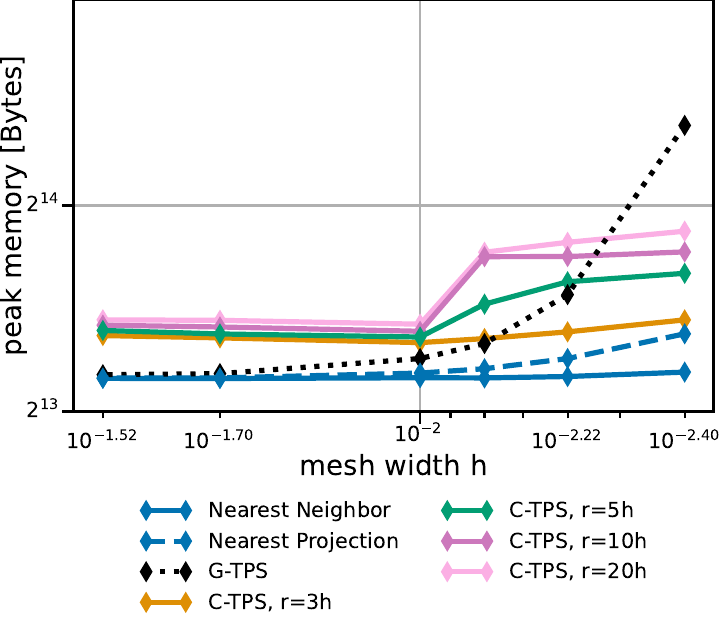}
  \includegraphics{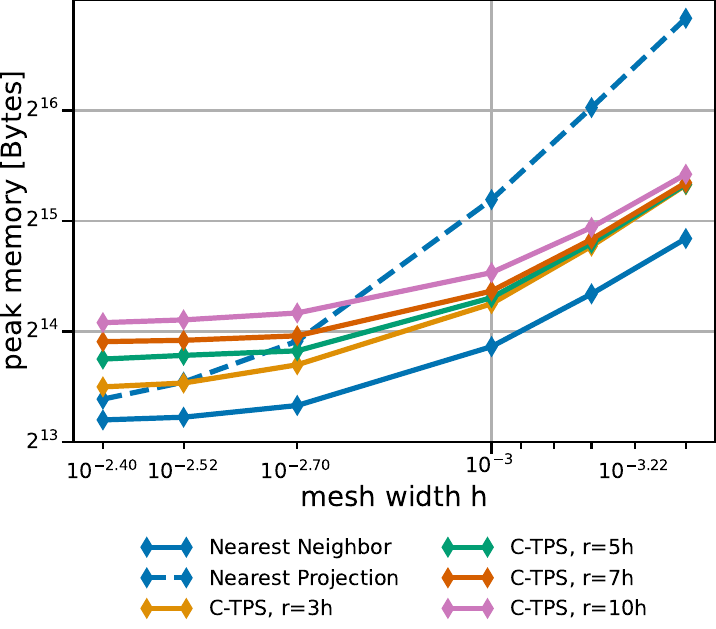}
  \caption{Comparison of the maximum overall memory consumption of various data mapping methods for coarse series (left) and fine series (right) from \autoref{tab:meshes}.
    Participant $S_2$ uses $h=0.009$ for the coarse series (left) and $h=0.0014$ for the fine series (right).
    The memory consumption is the maximum of all ranks of participant $S_2$, which is executing the mapping.
  }
  \label{fig:mapping_memory}
\end{figure}

Next, we compare the same methods in terms of compute time. Here, we have to distinguish between one-time preparation time (e.g., QR-factorization, matrix initialization in PETSc, or nearest-neighbor search) and recurrent mapping time in every mapping operation (e.g., back substitution or GMRES solve). \autoref{fig:mapping_time_coarse} and \autoref{fig:mapping_time_fine} give both data for various data mapping methods and the coarse and fine series, respectively. The quickly increasing preparation time of RBF mapping with G-TPS makes the method unpractical for finer meshes. For C-TPS, both the preparation time and the recurrent mapping time increases significantly with increasing support radius and decreasing mesh width. For fine meshes, larger support radii are thus discouraged despite their superior accuracy. For small cases, the overhead of the PETSc solver for RBF is more costly than the overall serial QR-factorization in G-TPS. Nearest-neighbor and nearest-projection mapping are both drastically cheaper than RBF data mapping, particularly in the recurrent mapping time. Finally, \autoref{fig:mapping_memory} compares the peak memory consumption of all data mappings. RBF mapping with G-TPS shows a drastic increase in memory with increasing mesh size. For the coarse series, all methods show the expected behavior: higher memory consumption for RBF than for nearest-neighbor and nearest-projection mapping and increasing memory consumption for RBF mapping using C-TPS with increasing support radius.
For the fine series, the nearest-project surpassed even the RBF methods, due to the additional cost of handling connectivity information.

We conclude that RBF data mapping with local basis functions is a powerful method. There is a natural trade-off between accuracy and compute effort when modifying the support radius. A good compromise is a support radius of $r=5h$ to $r=7h$. In the current implementation, C-TPS should be preferred over Gaussians as basis functions. RBF mapping with G-TPS should only be used for small mesh sizes. 
When RBF data mapping becomes too expensive, nearest-projection mapping is a good alternative except for very large out-of-memory cases. 
Scalability results for the mapping computation were recently published in~\cite{AminPerf2021}. In addition, we aim for a more in-depth analysis of mapping variants with further geometries in future work.

\subsection{Communication}
\label{ssec:com}

\begin{figure}[h!]
  \centering
  \begin{tikzpicture}
      \foreach \n in {0,1,2,3,4,5} {
        \node[inNode,inner sep=3pt] (a\n) at (0, \n) {};
      }
      \begin{scope}[every node/.style={outNode, inner sep=3pt}]
      \node (b0) at (2, 0.5) {};
      \node (b1) at (2, 1.5) {};
      \node (b2) at (2, 3.5) {};
      \node (b3) at (2, 4.5) {};
      \end{scope}

    \begin{scope}[line width=2pt]
      \foreach \n in {0,2,4} {
        \draw (a\n) ++(-0.5, -0.25) coordinate (au\n) -- ++(0,1.5) coordinate (ab\n);
      }
      \foreach \n in {0,2} {
        \draw (b\n) ++(0.5, -0.25) coordinate (bu\n) -- ++(0,1.5) coordinate (bb\n);
      }
    \end{scope}

    \path
      (a0) +(-0.5, 0.5) node[anchor=east] {$S_1^0$}
      (a2) +(-0.5, 0.5) node[anchor=east] {$S_1^1$}
      (a4) +(-0.5, 0.5) node[anchor=east] {$S_1^2$}
      (b0) +( 0.5, 0.5) node[anchor=west] {$S_2^0$}
      (b2) +( 0.5, 0.5) node[anchor=west] {$S_2^1$}
      ;

    \begin{pgfonlayer}{bg}
      \fill[gray!20] (au0) to[out=0,in=180,looseness=1.5] (bu0) -- (bb0) to[out=180,in=0] (ab0) -- cycle;
      \fill[gray!20] (au2) to[out=0,in=180] (bu0) -- (bb0) to[out=180,in=0] (ab2) -- cycle;
      \fill[gray!20] (au2) to[out=0,in=180] (bu2) -- (bb2) to[out=180,in=0] (ab2) -- cycle;
      \fill[gray!20] (au4) to[out=0,in=180] (bu2) -- (bb2) to[out=180,in=0,looseness=1.5] (ab4) -- cycle;
    \end{pgfonlayer}

    \draw[out=0,in=180,
      decoration={markings, mark=at position 0.5 with {\arrow[line width=1pt]{latex}}},
          every edge/.append style={postaction={decorate}},
      ] (a0) edge (b0) (a1) edge (b1) (a2) edge (b1) (a3) edge (b2) (a4) edge (b2) (a5) edge (b3) ;
  \end{tikzpicture}
  \caption{Communication initialization in preCICE.
    Given the distribution of vertices among the ranks of parallel participants $S_1^i$ and $S_2^j$,
    combined with a data mapping between the vertices shown in the middle,
    preCICE deduces the required communication pattern of ranks between participants $S_1$ and $S_2$, depicted as the gray connections.}
  \label{fig:comm}
\end{figure}
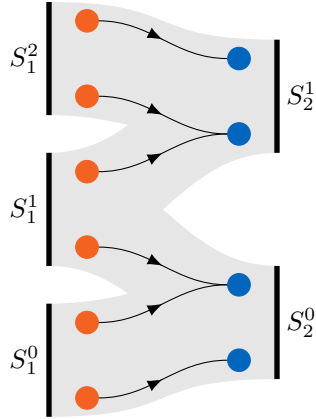

Besides coupling schemes and data mapping, the third feature pillar of preCICE is inter-code communication. For large-scale simulations on massively-parallel high-performance computing systems, efficient inter-code communication is a necessity. Employing any central instance not only deteriorates the communication performance, but can also be memory prohibitive when large amounts of data must be communicated. Therefore, preCICE implements fully parallel point-to-point communication~\cite{Uekermann2016, Bungartz2016_ExaFSA_LNCSE, Shukaev2015}. 
In the initialization phase, preCICE performs an analysis of the coupling mesh partitions and the defined data mappings of each connected pair of participants to find the list of required connections between the MPI ranks of either participant (cf.~\autoref{fig:comm}). 
To this end, bounding boxes around mesh partitions are compared in a first step, leading to preliminary communication channels. In a second step, actual mesh data is compared in a fully-parallel fashion~\cite{AminPerf2021}.

As communication backends, preCICE supports MPI and TCP/IP.
In general, communication via MPI is faster. However, TCP/IP communication is more robust and flexible, since not all MPI implementations support the necessary inter-code MPI functionality. For MPI-based communication, preCICE creates a single inter-code communicator including all involved ranks from both participants. To establish TCP/IP-based connections, on the other hand, each pair of connected ranks exchanges a connection token via the file system. To reduce the load on the file system, a hash-based scheme is used, which distributes the connection files uniformly across different directories~\cite{Lindner2020_ExaFSA, Lindner2019}. This minimizes the number of files per directory, and, thus, improves efficiency and increases the feasible number of connections.

\subsection{Getting and building preCICE}
\label{ssec:building}

After introducing the basic coupling methods implemented in preCICE in the last three sections, we now give an overview of various ways to get preCICE.
The GitHub repository\footnote{preCICE repository on GitHub: \url{www.github.com/precice/precice}} is the central platform for development,
issue tracking, and contributing. It provides the release timeline with release notes, automatic source archives, and build artifacts. However, the repository 
only contains the preCICE library and native (C and Fortran) language bindings. Additional derived software is hosted in separate repositories under the preCICE GitHub organization: 
adapter codes, tutorials, Python and MATLAB language bindings, and more.
All these components, together with the core library, are part of the \textit{preCICE distribution}\footnote{preCICE distribution: \url{https://precice.org/installation-distribution.html}}, a versioned and citable ecosystem of components that are meant to work together and are maintained by the preCICE developers. Everything presented in this paper refers to the version v2104.0 of the distribution~\cite{Chourdakis2021_Distribution}.

On the other hand, the preCICE library also depends on various other libraries for a range of features. See \autoref{table:dependencies} for an overview of dependencies and their associated features in preCICE.

\begin{table}[h!]
  \begin{tabular}{lllp{5cm}}\toprule
    Dependency       & Version & CMake Option                       & Features                                               \\\midrule
    Boost Geometry   & $\geq$ 1.65.1  & \textit{required}                  & Spacial index trees                                    \\
    Boost Container  & $\geq$ 1.65.1  & \textit{required}                  & Flat maps and sets                                     \\
    Boost Stacktrace & $\geq$ 1.65.1  & \textit{required}                  & Stacktrace information                                 \\
    Boost Log        & $\geq$ 1.65.1  & \textit{required}                  & Configurable logging                                   \\
    Boost Test       & $\geq$ 1.65.1  & \textit{required}                  & Base of testing framework                              \\
    Eigen            & 3       & \textit{required}                  & Mesh representation and radial basis function mapping. \\
    libxml           & 2       & \textit{required}                  & Parsing of XML config files.                    \\
    PETSc            & $\geq$ 3.6     & \incode{PRECICE_PETScMapping}     & Parallel RBF mapping.                \\
    Python           & $\geq$ 3.6     & \incode{PRECICE_PythonActions}    & User-defined actions                                   \\
    NumPy            & $\geq$ 1.18.1  & \incode{PRECICE_PythonActions}    & User-defined actions                                   \\
    MPI              & MPI-3   & \incode{PRECICE_MPICommunication} & MPI communication back-end                             \\
    \bottomrule
  \end{tabular}
  \caption{Dependencies of preCICE and associated features in preCICE.}
  \label{table:dependencies}
\end{table}

\begin{figure}[h!]
  \centering
  \includegraphics[width=\textwidth]{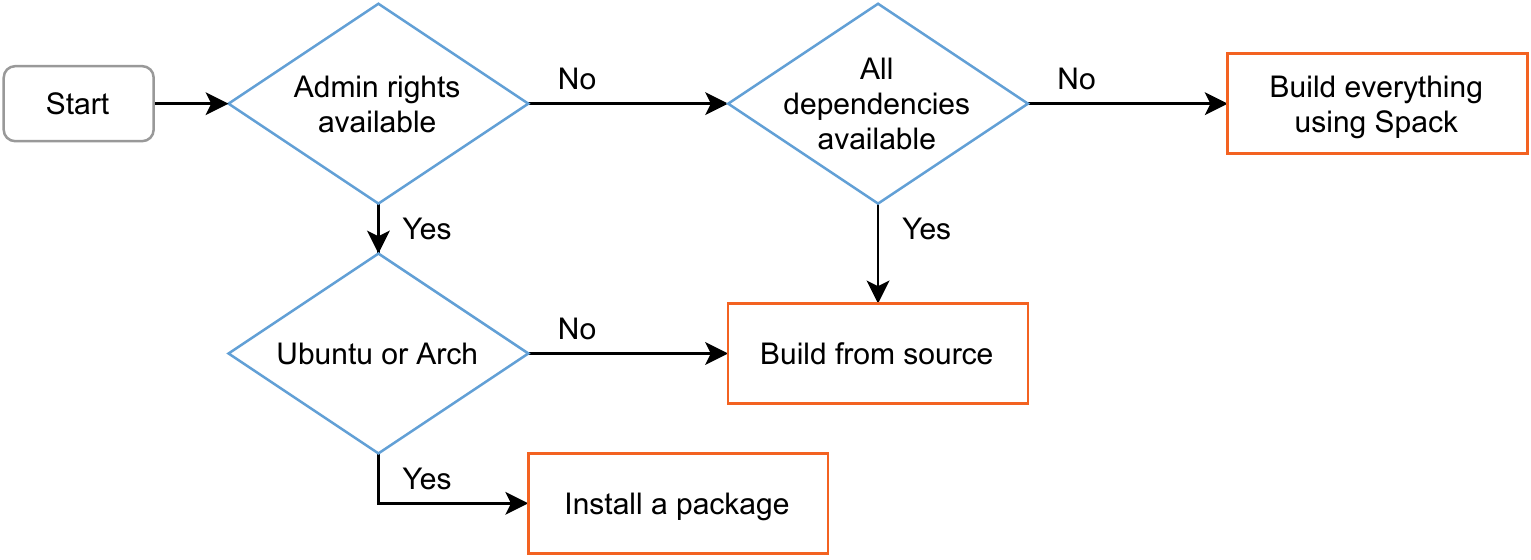}
  \caption{Decision graph and overview of the different installation methods for preCICE on Linux.
  On macOS, users can build using Spack, or install dependencies from Homebrew and build from source.
  On Windows, preCICE is available from MINGW (experimental), while it can also work inside the Windows Subsystem for Linux.
  Before deciding to install preCICE on their system, users can try a
  virtual machine containing all the software needed to run the preCICE tutorials.}
  \label{fig:install-decision}
\end{figure}

A decision graph for getting preCICE is shown in \autoref{fig:install-decision}. We describe the different options in the following.

\paragraph{Debian packages for preCICE on Ubuntu}
Due to the popularity of Ubuntu among preCICE users, we provide corresponding Debian packages.
We aim to support the latest two Ubuntu \textit{long term support} (LTS) releases, which is frequency-wise compatible with our
strategy to not release new major versions (breaking changes) more often than every two to three years.
The Debian packages contained in our GitHub releases allow one-click installation on supported platforms.
This avoids explicit dependency management by the user. The current Debian package is always generated for 
the latest Ubuntu release as well as the latest Ubuntu LTS.

\paragraph{Building preCICE using Spack}
In addition, we maintain a Spack~\cite{spack} package which allows to build the complete required software stack
from source code. 
This is essential to be able to test arbitrary combinations of dependency versions and different compilers.
We actively maintain a build recipe with common
configurations. Moreover, preCICE is a member of the Extreme-scale Scientific Development Kit (xSDK) \footnote{xSDK: \url{http://xsdk.info/}} since \incode{xsdk-0.5.0} (November 2019)\footnote{See {\scriptsize \url{https://github.com/xsdk-project/xsdk-policy-compatibility/blob/master/precice-policy-compatibility.md}} for more details on all policies fulfilled by preCICE.}, which promises
compatibility to other major scientific computing packages.

\paragraph{Building preCICE with CMake}
For other platforms, we provide an in-depth guide on how to build preCICE from source. Versioned code archives are available for every release.
For cross-platform build system configuration, preCICE leverages the industry standard CMake.
It allows users and developers to develop and build in their environment of choice.
The adoption of CMake simplifies package generation and the future adoption of Windows and macOS support.
macOS works out-of-the-box since preCICE v2.2. There are several ways to support Windows:
Since v1.x, there is community support via MinGW.
Windows users that rely on WSL (Windows Subsystem for Linux) can install preCICE there normally.
We are currently preparing native Windows support (MSVC compiler), in addition. 

\paragraph{preCICE tutorials virtual machine image}
Before running their first coupled simulation, a user needs to install not only the preCICE library, but also a minimum set of adapters and third-party solvers.
This can become even more complicated if the user does not already work on a platform compatible with all components.
To lower the entry barrier, we provide a virtual machine (VM) image with all components needed to run the preCICE tutorials.
We create\footnote{preCICE VM sources: \url{https://github.com/precice/vm}} and distribute\footnote{preCICE Vagrant box: \url{https://app.vagrantup.com/precice/precice-vm}\\documentation: \url{https://precice.org/installation-vm.html}}
this image as a Vagrant\footnote{Vagrant: \url{https://github.com/hashicorp/vagrant}} Box,
which is currently available for VirtualBox, but could easily be packaged for other providers as well.
We chose Vagrant instead of a provider-specific system, as Vagrant allows us to highly automate the box generation,
integrates with the host system automatically (SSH access, shared folders),
provides infrastructure to distribute the box, and works with various host platforms and virtualization providers.
We chose a VM instead of a container-based system, as virtual machines provide access to a full graphical environment by default
and many users in our community already have experience working with virtual machines, but not with containers.

\subsection{Application programming interface and configuration} 
\label{ssec:api}

Now that we know which coupling methods preCICE offers and how to get preCICE, we show in this section how preCICE can actually be used.
Even though preCICE is a C++ library, it also supports the  software programming languages most widely-used in simulation, as shown in \autoref{table:bindings}.
Alongside the native application programming interface (API) for C++, preCICE provides C and Fortran bindings by default.
The API for other languages is provided via independent projects, as they follow different release cycles, different project management, and different developer and installation procedures. Python bindings are based on Cython, installable via pip from PyPI. MATLAB bindings~\cite{Volland2019} are based on the MEX interface and Julia bindings are currently in the prototype phase. Finally, we develop an independent Fortran module for easier integration of preCICE into Fortran codes. The architecture and relation between these projects is described in the documentation\footnote{API documentation overview: \url{https://precice.org/couple-your-code-api.html}}. All language bindings also provide so-called \textit{solverdummies} as example codes and provide \incode{pkg-config} and \incode{CMakeConfig} files for integration into other projects.

\begin{table}[h!]
  \begin{tabular}{lll}\toprule
    Language       & Location                         & Installation                                   \\\midrule
    C++            & \github{precice/precice}         & \textit{native API}                            \\
    C              & \github{precice/precice}         & \inbash{cmake -DPRECICE_ENABLE_C=ON .}       \\
    Fortran        & \github{precice/precice}         & \inbash{cmake -DPRECICE_ENABLE_FORTRAN=ON .} \\
    Fortran Module & \github{precice/fortran-module}  & \inbash{make}                                  \\
    Python         & \github{precice/python-bindings} & \inbash{pip install pyprecice@2.0.2}           \\
    MATLAB         & \github{precice/matlab-bindings} & MATLAB script                                  \\
    Julia          & \github{precice/julia-bindings}  & Julia package (experimental)                   \\ 
    \bottomrule
  \end{tabular}
  \caption{Programming languages supported by preCICE. CMake options for C and Fortran bindings are set to ON by default.}
  \label{table:bindings}
\end{table}

To introduce the API of preCICE, we use an example: we develop an adapter for a fluid solver written in Python to couple it to an already adapted solid solver for fluid-structure interaction (FSI). Mathematically, we realize a Dirichlet-Neumann coupling: we use the kinematic interface condition as Dirichlet boundary condition in the fluid solver and the dynamic interface condition as Neumann boundary condition in the solid solver. 
Thus, concerning coupling data, we receive the deformation of the solid from the solid solver as displacement values at the coupling interface and we return forces on the coupling interface to the solid solver. 
This example problem is representative for many preCICE users: an existing (in-house) fluid solver should be coupled to an off-the-shelf solid solver, which is already adapted for preCICE. 
The simplified code of the uncoupled fluid solver is depicted in \autoref{code:original}. \incode{u} is the current solution, for example velocity and pressure values. We use an adaptive time step size, computed in line 3, and solve one time step in line 4.

\begin{listing}[h!]
\caption{Original uncoupled fluid solver in Python}
\small
\label{code:original}
\begin{minted}[mathescape,linenos,numbersep=5pt,gobble=0,frame=none,framesep=20mm,escapeinside=||,breaklines]{python}
u = initialize_solution()
while t < t_end: # main time loop
    dt = compute_adaptive_dt() 
    u = solve_time_step(dt, u) # returns new solution
    t = t + dt    
\end{minted}
\end{listing}

\paragraph{Creating a handle to preCICE} \autoref{code:example} shows the fully-coupled fluid solver. The preCICE API is used at multiple locations, which we explain in the following paragraphs. For the sake of simplicity, we do not develop a general stand-alone adapter, but directly use the preCICE API in the fluid code -- meaning, we develop an \textit{adapted code}. In \autoref{sec:adapters}, we give an overview of several \textit{real} adapters. As the fluid code is written in Python, we make use of the Python bindings of preCICE: preCICE is imported in line 1. In line 3, the solver interface of preCICE is created. We pass the name of the solver and the preCICE configuration file. The latter defines the overall coupling topology (who is coupled to whom) and the used coupling methods (acceleration, data mapping, communication, etc.). We come back to this file later. Moreover, for parallel coupled codes, we need to give the current parallel rank (here 0) and the number of ranks (here 1) to preCICE.
The solver interface is initialized in line 13. Here, preCICE performs several first steps, such as setting up internal data structures and creating communication channels. In the end, the solver interface is finalized in line 42. Internal data structures are torn down and communication channels are closed.

\paragraph{Coupling meshes} Coupling data and coupling meshes are referred to by IDs, which are collected in lines 5 to 8.
The coupling mesh is defined before the initialization in line 11. preCICE treats coupling meshes as (unstructured) clouds of vertices, arranged in two-dimensional arrays of size vertices by dimension. Certain features of preCICE (e.g., nearest-projection data mapping) require mesh connectivity in addition. To this end, edges, triangles, and quads can optionally be defined, a step which we do not show in the example. 
The control of the end of the simulation is handed over to preCICE in line 17 to steer a synchronized end of all participants.
On the coupling mesh, coupling data structures are accessed in lines 24 and 32. The example uses specific calls for the vector-valued displacement and force values. The displacement values are used as Dirichlet boundary condition, here depicted as additional input of \inpython{solve_time_step} in line 28. The force values are computed from the current solution by means of a helper function in line 31.

\begin{listing}[h!]
\caption{An adapted fluid solver written in Python. While preCICE is a C++ library, bindings for C++, C, Fortran, Python, and MATLAB make it possible to couple a large variety of participants in a minimally invasive way.}
\small
\label{code:example}
\begin{minted}[mathescape,linenos,numbersep=5pt,gobble=0,frame=none,framesep=20mm,escapeinside=||,breaklines]{python}
import precice

|{\color{my_orange}interface}| = precice.Interface("Fluid", "precice-config.xml", 0, 1)

mesh_id = interface.get_mesh_id("Fluid-Mesh")

displ_id = interface.get_data_id("Displacement", mesh_id)
force_id = interface.get_data_id("Force", mesh_id)

positions = ... #define interface mesh, 2D array with shape (n, dim)
vertex_ids = |{\color{my_orange}interface}|.set_mesh_vertices(mesh_id, positions)

precice_dt = |{\color{my_orange}interface}|.initialize()

u = initialize_solution()

while |{\color{my_orange}interface}|.is_coupling_ongoing(): # main time loop
        
    if |{\color{my_orange}interface}|.is_action_required(precice.action_write_iteration_checkpoint()):
        u_checkpoint = u
        interface.mark_action_fulfilled(precice.action_write_iteration_checkpoint())    

    # returns 2D array with shape (n, dim)
    displacements = |{\color{my_orange}interface}|.read_block_vector_data(displ_id, vertex_ids)

    dt = compute_adaptive_dt() 
    dt = min(precice_dt, dt)
    u = solve_time_step(dt, u, displacements) # returns new solution
    
    # returns 2D array with shape (n, dim)
    forces = compute_forces(u) 
    |{\color{my_orange}interface}|.write_block_vector_data(force_id, vertex_ids, forces)
    
    precice_dt = |{\color{my_orange}interface}|.advance(dt)

    if |{\color{my_orange}interface}|.is_action_required(precice.action_read_iteration_checkpoint()):
        u = u_checkpoint
        |{\color{my_orange}interface}|.mark_action_fulfilled(precice.action_read_iteration_checkpoint())
    else: # continue to next time step 
        t = t + dt    

|{\color{my_orange}interface}|.finalize()

\end{minted}
\end{listing}

\begin{figure}[h!]
\centering
\includegraphics[width=\textwidth]{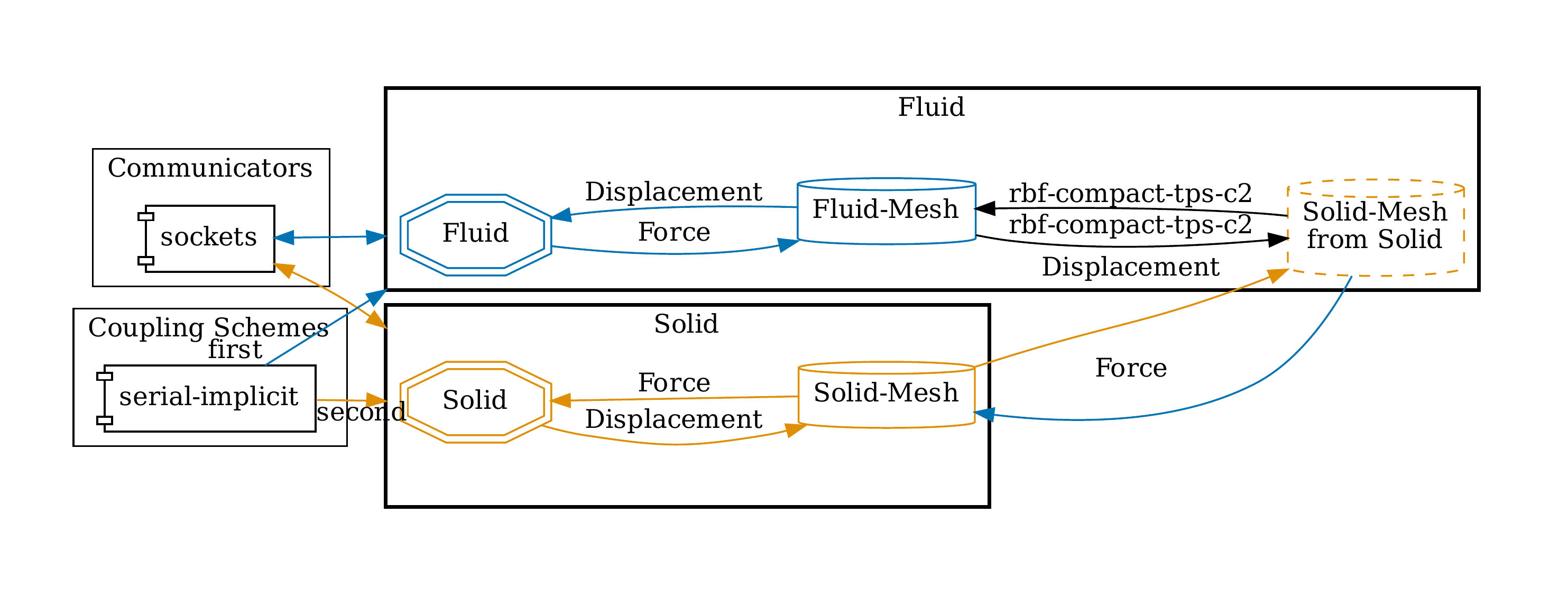}
\caption{
\label{fig:config}
Auto-generated visualization of the preCICE configuration of \autoref{code:config} using the \textit{Config Visualizer} of preCICE.}
\end{figure}

\paragraph{Configuration} \autoref{code:config} gives an excerpt of a fitting preCICE configuration for our FSI example. The dimension of the scenario is specified in line 1 and can be either two or three. Two participants, \incode{Fluid} and \incode{Solid}, are configured. \incode{Fluid} uses the mesh from \incode{Solid} in line 13. This way, we can define data mappings between both meshes in lines 16 and 17. Here, we use RBF data mappings with compact thin-plate splines as basis functions. In line 22, we configure a TCP/IP sockets connection between both participants. Finally, in lines 24 to 31, a serial implicit coupling between both participants with an IQN-ILS acceleration is defined. \incode{Fluid} is the first participant, meaning that it starts each iteration. preCICE comes with a standalone Python tool called \textit{Config Visualizer}\footnote{preCICE config visualizer: \url{https://github.com/precice/config-visualizer}}, which helps understanding and debugging preCICE configuration files. The tool generates graphviz dot files\cite{Ellson03graphviz}, which can be, for example, converted to PDF. The generated PDF output is shown in \autoref{fig:config} for the example configuration.

\begin{listing}[h!]
\caption{Excerpt of a preCICE configuration file. Two participants \incode{Fluid} and \incode{Solid} are coupled.}
\small
\label{code:config}
\begin{minted}[mathescape,linenos,numbersep=5pt,gobble=0,frame=none,framesep=20mm,escapeinside=||,breaklines]{xml}
<solver-interface dimensions="3">   
  <data:vector name="Force"/>
  <data:vector name="Displacement"/> 
 
  <mesh name="Fluid-Mesh">
    <use-data name="Displacement"/>
    <use-data name="Force"/>
  </mesh>
  <mesh name="Solid-Mesh"> ... </mesh>

  <participant name="Fluid">
    <use-mesh name="Fluid-Mesh" provide="yes"/>
    <use-mesh name="Solid-Mesh" from="Solid"/>
    <write-data name="Force" mesh="Fluid-Mesh"/>
    <read-data name="Displacement" mesh="Fluid-Mesh"/>
    <mapping:rbf-compact-tps-c2 from="Fluid-Mesh" constraint="conservative" .../>
    <mapping:rbf-compact-tps-c2 from="Solid-Mesh" constraint="consistent" .../>
  </participant>
    
  <participant name="Solid"> ... </participant>
    
  <m2n:sockets from="Fluid" to="Solid" />
    
  <coupling-scheme:serial-implicit>
    <participants first="Fluid" second="Solid"/>
    <time-window-size value="1e-3"/> 
    <exchange data="Force" mesh="Solid-Mesh" from="Fluid" .../> 
    <exchange data="Displacement" mesh="Solid-Mesh" from="Solid" .../>
    ...
    <acceleration:IQN-ILS> ... </acceleration:IQN-ILS>
  </coupling-scheme:serial-implicit>
</solver-interface>
\end{minted}
\end{listing}

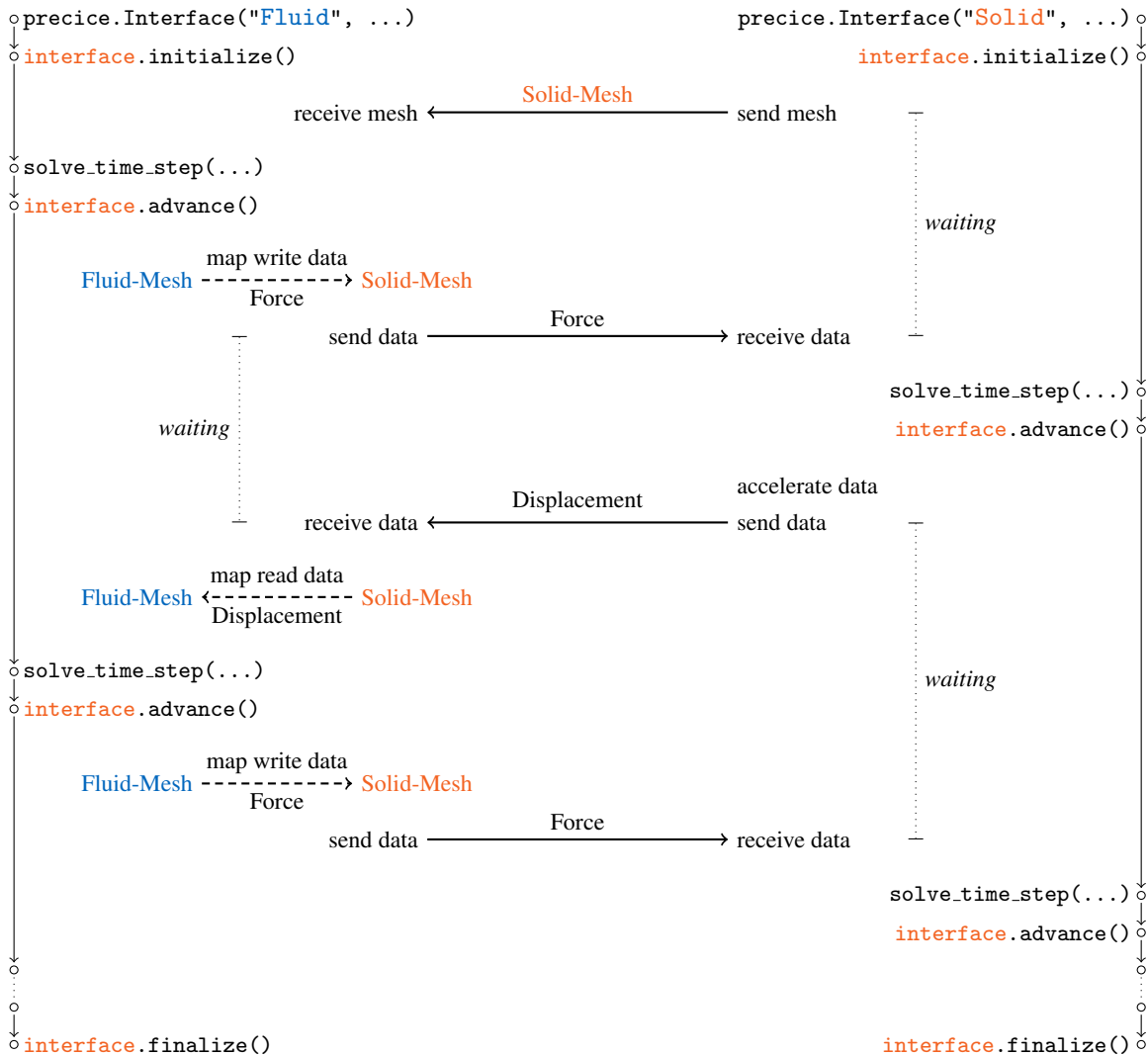
\begin{figure}[h!]
\begin{tikzpicture}

 \coordinate (FE) at (0,0.25);
 \coordinate (SE) at (15,0.25);
 \coordinate (F0) at (0,14);
 \coordinate (S0) at (15,14);
 \draw (F0) circle (1.5pt);
 \draw (FE) circle (1.5pt);
 \draw (S0) circle (1.5pt);
 \draw (SE) circle (1.5pt);
 
 \node[right] at (F0) {\small \texttt{precice.Interface("{\normalsize\color{blue1}{Fluid}}", \ldots)}};
 \node[left] at (S0) {\small \texttt{precice.Interface("{\normalsize\color{my_orange}{Solid}}", \ldots)}};

\coordinate (F2) at (0,13.5);
\coordinate (S2) at (15,13.5);
\node[right] at (F2) {\small \texttt{{\color{my_orange}{interface}}.initialize()}};
\node[left] at (S2) {\small \texttt{{\color{my_orange}{interface}}.initialize()}};
\draw (F2) circle (1.5pt);
\draw (S2) circle (1.5pt);

\coordinate (F_mesh) at (5.5,12.75);
\node[left] at (F_mesh) {\small receive mesh};
\coordinate (S_mesh) at (9.5,12.75);
\node[right] at (S_mesh) {\small send mesh};
\draw[->,thick] (S_mesh) -- (F_mesh) node[midway,above] {\small \color{my_orange} Solid-Mesh};

\coordinate (F3) at (0,12);
\coordinate (S3) at (15,9);
\node[right] at (F3) {\small \texttt{solve\_time\_step(\ldots)}};
\node[left] at (S3) {\small \texttt{solve\_time\_step(\ldots)}};
\draw (F3) circle (1.5pt);
\draw (S3) circle (1.5pt);

\coordinate (F4) at (0,11.5);
\coordinate (S4) at (15,8.5);
\node[right] at (F4) {\small \texttt{{\color{my_orange}{interface}}.advance()}};
\node[left] at (S4) {\small \texttt{{\color{my_orange}{interface}}.advance()}};
\draw (F4) circle (1.5pt);
\draw (S4) circle (1.5pt);

\coordinate (F_map1) at (2.5,10.5);
\node[left] at (F_map1) {\small \color{blue1} Fluid-Mesh};
\coordinate (F_map2) at (4.5,10.5);
\node[right] at (F_map2) {\small \color{my_orange} Solid-Mesh};
\draw[->, thick, densely dashed] (F_map1) -- (F_map2) node[midway,above] {\small map write data} node[midway,below] {\small Force};

\draw[|-|, dotted] (12,12.75) to node[midway,right] {\small \textit{waiting}} (12,9.75);

\coordinate (F_com1) at (5.5,9.75);
\node[left] at (F_com1) {\small send data};
\coordinate (S_com1) at (9.5,9.75);
\node[right] at (S_com1) {\small receive data};
\draw[->,thick,out=0,in=180] (F_com1) to node[midway,above] {\small  Force} (S_com1);

\coordinate (F_com2) at (5.5,7.25);
\node[left] at (F_com2) {\small receive data};
\coordinate (S_com2) at (9.5,7.25);
\node[right] at (S_com2) {\small send data};
\draw[->,thick, out=180,in=0] (S_com2) to node[midway,above] {\small   Displacement} (F_com2);

\draw[|-|, dotted] (3,9.75) to node[midway,left] {\small \textit{waiting}} (3,7.25);

\node[right] at (9.5,7.75) {\small accelerate data};

\coordinate (F_map3) at (2.5,6.25);
\node[left] at (F_map3) {\small \color{blue1} Fluid-Mesh};
\coordinate (F_map4) at (4.5,6.25);
\node[right] at (F_map4) {\small \color{my_orange} Solid-Mesh};
\draw[->, thick, densely dashed] (F_map4) -- (F_map3) node[midway,above] {\small map read data} node[midway,below] {\small Displacement};

\coordinate (F5) at (0,5.25);
\coordinate (S5) at (15,2.25);
\node[right] at (F5) {\small  \texttt{solve\_time\_step(\ldots)}};
\node[left] at (S5) {\small \texttt{solve\_time\_step(\ldots)}};
\draw (F5) circle (1.5pt);
\draw (S5) circle (1.5pt);

\coordinate (F6) at (0,4.75);
\coordinate (S6) at (15,1.75);
\node[right] at (F6) {\small \texttt{{\color{my_orange}{interface}}.advance()}};
\node[left] at (S6) {\small \texttt{{\color{my_orange}{interface}}.advance()}};
\draw (F6) circle (1.5pt);
\draw (S6) circle (1.5pt);

\coordinate (F_map5) at (2.5,3.75);
\node[left] at (F_map5) {\small \color{blue1} Fluid-Mesh};
\coordinate (F_map6) at (4.5,3.75);
\node[right] at (F_map6) {\small \color{my_orange} Solid-Mesh};
\draw[->, thick, densely dashed] (F_map5) -- (F_map6) node[midway,above] {\small map write data} node[midway,below] {\small Force};

\coordinate (F_com3) at (5.5,3);
\node[left] at (F_com3) {\small send data};
\coordinate (S_com3) at (9.5,3);
\node[right] at (S_com3) {\small receive data};
\draw[->,thick,out=0,in=180] (F_com3) to node[midway,above] {\small   Force} (S_com3);

\draw[|-|, dotted] (12,7.25) to node[midway,right] {\small \textit{waiting}} (12,3);

\coordinate (F7) at (0,1.25);
\coordinate (S7) at (15,1.25);
\draw (F7) circle (1.5pt);
\draw (S7) circle (1.5pt);

\coordinate (F8) at (0,0.75);
\coordinate (S8) at (15,0.75);
\draw (F8) circle (1.5pt);
\draw (S8) circle (1.5pt);

\node[right] at (FE) {\small \texttt{{\color{my_orange}{interface}}.finalize()}};
\node[left] at (SE) {\small \texttt{{\color{my_orange}{interface}}.finalize()}};

\draw [->, shorten <=.1cm, shorten >=.1cm] (F0) to (F2);
\draw [->, shorten <=.1cm, shorten >=.1cm] (F2) to (F3);
\draw [->, shorten <=.1cm, shorten >=.1cm] (F3) to (F4);
\draw [->, shorten <=.1cm, shorten >=.1cm] (F4) to (F5);
\draw [->, shorten <=.1cm, shorten >=.1cm] (F5) to (F6);
\draw [->, shorten <=.1cm, shorten >=.1cm] (F6) to (F7);
\draw [dotted, shorten <=.1cm, shorten >=.1cm] (F7) to (F8);
\draw [->, shorten <=.1cm, shorten >=.1cm] (F8) to (FE);
\draw [->, shorten <=.1cm, shorten >=.1cm] (S0) to (S2);
\draw [->, shorten <=.1cm, shorten >=.1cm] (S2) to (S3);
\draw [->, shorten <=.1cm, shorten >=.1cm] (S3) to (S4);
\draw [->, shorten <=.1cm, shorten >=.1cm] (S4) to (S5);
\draw [->, shorten <=.1cm, shorten >=.1cm] (S5) to (S6);
\draw [->, shorten <=.1cm, shorten >=.1cm] (S6) to (S7);
\draw [dotted, shorten <=.1cm, shorten >=.1cm] (S7) to (S8);
\draw [->, shorten <=.1cm, shorten >=.1cm] (S8) to (SE);
\end{tikzpicture}
\caption{
\label{fig:schedule}
Overall flow of coupling steps resulting from the preCICE configuration of \autoref{code:config}.
The serial coupling scheme leads to a staggered execution of both participants, one after the other. Both participants wait in \inpython{initialize} and \inpython{advance} for synchronization. We assume identical time step sizes in both participants.
}
\end{figure}

\paragraph{Coupling flow} The actual coupling between participants happens within the single preCICE API function \inpython{advance}, line 34 in \autoref{code:example}. This includes communication, mapping, and acceleration of coupling data -- whatever methods are defined in the preCICE configuration. To better understand the order and relation of individual coupling steps, \autoref{fig:schedule} depicts the overall coupling flow when using the example configuration of \autoref{code:config}.
For this visualization, we further assume that both participants use identical time step sizes.
During the initialization, \incode{Solid-Mesh} is sent from \incode{Solid} to \incode{Fluid}. 
A serial coupling scheme leads to a staggered execution of both participants: one after the other. This implies, in particular, that the behavior of both participants within the preCICE API functions cannot be symmetric.
In the example, \incode{Fluid} is the first participant of the coupling scheme. 
This means that, after the first time step of \incode{Fluid}, the first \inpython{advance} sends force values to \incode{Solid}, as can be seen in the figure. This coupling data is, however, already received in \inpython{initialize} of \incode{Solid}, such that the solver can use it in its first time step. The first displacement values are then sent at the start of the first \inpython{advance} of \incode{Solid} and received at the end of the first advance of \incode{Fluid}.  
Data is mapped in both direction within \inpython{advance} of \incode{Fluid}. Convergence acceleration of the coupling iteration is always executed in \inpython{advance} of the second participant, here \incode{Solid}.
Please note that a different preCICE configuration could lead to a completely different order and relation of steps: the roles of first and second could be swapped, one or both data mappings could be computed on \incode{Solid}, or the serial coupling scheme could be replaced by a parallel one, to only name a few choices. All these changes can be configured at runtime. The adapted fluid code in \autoref{code:example} remains unchanged -- and would remain unchanged even if \incode{Solid} would be coupled with a third participant.

\paragraph{Timestepping} So far, we assumed  that the coupled participants use matching time step sizes. preCICE is, however, also able to handle non-matching time step sizes. 
Then, data is only exchanged at the end of each time window, defined in the preCICE configuration, line 26 of \autoref{code:config}. Alternatively, the time window size can also be imposed by the first participant.
If a solver uses a smaller time step size than the time window size, it subcycles within the time window. This means, in particular, that the same coupling data is used throughout the time window, which can reduce the time discretization order of the coupled codes. We are currently working on a higher order time representation of coupling data to sample from, a coupling procedure known as waveform iteration~\cite{QNWI}.
To allow preCICE to track the time of a solver, the current time step size needs to be passed to preCICE in \incode{advance}, cf.~line 34 in \autoref{code:example}. preCICE then returns the remaining time within the current time window, which the coupled solver has to respect. Therefore, in line 27, the solver's time step size is restricted, if required.

\paragraph{Implicit coupling}
We still need to explain how implicit coupling is realized. Please remember that by implicit coupling we mean the repetition of time windows until sufficient convergence of coupling data (cf.~\autoref{ssec:cplscheme}). To this end, a coupled solver needs to be able to move backwards in time, which we realize by writing and reading checkpoints of the complete internal solver state (lines 20 and 37 in \autoref{code:example}). Writing checkpoints is required when entering time windows for the first time and reading checkpoints is required at the end of a time window, whenever convergence is not achieved. As the solver does not know anything about the coupling scheme, preCICE tells the solver when it is time to write and read checkpoints in lines 19 and 36. At the end of the loop body, time is only increased when convergence is achieved, in line 40. Please note that, with this checkpointing mechanism, nested time and coupling loops are not necessary, but everything can be handled within one \incode{while} loop.

\paragraph{Help and further information}
We cannot explain all preCICE API functions and all configuration options with this single example. Please also consider the official user documentation\footnote{preCICE documentation: \url{https://precice.org/docs.html}}, which includes complete API and configuration references.
preCICE supports adapter development by extensive sanity checks of correct API usage -- a simple example: \texttt{advance} cannot be called before \inpython{initialize}. Moreover, the preCICE configuration is checked against the configuration reference and extensive logging is configurable.

 \section{Official adapters}
\label{sec:adapters}

A library such as preCICE can only live as part of an application (a solver) that calls it.
To call preCICE, the solver needs to contain code that knows how to interact with the library.
We saw in \autoref{sec:library} that this additional code is short, but the user should be able to start setting up
a coupled simulation at the level of describing a scenario, not at the level writing code for each of the involved solvers. 

To lower the entry barrier and to make sure that the majority of users can keep using popular solvers
with the latest versions of preCICE, we have developed a set of official adapters, which we host and maintain in their
own repositories under the preCICE GitHub organization\footnote{All preCICE repositories on GitHub: \url{https://github.com/precice/}}.
This allows each project to follow a fitting development cycle and makes it easier for the community to contribute and to adopt projects upstream.
As the collection of adapters grows, such community contributions
are crucial, not only in fixes and features, but also in assuming maintainer roles.

We present here all mature official adapters to date. All the solvers discussed in this section are free/open-source projects,
a fundamental property that greatly facilitates the adapter development and distribution.
For free/open-source solvers, adapters can have the form of (i) in-place source code modifications, of (ii) calls to an additional adapter class,
or of (iii) runtime plugins, wherever supported. In contrast to adapters for open-source solvers, coupling of closed-source solvers usually entails
interacting through a wrapper, API, or control files, architectures which potentially cancel fundamental features of preCICE.

We begin with OpenFOAM and SU2, two solvers primarily used for simulating fluids.
We continue with CalculiX and code\_aster, two solvers primarily used for simulating solids.
We then discuss FEniCS, deal.II, and Nutils, general FEM frameworks for which we provide various coupled examples.
At the end of the section, we list further adapters maintained by the community.
  
\subsection{OpenFOAM}

OpenFOAM\footnote{OpenFOAM website (OpenCFD): \url{https://www.openfoam.com/}. Several alternative versions/forks exist.} is a finite volume toolbox and collection of solvers primarily for CFD simulations~\cite{Weller1998_OpenFOAM}.
The OpenFOAM adapter\footnote{OpenFOAM adapter documentation: \url{https://www.precice.org/adapter-openfoam-overview.html}} is currently the most frequently used of the listed adapters and OpenFOAM represents the fluid solver in most of our tutorial cases.
It is also the adapter with the highest number of contributions in the context of student and research projects~\cite{CheungYau2016, Chourdakis2017, Risseeuw2019, Chourdakis2019_OFW14}.
The adapter is being actively developed and more features have been added in the past years by multiple contributors. A separate reference publication for the OpenFOAM-preCICE adapter is in preparation.

On the technical side, the adapter is an OpenFOAM function object, to which OpenFOAM can link at runtime.
Function objects are \emph{plug-ins} that OpenFOAM uses mainly for optional post-processing tools.
Implementing the adapter in this way allows using the adapter with any standard or in-house OpenFOAM solver (each being a stand-alone application) that supports function objects, without modifying the code of the solver~\cite{Chourdakis2017}.
The separation between the solver and the adapter has facilitated development and increased user adoption, such that we now aim for this model wherever possible. We support the latest versions of the major OpenFOAM variants,
including v1706--v2106 (ESI/OpenCFD, main adapter branch) and 4.0--8 (The OpenFOAM Foundation, version-specific branches).
The adapter can be built from source using the WMake build system of OpenFOAM and installed into the \incode{FOAM_USER_LIBBIN} directory.

On the application side, the adapter supports conjugate heat transfer (CHT), fluid-structure interaction (FSI), and fluid-fluid coupling. 
In terms of CHT, it can read and write temperature, heat flux, sink temperature, and heat transfer coefficient, 
allowing not only for Dirichlet-Neumann, but also for Robin-Robin coupling.
As each OpenFOAM boundary condition supports different operations, the user needs to set compatible boundary conditions for
each interface: \incpp{fixedValue} for Dirichlet boundaries, \incpp{fixedGradient} for Neumann boundaries, or \incpp{mixed} for Robin boundaries.
The adapter computes the heat flux in a similar way as the \incode{wallHeatFlux} function object distributed with OpenFOAM.
This requires the adapter to distinguish solvers into types (compressible, incompressible, and basic),
as the underlying classes do not share a common interface.
The refactored transport models class hierarchy introduced in OpenFOAM 8\footnote{OpenFOAM 8 Release Notes: \url{https://openfoam.org/version/8/}}
eliminates the need for a distinction between compressible and incompressible solvers,
but, at the same time, such breaking changes in the public interfaces make supporting multiple OpenFOAM versions with the same code even more challenging.
For this reason, we are planning to separate the functional from the application-specific components of the adapter even further
and delegate part of the work to OpenFOAM in future work.

In terms of FSI, the adapter can read absolute and relative displacements (defined on either face nodes or face centers), while it can write forces and stresses (on face centers).
In order to read displacements, the reader needs to set a \incode{fixedValue} boundary condition for the displacement, as well as a \incode{movingWallVelocity}
for the velocity. At least the mesh motion solver \incode{displacementLaplacian} is known to work.
Again, a distinction between compressible and incompressible solvers is needed.

The adapter also supports fluid-fluid coupling, reading and writing pressure, velocity, as well as their gradients.
This is an area of active research and further development.

In addition to providing face nodes/centers to preCICE, the adapter can also construct and provide face edges and triangles,
thus supporting the nearest-projection mapping feature of preCICE.

The coupling fields, patch names, participant name, path to the preCICE configuration file, and more are configured
in the adapter configuration file \inbash{system/preciceDict}, an OpenFOAM dictionary. In addition to that, the user needs to specify the adapter function object in the \inbash{system/controlDict} and set compatible boundary conditions for any fields to be read.
Several tutorial cases are available, using the solvers pimpleFoam, buoyantPimpleFoam, buoyantSimpleFoam, and laplacianFoam.
There are also several examples in which the preCICE community has used the adapter (as-is or modified)
with further standard and in-house OpenFOAM solvers, including cases with compressible multiphase flow~\cite{Seufert2019_COUPLED}
and cases with volume coupling~\cite{Rousset2020_preCICE2020, Arya2020_preCICE2020, scheiblhofer2019coupling}.

The code is available on GitHub\footnote{OpenFOAM adapter on GitHub: \url{https://github.com/precice/openfoam-adapter}, GPLv3} under the GPLv3 license,
the same license as OpenFOAM. The code contains also comments with instructions on extending it.

The OpenFOAM community is currently developing (and has already done so in the past) very important contributions in bringing multi-physics
simulations to OpenFOAM. Prominent examples include the standard CHT solver chtMultiRegionFoam
\footnote{OpenFOAM User Guide -- chtMultiRegionFoam: \url{https://www.openfoam.com/documentation/guides/latest/doc/guide-applications-solvers-heat-transfer-chtMultiRegionFoam.html}.}
and the FSI solvers fsiFoam~\cite{Tukovic2018_fsiFoam} and solids4Foam~\cite{Cardiff2018_solids4Foam}.
These projects solve the respective multi-physics problem monolithically, at least software-wise: they implement both single-physics domains inside OpenFOAM, compiled in the same executable.
In contrast, the OpenFOAM-preCICE adapter provides additional flexibility to couple OpenFOAM with any other solver via preCICE.
Other projects also apply the partitioned approach to extend OpenFOAM with the functionality of other codes, including
OpenFPCI (ParaFEM)~\cite{Hewitt2019_OpenFPCI},
EOF-Library (Elmer)~\cite{Vencels2019_EOF}, and ATHLET-OpenFOAM coupling~\cite{Herb2014_ATHLET}.
OpenFOAM has previously also been coupled with preCICE using independent (unofficial) adapters in the theses of Kevin Rave~\cite{Rave2017_thesis} (CHT) and David Schneider~\cite{Schneider2018_thesis} (FSI),
as well as in the project FOAM-FSI of David Blom for foam-extend\footnote{FOAM-FSI on GitHub: \url{https://github.com/davidsblom/FOAM-FSI}}.
The official OpenFOAM-preCICE adapter differs in providing a general-purpose adapter for preCICE for a wide range of users and use cases.

\subsection{SU2} 

SU2\footnote{SU2 website: \url{https://su2code.github.io/}} (Stanford University Unstructured) is a finite volume solver which provides compressible and incompressible solver variants for CFD~\cite{SU2-paper}. The SU2 adapter~\cite{AlexRusch2016} supports SU2 v6.0 ``Falcon'' and contributions from the community are particularly welcome in this project\footnote{SU2 adapter on GitHub: \url{https://github.com/precice/su2-adapter}, LGPLv3}.

As SU2 is written in C++, the adapter directly uses the C++ API of preCICE. The API calls are provided by an adapter class, which is utilized in the SU2 solver files (e.g., in \incode{SU2_CFD.cpp}). An installation script copies the modified, version-specific files to specific locations in the SU2 source code, which is then built normally.

The adapter is designed for FSI applications and supports reading forces and writing absolute or relative displacements. The adapter is configured via additional options in the native configuration file of SU2 and the modified solver can be executed with or without enabling preCICE. The user can set the name of the marker which identifies the FSI interface in the geometry file of SU2 and can run simulations with multiple coupling interfaces.

Similarly to the adapter, SU2 is also used as a solver in other coupling projects, for example CUPyDO~\cite{CUPyDO-THOMAS201969}, where the CUPyDO coupler calls SU2 via a Python wrapper. In addition to external coupling options, SU2 offers monolithic capabilities for multi-physics simulation such as FSI and CHT~\cite{SU2-multiphysics, SU2-FSI-monolithic}.

\subsection{CalculiX} 

CalculiX\footnote{CalculiX website: \url{http://www.calculix.de/}} is an open-source FEM code~\cite{Dhondt2004_CalculiX}. CalculiX offers a variety of solvers and the CalculiX adapter\footnote{CalculiX adapter on GitHub: \url{https://github.com/precice/calculix-adapter}, GPLv2} enables coupling some of these solvers via preCICE. The adapter supports the \emph{dynamic linear geometric} and the \emph{dynamic nonlinear geometric} solvers of CalculiX for coupled FSI problems, as well as \emph{static thermal} and \emph{dynamic thermal} solvers for CHT problems. The CalculiX adapter~\cite{CheungYau2016, Uekermann2017_Adapters} is compatible with CalculiX 2.16, it is regularly updated for new CalculiX releases, and maintains support for older versions in version-specific branches.

The adapter directly modifies the source code of CalculiX and produces a stand-alone executable \inbash{ccx_preCICE}, which can be used both for coupled and for CalculiX-only simulations: the flag \inbash{-precice-participant <name>} enables the preCICE adapter. All preCICE-related functionality is provided in additional source files supplied with the adapter.

The adapter is configured through a YAML file that specifies the coupling interface names, coupling data variable types, and type of interface (mesh nodes, or mesh nodes with connectivity).
 It can be used with both linear and quadratic tetrahedral (C3D4 and C3D10) and hexahedral (C3D8 and C3D20) solid elements, as well as S3 and S6 tetrahedral shell elements. The adapter also supports nearest-projection mapping. The latter requires a second mesh file (\inbash{.sur}) defining the interface mesh connectivity. The configuration is described in the CalculiX adapter documentation\footnote{CalculiX adapter documentation: \url{https://precice.org/adapter-calculix-overview.html}}.

CalculiX is written in C and Fortran. However, all preCICE functionality is incorporated using the C bindings of preCICE. To perform coupled simulations with CalculiX in parallel on shared-memory systems, the adapter treats CalculiX as a serial participant, while the CalculiX linear solver is executed in parallel. More details on how to run CalculiX in parallel is available in the CalculiX user manual~\cite{CalculiX_docs_217}.

\subsection{code\_aster}

code\_aster\footnote{code\_aster website: \url{https://www.code-aster.org/}} is an FEM code in Fortran (with a Python API) developed by EDF France,
offering solvers for heat transfer, structural analysis, and more, with one of the main applications being nuclear power engineering.
The code\_aster adapter~\cite{CheungYau2016}\footnote{code\_aster adapter documentation: \url{https://www.precice.org/adapter-code_aster.html}}
is compatible code\_aster 14.4 and 14.6, while it is being maintained to work with the latest versions of preCICE.

On the technical side, the adapter is a single \inbash{adapter.py} file providing methods that are used in an example
\inbash{adapter.comm} command file for CHT simulations. As a Python code, the adapter depends on the preCICE Python bindings. It can be installed by copying the adapter file into the \inbash{ASTER_ROOT/14.4/lib/aster/Execution/} directory.

On the application side, the adapter can currently read and write sink temperature and heat transfer coefficient, thus supporting Robin-Robin coupling for CHT.
The preCICE tutorials include such a Robin-Robin CHT case with code\_aster and the steady-state fluid solver buoyantSimpleFoam.
code\_aster can be programmed by the user using Python and code\_aster command files, \inbash{.comm}, which are included in predefined file unit numbers.
The adapter is configured using a command file \inbash{config.comm}, included as \inpython{UNITE=90} by the \inbash{adapter.comm}.
The case is defined in the command file \inbash{def.comm}, which is included as \inpython{UNITE=91}.

The code is available on GitHub\footnote{code\_aster adapter on GitHub: \url{https://github.com/precice/code_aster-adapter}, GPLv2}
under the GPLv2 license and can be easily extended by adding more coupling fields in \inbash{adapter.py}
and providing their names as arguments to the method \inpython{adapter.writeCouplingData()}.

\subsection{FEniCS}

FEniCS is an open-source general-purpose FEM package with a high-level Python interface~\cite{AlnaesBlechta2015a}. FEniCS does not provide ready-to-use solvers, but instead provides a broad range of tools for solving partial differential equations with a high level of abstraction. A wide variety of examples is provided in the FEniCS project to illustrate its usage~\cite{Langtangen2016}. The FEniCS-preCICE adapter facilitates coupling of FEniCS-based solvers using preCICE. We give here a short overview of the adapter, for which you can read more details in the FEniCS-preCICE reference paper~\cite{rodenberg2021fenicsprecice}.

The FEniCS adapter~\cite{Monge2018, Hertrich2019} provides high-level functionalities that the users can incorporate in their FEniCS-based solvers. For using the adapter, a user needs to understand partitioned coupling and the general approach of preCICE, but does not need to understand how to use the generic preCICE API with FEniCS data structures.

The adapter is configured using a JSON file. Afterwards, the adapter is initialized by providing a FEniCS \inpython{Mesh} and a \inpython{SubDomain} to define the coupling boundary. Connectivity information is automatically obtained from the \inpython{Mesh} and nearest projection mapping is directly supported. The user needs to define FEniCS \inpython{FunctionSpace} objects to provide information on the data being coupled.

For data exchange, the adapter offers a simple function \inpython{adapter.write_data(solution)}. This function samples a given solution on the previously defined coupling mesh and writes the samples to preCICE. The function \inpython{coupling_data = adapter.read_data()} returns the interface data provided by preCICE. The adapter provides two possibilities to transform this raw \inpython{coupling_data} to a boundary condition that can be used in FEM: (1) An \inpython{Expression} can be generated and used as a functional representation of provided \inpython{coupling_data} via interpolation or (2) a \inpython{PointSource} can be generated to apply point-wise loads. Both approaches have their respective use-cases for FEniCS users (see~\cite{rodenberg2021fenicsprecice}), but a user can also use the raw \inpython{coupling_data} to create boundary conditions depending on the individual requirements. Finally, the adapter also provides convenient tools for checkpointing and steering.

The adapter supports the built-in parallelism of FEniCS and uses its domain decomposition. If the coupling interfaces are decomposed over multiple ranks, the adapter implements additional inter-process communication at the interface between two ranks.

Many packages similar to FEniCS exist, such as firedrake~\cite{Rathgeber2016}, or the FEniCS successor \mbox{FEniCS-X}\footnote{DOLFINx (basis of FEniCS-X) on GitHub: \url{https://github.com/FEniCS/dolfinx}}. The adapter is not designed to work with these packages, but it can serve as a template for the development of specialized adapters. Possible contributions to the FEniCS adapter include the extension to other coupling physics, such as electromagnetic applications.

The adapter is distributed under the LGPLv3.0 license on PyPI\footnote{Package \inpython{fenicsprecice} on PyPI: \url{https://pypi.org/project/fenicsprecice/}}. If preCICE is installed on the system, the latest version of the adapter can be installed via pip. With FEniCS being a Python-based package, the adapter depends on the preCICE Python bindings, which are automatically installed with the adapter. The source code of the adapter is available on GitHub\footnote{FEniCS adapter on GitHub: \url{https://github.com/precice/fenics-adapter}, LGPLv3.0} and user documentation can be found on the preCICE website\footnote{FEniCS adapter documentation: \url{https://www.precice.org/adapter-fenics.html}}.

Related work to solve multi-physics problems with FEniCS includes, for example, the monolithic fluid-structure interaction solver turtleFSI~\cite{Bergersen2020} written in FEniCS, as well as \mbox{FENICS-HPC}~\cite{FENICS-HPC}. FEniCS extensions such as multiphenics\footnote{multiphenics website: \url{https://mathlab.sissa.it/multiphenics}} have also been developed to promote prototyping of multi-physics problems.

\subsection{deal.II} 

deal.II~\cite{dealII92,dealii2019design} is a general-purpose FEM library written in C++. Similar to FEniCS, deal.II does not include ready-to-use solvers, but allows users to write their own application codes by providing an easy-to-use interface to complex FEM-specific data structures and algorithms. The library
provides state-of-the art numerical techniques and their implementations leverage distributed memory computations, vectorization, threading and matrix-free implementations, which have been proven to scale up to whole supercomputers~\cite{dealiiDistributed,kronbichler2018performance}.

While deal.II is a general-purpose library, the deal.II adapter focuses on a subset of relevant applications and features. Instead of trying to provide a general-purpose deal.II adapter, we provide examples for users that want to develop their own preCICE-enabled solvers with deal.II.
These examples\footnote{deal.II adapter on GitHub: \url{https://github.com/precice/dealii-adapter}, LGPLv3} show
linear and non-linear elastic solid mechanics codes in a coupled FSI scenario. From a user perspective, these coupled codes are ready-to-use without detailed knowledge of deal.II itself, but can also provide a starting point for own application-specific adapter developments.
Similarly to deal.II and preCICE, the examples are built using CMake and a parameter file controls solver-specific preCICE settings. Simple meshes can directly be defined in the source code, whereas external meshes can be loaded at runtime.

In addition to these examples, we have created a very basic stand-alone one-way coupling example and contributed it to the deal.II code-gallery\footnote{preCICE example contributed to the deal.II code-gallery:\\ \url{https://dealii.org/developer/doxygen/deal.II/code\_gallery\_coupled\_laplace\_problem.html}}. In this example, a Laplace problem is coupled to a time-dependent C++ boundary condition code. The example is meant to serve as a first impression of how the preCICE API looks and how to use it along with deal.II.

As the low-level design of deal.II offers a lot of freedom of implementation approaches, multi-physics simulations have also been implemented in other ways~\cite{ExaDG2020, wick2013}. 

\subsection{Nutils} 

Similar to FEniCS and deal.II, Nutils~\cite{Nutils} is also a general-purpose FEM library. 
Missing capabilities for distributed computing and the fact that Nutils is purely written in Python, including matrix assembly, makes the library somehow less performant than alternatives. The powerful and intuitive API of Nutils, however, allows for radically-fast prototyping. These points make Nutils a perfect option for the \textit{cheaper}, but possibly more complex participant of a coupled simulation, or for testing new coupling approaches.
A first partitioned heat conduction example coupling two Nutils participants was developed and validated within only a half day of work and is available as a preCICE tutorial (cf.~\autoref{sec:tutorials}). 
In general, coupling a new Nutils application code is a rather simple task and can be realized best by copying and adapting existing examples.
Defining coupling meshes and accessing coupling data is a particularly simple task  as illustrated in~\autoref{code:nutils}.
Therefore, in contrast to (for example) FEniCS, developing a general stand-alone Nutils-preCICE adapter is not necessary. 

\begin{listing}[h!]
\caption{A simplified coupled Nutils code. Due to the rich and flexible API of Nutils, defining coupling meshes and accessing coupling data are simple tasks, rendering a stand-alone Nutils-preCICE adapter unnecessary.}
\small
\label{code:nutils}
\begin{minted}[mathescape,linenos,numbersep=5pt,gobble=0,frame=none,framesep=20mm,escapeinside=||,breaklines]{python}
import precice, nutils

domain = ... # define Nutils domain
ns.u = ... # Nutils solution u in namespace ns 
|{\color{my_orange}interface}| = precice.Interface("FluidSolver", "precice-config.xml", 0, 1)
[...]
# defining a coupling mesh
coupling_boundary = domain.boundary['top']
coupling_sample = coupling_boundary.sample('gauss', degree=2)
vertices = coupling_sample.eval(ns.x)
vertex_ids = |{\color{my_orange}interface}|.set_mesh_vertices(meshID, vertices)

# instead of Gauss points, we can also couple at (sub-sampled) cell vertices
coupling_sample = couplinginterface.sample('uniform', 4) # 4 sub samples per cell

# or volume coupling
coupling_sample = domain.sample('gauss', degree=2)
 
# reading coupling data and applying as boundary condition
read_data = |{\color{my_orange}interface}|.read_block_scalar_data(read_data_id, vertex_ids)
read_function = coupling_sample.asfunction(read_data)
sqr = coupling_sample.integral((ns.u - read_function)**2) 
constraints = nutils.solver.optimize(sqr, ...) # for a Dirichlet BC

# writing data
write_data = coupling_sample.eval('u' @ ns, ...)
|{\color{my_orange}interface}|.write_block_scalar_data(write_data_id, vertex_ids, write_data)
\end{minted}
\end{listing}

In recent years, several examples have been realized. The preCICE documentation gives an up-to-date overview\footnote{Nutils adapter documentation: \url{https://precice.org/adapter-nutils.html}}. The preCICE tutorials (cf.~\autoref{sec:tutorials}) include a partitioned heat conduction case, Nutils as solid participant within a CHT scenario, and Nutils as fluid participant within an FSI scenario. The latter models the incompressible Navier-Stokes equations in an arbitrary-Lagrangian-Eulerian framework and uses a fully-second-order time integration~\cite{QNWI}. 
Moreover, a 1D compressible fluid solver in Nutils is coupled to a 3D OpenFOAM solver in~\cite{Chourdakis2019_OFW14}. Lastly, current work focuses on coupling a fracture mechanics solver in Nutils to an electro-chemistry corrosion model in FEniCS. A first prototype of the Nutils participant is available on GitHub\footnote{Coupled fracture mechanics Nutils solver: \url{https://github.com/uekerman/Coupled-Brittle-Fracture}, MIT}. 

\subsection{Further adapters} 

The aforementioned are not the only adapters published in the preCICE GitHub organization.
The organization also includes a few less-actively maintained projects, which mainly serve as
starting points for anyone who wants to build a more complete solution.
If this applies to you, we would appreciate your feedback and contributions, especially in tutorial
cases and maintenance.

\begin{itemize}
  \item ANSYS Fluent: Intended for the fluid part in FSI
  and implemented as a so-called \textit{user-defined function} plug-in.
  This adapter~\cite{Gatzhammer2015,Hertrich2019} is currently experimental\footnote{Fluent adapter on GitHub: \url{https://github.com/precice/fluent-adapter}, GPLv3}.
  \item COMSOL Multiphysics: Intended for the structure part in FSI. Similarly to Fluent,
  this is one of the earliest adapters and it is currently not actively maintained.
  \item MBDyn: Intended for the structure part in FSI. Contributed by the TU Delft Wind Energy group and irregularly extended by the community (e.g., in~\cite{Cocco2020}).
  The adapter repository\footnote{MBDyn adapter on GitHub: \url{https://github.com/precice/mbdyn-adapter}, GPLv3} includes a tutorial case which simulates 3D cavity flow with a flexible bottom surface in which MBDyn is coupled to OpenFOAM.
  \item LS-DYNA: Intended for the structure part of CHT. Not a ready-to-use adapter, but rather a detailed description on how to create an actual LS-DYNA adapter. Contributed by the LKR group at the Austrian Institute of Technology~\cite{scheiblhofer2019coupling}.
  \item Elmer FEM: Intended for the structure part in FSI. Currently under development in a student project\footnote{Elmer adapter on GitHub: \url{https://github.com/HishamSaeed/elmer-adapter} (under development)}.
\end{itemize}

Apart from these codes, you can also find a list of community-developed projects in \autoref{sec:community}.

 \section{Illustrative examples}\label{sec:tutorials}

After installing preCICE, a user typically wants to run a first coupled example case as close to their application and preferred solvers as possible. Such an example needs to be simple enough to follow without significant expertise in any of the involved solvers, but yet full-featured in terms of coupling. With this in mind, we offer a collection of tutorial cases hosted on \github{precice/tutorials}, with step-by-step guides in the preCICE documentation\footnote{preCICE tutorials documentation: \url{https://precice.org/tutorials.html}}. Such a tutorial consists of all the required instructions and configuration files necessary to run the coupled simulation, as well as convenience scripts to run, visualize, and cleanup each case. The same cases and scripts are also used in the preCICE system tests (cf.~\autoref{ssec:system_tests}) and all tutorials follow a consistent structure and naming scheme, described in the contributing guidelines\footnote{preCICE contributing guidelines: \url{https://precice.org/community-contribute-to-precice.html}}.

An important design decision of these tutorials is that every combination of the available solvers should work and give reasonably similar results,
demonstrating the plug-and-play concept of preCICE. This lets the user start from a case as close to their target as possible
and then potentially replace one of the participants with their own solver, maintaining a reference to compare with.
To achieve this goal, we had to develop features that may otherwise seem unnatural.
One of the most prominent such features is that the adapters for OpenFOAM and CalculiX (natively 3D solvers) can also work
in a \emph{2D mode}, coupling only lines of face points instead of surfaces. This automatic 2D mode works either by applying
additional interpolation to map points from the mesh nodes to the face centers, or directly switching to data available at the face centers
or to data on a predefined plane. Even if the domain in that case appears to have an out-of-plane thickness, this is only there
for the solver configuration and not important for the coupling.

Historically, the collection has grown with contributions from the community. As such, the tutorials should currently be seen as individual examples showcasing particular applications and features, rather than as a structured cookbook. We focus here on two representative cases that are available for a wide range of solvers: a CHT and an FSI example. The complete collection of tutorials is listed in~\autoref{fig:tutorials-list}.

\begin{figure}[h!]
\centering
\small
\begin{framed}
\begin{minipage}{\textwidth}
  \begin{minipage}{0.5\textwidth}
    \begin{itemize}[leftmargin=*]
      \item \textbf{Quickstart:} A 2D case coupling a channel flow in OpenFOAM with an in-house rigid body motion solver in C++.
      This is meant as the single first tutorial a new user should try, requiring as few components as possible.
      \item \textbf{Flow in a channel with a perpendicular flap:} A 2D FSI tutorial with OpenFOAM, SU2, deal.II, FEniCS, Nutils, and CalculiX.
      \item \textbf{Flow over a heated plate:} A 2D CHT tutorial with OpenFOAM, FEniCS, and Nutils.
      \item \textbf{Partitioned heat conduction:} A 2D heat conduction tutorial with FEniCS, Nutils, and OpenFOAM.
      \item \textbf{Turek-Hron FSI3:} An FSI tutorial based on the benchmark described at~\cite{turek2006proposal}, with OpenFOAM and deal.II.
      \item \textbf{Multiple perpendicular flaps:} A three-field FSI tutorial (fully implicit coupling, transient) with OpenFOAM and deal.II.
      \item \textbf{3D elastic tube:} A 3D FSI tutorial with OpenFOAM and CalculiX.
    \end{itemize}  
  \end{minipage}\hfill
  \begin{minipage}{0.5\textwidth}
    \begin{itemize}
      \item \textbf{1D elastic tube:} A 1D FSI tutorial with toy solvers in Python and C++.
      \item \textbf{Flow over a heated plate -- nearest projection:} A nearest-projection mapping variant of the \emph{flow over a heated plate} tutorial, with OpenFOAM.
      \item \textbf{Flow over a heated plate -- steady state:} A steady-state variant of the \emph{flow over a heated plate} tutorial, with OpenFOAM and code\_aster.
      \item \textbf{Heat exchanger:} A three-field CHT tutorial (explicit coupling, steady state) with OpenFOAM and CalculiX.
      \item \textbf{Partitioned heat conduction -- complex setup:} A variant of the \emph{partitioned heat conduction} tutorial with a more complex geometry, with FEniCS.
      \item \textbf{Partitioned beam:} A structure-structure coupling tutorial with CalculiX.
      \item \textbf{Partitioned pipe:} A fluid-fluid coupling tutorial with OpenFOAM.
    \end{itemize}
  \end{minipage}
\end{minipage}
\end{framed}
\caption{List of available tutorial cases}
\label{fig:tutorials-list}
\end{figure}

\subsection{Flow over a heated plate CHT tutorial}

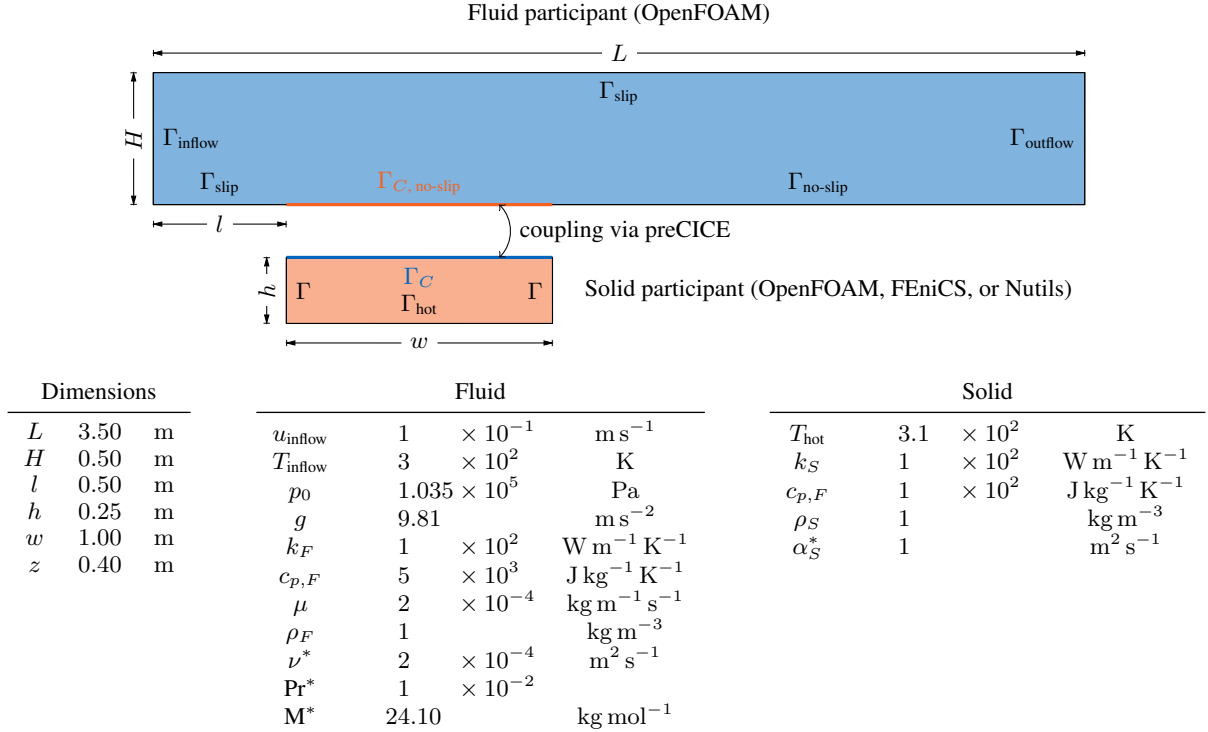
\begin{figure}[h!]
\begin{center}
\strut\vspace*{-\baselineskip}\newline
\begin{tikzpicture}[scale=3.5, every node/.style={font=\small}]

\coordinate(origin) at (0,0);

\coordinate(lengthPlate) at (1,0);
\coordinate(thicknessPlate) at (0,-0.25);

\coordinate(leftTopPlate) at ($(origin) + (0,-.2)$);
\coordinate(rightTopPlate) at ($(leftTopPlate)+(lengthPlate)$);
\coordinate(leftBottomPlate) at ($(leftTopPlate)+(thicknessPlate)$);
\coordinate(rightBottomPlate) at ($(rightTopPlate)+(thicknessPlate)$);

\coordinate(channelWidth) at (0,0.5);
\coordinate(channelPrePlate) at (-0.5,0);
\coordinate(channelPostPlate) at (2,0);

\coordinate(leftBottomInflow) at ($(origin)+(channelPrePlate)$);
\coordinate(leftTopInflow) at ($(leftBottomInflow)+(channelWidth)$);

\coordinate(rightBottomOutflow) at ($(origin)+(lengthPlate)+(channelPostPlate)$);
\coordinate(rightTopOutflow) at ($(rightBottomOutflow)+(channelWidth)$);

\draw[fill=participantBcolor!50](leftBottomInflow) rectangle (rightTopOutflow);
\coordinate(midChannel) at ($.5*(leftTopInflow)+.5*(rightTopOutflow)$);
\node[above = .5cm of midChannel]{Fluid participant (OpenFOAM)};

\draw[fill=participantAcolor!50](leftTopPlate) rectangle (rightBottomPlate);
\coordinate(midPlate) at ($.5*(rightBottomPlate)+.5*(rightTopPlate)$);
\node[right = .3cm of midPlate, anchor=west]{Solid participant (OpenFOAM, FEniCS, or Nutils)};

\draw[very thick, participantBcolor](leftTopPlate) --  node[below, black, align=center]{\textcolor{participantBcolor}{$\Gamma_C$}} (rightTopPlate);  \draw[very thick, participantAcolor](origin) -- node[above, black, align=center]{\textcolor{participantAcolor}{$\Gamma_{C\text{, no-slip}}$}} ($(origin)+(lengthPlate)$);  \draw[<->] ($(origin)+(lengthPlate)+(-.2,0)$) to[in=45, out=-45] node[right]{coupling via preCICE}($(rightTopPlate)+(-.2,0)$);
\draw(leftBottomPlate) -- node[above, black]{$\Gamma_\text{hot}$} (rightBottomPlate);  \draw(leftTopPlate) -- node[right, black]{$\Gamma$} (leftBottomPlate);  \draw(rightTopPlate) -- node[left, black]{$\Gamma$} (rightBottomPlate);  

\draw(leftBottomInflow) -- node[above, black, align=center]{$\Gamma_\text{slip}$} (origin);  \draw(rightBottomOutflow) -- node[above, black, align=center]{$\Gamma_\text{no-slip}$} ($(origin)+(lengthPlate)$);  \draw(leftTopInflow) -- node[below, black, align=center]{$\Gamma_\text{slip}$} (rightTopOutflow);  \draw(leftTopInflow) -- node[right, black, align=right]{$\Gamma_\text{inflow}$} (leftBottomInflow);  \draw(rightBottomOutflow) -- node[left, black, align=left]{$\Gamma_\text{outflow}$} (rightTopOutflow);  

\dimline[extension start length=0, extension end length=0] {($(leftBottomInflow)+(-.21em,0)$)}{($(leftTopInflow)+(-.21em,0)$)}{$H$};
\dimline[extension start length=0, extension end length=0] {($(leftTopInflow)+(0,.21em)$)}{($(rightTopOutflow)+(0,.21em)$)}{$L$};
\dimline[extension start length=0, extension end length=0] {($(leftBottomInflow)-(0,.21em)$)}{($(origin)-(0,.21em)$)}{$l$};

\dimline[extension start length=0, extension end length=0] {($(leftBottomPlate)-(.21em,0)$)}{($(leftTopPlate)-(.21em,0)$)}{$h$};
\dimline[extension start length=0, extension end length=0] {($(leftBottomPlate)-(0,.21em)$)}{($(rightBottomPlate)-(0,.21em)$)}{$w$};

\end{tikzpicture}
 \end{center}

\begin{minipage}[t]{\textwidth}
\begin{minipage}[t]{.2\textwidth}
\begin{footnotesize}
\begin{center}
\begin{tabular}[t]{c S[table-format=1.2] s}
\multicolumn{3}{c}{Dimensions}\\
\midrule
$L$ & 3.50 & \meter\\
$H$ & 0.50 & \meter\\
$l$ & 0.50 & \meter\\
$h$ & 0.25 & \meter\\
$w$ & 1.00 & \meter\\
$z$ & 0.40 & \meter
\end{tabular}
\end{center}
\end{footnotesize}
\end{minipage}
\begin{minipage}[t]{.4\textwidth}
\begin{footnotesize}
\begin{center}
\begin{tabular}[t]{c S[table-format=4.3e2] s}
\multicolumn{3}{c}{Fluid}\\
\midrule
$u_\text{inflow}$ & 1e-1 & \meter\per\second\\
$T_\text{inflow}$ & 3e2  & \kelvin\\
$p_0$ & 1.035e5 & \pascal\\
$g$        & 9.81 & \meter\per\second\squared\\  $k_F$      & 1e2  & \watt\per\meter\per\kelvin\\ $c_{p,F}$  & 5e3  & \joule\per\kilogram\per\kelvin\\  $\mu$      & 2e-4 & \kilogram\per\meter\per\second\\  $\rho_F$   & 1  & \kilogram\per\meter\cubed \\ $\nu^*$ & 2e-4 & \meter\squared\per\second\\
$\text{Pr}^*$& 1e-2 & \\  $\text{M}^*$& 24.10 & \kilogram\per\mol \\  \end{tabular}
\end{center}
\end{footnotesize}
\end{minipage}
\begin{minipage}[t]{.4\textwidth}
\begin{footnotesize}
\begin{center}
\begin{tabular}[t]{c S[table-format=4.3e2] s}
\multicolumn{3}{c}{Solid}\\
\midrule
$T_\text{hot}$ & 3.1e2 & \kelvin\\
$k_S$          & 1e2   & \watt\per\meter\per\kelvin \\  $c_{p,F}$        & 1e2 & \joule\per\kilogram\per\kelvin\\
$\rho_S$         & 1 & \kilogram\per\meter\cubed \\  $\alpha_S^*$       & 1     & \meter\squared\per\second\\  \end{tabular}
\end{center}
\end{footnotesize}
\end{minipage}
\end{minipage}

\caption{Flow over a heated plate CHT tutorial: The setup is depicted at the top, geometric and physical parameters are listed in the tables at the bottom of the figure. The fluid participant reads heat flux at the interface $\Gamma_C$, while the solid participant reads temperature. The boundary values for the inflow ($\Gamma_\text{inflow}$), outflow ($\Gamma_\text{outflow}$), and the hot bottom of the plate ($\Gamma_\text{hot}$) are listed in the tables at the bottom of the figure. All other boundaries are insulated. $u_\text{inflow}$, $T_\text{inflow}$: velocity and temperature at the inflow boundary. $p_0$: ambient pressure at all boundaries of the fluid. $T_\text{hot}$: temperature at the bottom of the plate. $g$: acceleration due to gravity. $k_F$ and $k_S$: thermal conductivity of the fluid and solid. $\rho_F$ and $\rho_S$: density of the fluid and solid. $c_{p,F}$ and $c_{p,S}$: specific heat capacity of the fluid and solid. $\alpha_S^* = k_S / (\rho_S c_{p,S})$: thermal diffusivity of the solid. $\mu$: dynamic viscosity. $\nu^* = \mu / \rho_F$: kinematic viscosity. $\text{Pr}^* = c_{p,F} \mu / k_F$: Prandtl number of the fluid. $M^* = \rho_F R T_\text{inflow} / p_0$: Molar mass of the fluid with $R$ being the gas constant. $z$ is the out-of-plane thickness: even if the coupled case is described as 2D, OpenFOAM is still a 3D solver. Quantities marked with a $^*$ are derived quantities.}
\label{fig:cht_setup}
\end{figure}

\begin{figure}[h!]
\begin{center}
\includegraphics[width=\textwidth]{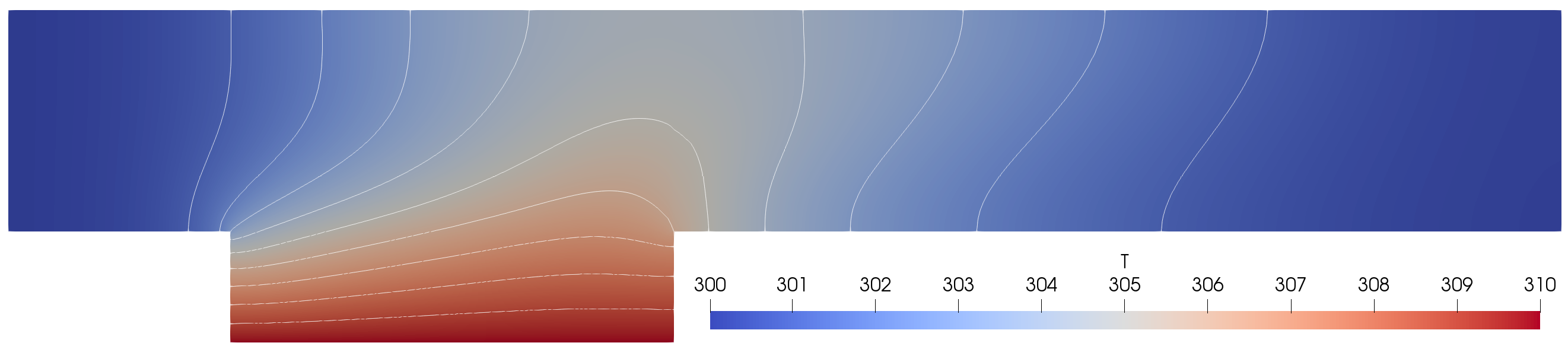}
\end{center}
\caption{Flow over a heated plate CHT tutorial: Isothermal lines for the OpenFOAM-FEniCS combination at time $t=10s$. The lines are continuous and smooth across the interface. Similar results are observed for all other solver combinations.}
\label{fig:cht_temperature}
\end{figure}

This tutorial consists of a simple CHT scenario with a relatively low runtime. Since experimental data is available by Vynnycky et al.~\cite{vynnycky1998forced} the scenario serves often as a validation case for CHT simulations (\cite{Birken2010, birken2015fast}). The case consists of a 2D channel flow, coupled at its bottom with a 2D heated solid plate (cf.~\autoref{fig:cht_setup}).
As heat is conducted across the solid plate, the temperature of the flow region above and downstream the plate increases, as shown in~\autoref{fig:cht_temperature}. We discuss here the transient variant of this tutorial\footnote{Documentation of this CHT tutorial: \url{https://precice.org/tutorials-flow-over-heated-plate.html}}.

The fluid participant is the compressible OpenFOAM solver buoyantPimpleFoam.
For the solid participant, the user can choose among the OpenFOAM solver laplacianFoam and heat conduction solver examples based on FEniCS or Nutils.
In the case of laplacianFoam, we compute the heat flux assuming a constant heat conductivity $k_S$, which is additionally specified in the OpenFOAM adapter. In the case of FEniCS, the solver was developed as part of~\cite{QNWI} based on a heat equation example from~\cite{Langtangen2016}. A detailed description of the solver can be found in~\cite{rodenberg2021fenicsprecice}. In case of Nutils, we provide a similar example.

All of the possible combinations ($\{\text{OpenFOAM}\} \times \{\text{OpenFOAM, FEniCS, Nutils}\}$) use the same preCICE configuration file
and the user can select any combination at runtime. The solid participant solvers write heat flux values and apply a Dirichlet boundary condition by reading temperature values at the coupling interface. Accordingly, the fluid OpenFOAM participant writes temperature values and applies a Neumann boundary condition by reading heat flux values at the coupling interface. By default, the tutorial is configured with a serial-implicit coupling scheme in combination with Aitken under-relaxation and nearest-neighbor mappings.

\begin{figure}[h!]
\begin{center}
\includegraphics[width=0.9\textwidth]{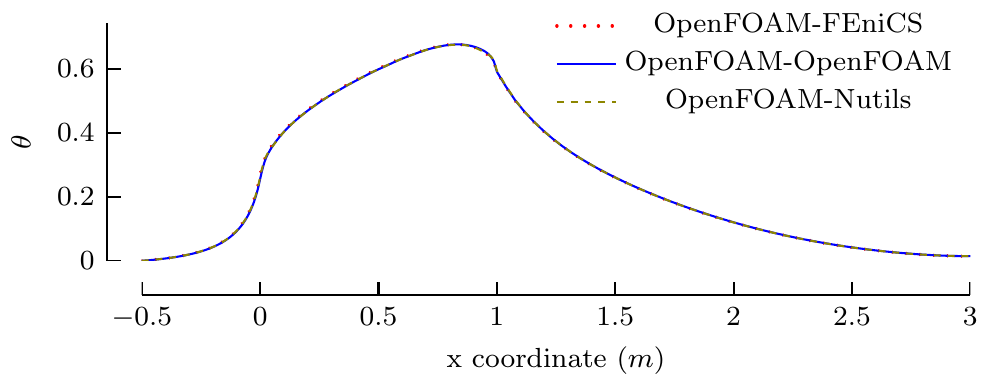}
\end{center}
\caption{Flow over a heated plate CHT tutorial: Comparison of non-dimensional temperature values $\theta$ at time $t=10s$  along a line $0.01m$ above the bottom of the channel for different combinations of solvers. $x\in\left[-0.5,0\right]$ describes the region of the channel upstream of the plate, $x\in\left[0,1\right]$ the region where the channel and the plate are coupled and $x\in\left[1,3\right]$ the downstream region.}
\label{fig:cht_results}
\end{figure}

A quantity that is commonly monitored in this scenario is the non-dimensional temperature $\theta = (T-T_\infty)/(T_h-T_\infty)$
along the interface. \autoref{fig:cht_results} depicts identical $\theta$ profiles for all solver combinations. Furthermore, the isothermal contour plot of the unified fluid-solid domain (cf. \autoref{fig:cht_temperature}) is continuous and smooth across the coupling interface.
Note that a quantitative comparison to~\cite{vynnycky1998forced} is not possible, as our cases describe flow inside a channel and not an open flow.

\subsection{Flow in a channel with an elastic perpendicular flap FSI tutorial} 

The most common use case of preCICE is FSI.
Often, one of the first case users aim to run is the Turek-Hron FSI3 benchmark~\cite{turek2006proposal}.
However, this case needs significant computational resources and specifications that are
not trivial to achieve with every solver out-of-the-box (e.g., parabolic inlet velocity profile in OpenFOAM).
A very common alternative is that of an elastic flap anchored at the bottom of a 2D channel flow as depicted in \autoref{fig:fsi_setup} and further described in the preCICE documentation\footnote{Documentation of this FSI tutorial: \url{https://precice.org/tutorials-perpendicular-flap.html}}.

\begin{figure}[h!]
\begin{center}
\begin{minipage}[t]{.55\textwidth}
\strut\vspace*{-\baselineskip}\newline
\begin{tikzpicture}[scale=1, every node/.style={font=\small}]

\coordinate(origin) at (0,0);

\coordinate(lengthFlap) at (0,1);
\coordinate(thicknessFlap) at (0.1,0);

\coordinate(leftTopFlap) at ($(origin) + (0,-2em)$);
\coordinate(rightTopFlap) at ($(leftTopFlap)+(thicknessFlap)$);
\coordinate(leftBottomFlap) at ($(leftTopFlap)-(lengthFlap)$);
\coordinate(rightBottomFlap) at ($(rightTopFlap)-(lengthFlap)$);

\coordinate(channelWidth) at (0,4);
\coordinate(channelPreFlap) at (-2.95,0);
\coordinate(channelPostFlap) at (2.95,0);

\coordinate(leftBottomInflow) at ($(origin)+(channelPreFlap)$);
\coordinate(leftTopInflow) at ($(leftBottomInflow)+(channelWidth)$);

\coordinate(rightBottomOutflow) at ($(origin)+(thicknessFlap)+(channelPostFlap)$);
\coordinate(rightTopOutflow) at ($(rightBottomOutflow)+(channelWidth)$);

\draw[fill=participantBcolor!50](leftBottomInflow) -- (origin) -- ($(origin)+(lengthFlap)$) -- ($(origin)+(thicknessFlap)+(lengthFlap)$) -- ($(origin)+(thicknessFlap)$) -- (rightBottomOutflow) -- (rightTopOutflow) -- (leftTopInflow) -- cycle;
\node[align=center, anchor=south, yshift=0.5em] at ($.25*(leftBottomInflow)+.25*(leftTopInflow)+.25*(rightBottomOutflow)+.25*(rightTopOutflow)$){Fluid participant\\ \textit{OpenFOAM, SU2, or Nutils}};
\draw[fill=participantAcolor!50](leftTopFlap) -- (rightTopFlap) -- (rightBottomFlap) -- (leftBottomFlap) -- cycle;
\node[anchor=north west, align=left](FEniCSLabel) at ($.5*(leftBottomFlap)+.5*(rightBottomFlap)+(-1em,-2em)$){Solid participant\\ \textit{CalculiX, deal.II, or FEniCS}};

\draw[participantBcolor,very thick](leftBottomFlap) -- (leftTopFlap) -- (rightTopFlap) -- node[right]{$\Gamma_\text{C}$} (rightBottomFlap);  \coordinate(midStructure) at ($.25*(leftBottomFlap)+.25*(leftTopFlap)+.25*(rightTopFlap)+.25*(rightBottomFlap)$);

\path [name path=A--B] ([xshift=-3em]FEniCSLabel.north) to[out=60, in=-30] ([yshift=-.5em]midStructure);
\path [name path=C--D](rightTopFlap) -- (rightBottomFlap);
\path [name intersections={of=A--B and C--D,by=E}];
\draw[->] ([xshift=-3em]FEniCSLabel.north) to[out=60, in=-30] (E);

\draw[participantAcolor,very thick](origin) -- ($(origin)+(lengthFlap)$) -- ($(origin)+(lengthFlap)+(thicknessFlap)$) -- node[right]{$\Gamma_\text{C}$} ($(origin)+(thicknessFlap)$);  \draw(leftBottomFlap) -- node[below, black]{$\Gamma_\text{fixed}$} (rightBottomFlap);  

\draw(leftBottomInflow) -- node[above]{$\Gamma_\text{no-slip}$} (origin);
\draw(rightBottomOutflow) -- node[above]{$\Gamma_\text{no-slip}$} ($(origin)+(thicknessFlap)$); \draw(leftTopInflow) -- node[below]{$\Gamma_\text{no-slip}$} (rightTopOutflow);  \draw(leftTopInflow) -- node[right]{$\Gamma_\text{inflow}$}(leftBottomInflow);  \draw(rightBottomOutflow) -- node[left]{$\Gamma_\text{outflow}$}(rightTopOutflow);

\dimline[extension start length=0, extension end length=0] {($(leftBottomInflow)+(-0.75em,0)$)}{($(leftTopInflow)+(-0.75em,0)$)}{$H$};
\dimline[extension start length=0, extension end length=0] {($(leftTopInflow)+(0,0.75em)$)}{($(rightTopOutflow)+(0,0.75em)$)}{$L$};
\dimline[extension start length=0, extension end length=0] {($(leftBottomInflow)-(0,0.75em)$)}{($(origin)-(0,0.75em)$)}{$l$};
\dimline[extension start length=0, extension end length=0, label style={fill=participantBcolor!50}] {($(origin)-(0.75em,0)$)}{($(origin)+(lengthFlap)-(0.75em,0)$)}{$h$};
\dimline[extension start length=0, extension end length=0, label style={fill=participantBcolor!50, above=0.6ex}, line style={arrows=dimline reverse-dimline reverse}] {($(origin)+(lengthFlap)+(0,0.75em)$)}{($(origin)+(lengthFlap)+(thicknessFlap)+(0,0.75em)$)}{$w$};
     
\foreach \i in {0,...,5}{\draw[->,participantBcolor] ([yshift=\i * 0.2cm, xshift = -.5cm]leftBottomFlap) -- ([yshift=\i * .2cm]leftBottomFlap);}
      \path([xshift = -.25cm]leftTopFlap) -- node[above, participantBcolor]{$F$} (leftTopFlap);
      
\draw[<->] ($(origin)+(thicknessFlap)+.25*(lengthFlap)+(.1em,0)$) to[in=30, out=-30] node[right]{coupling via preCICE}($.25*(rightBottomFlap)+.75*(rightTopFlap)+(.1em,0)$);   

\node[left = of midStructure, anchor=north east,yshift=1em]{
\begin{tabular}[t]{c S[table-format=1.2] s}
\multicolumn{3}{c}{Dimensions}\\
\midrule
$L$ & 6.00 & \meter\\
$H$ & 4.00 & \meter\\
$l$ & 2.95 & \meter\\
$h$ & 1.00 & \meter\\
$w$ & 0.10 & \meter\\
$z$ & 1.00 & \meter
\end{tabular}
};

\end{tikzpicture}
 \end{minipage}
\hfill
\begin{minipage}[t]{.39\textwidth}
\begin{tabular}[t]{c S[table-format=6.2e2] s}
\multicolumn{3}{c}{Fluid: compressible Euler}\\
\midrule
$\text{Ma}_\infty$ & 1e-2 & \\  $p_\infty$ & 101325 & \pascal \\ $T_\infty$ & 288.15 & \kelvin  \\[.5em]
\multicolumn{3}{c}{Fluid: incompressible Navier-Stokes}\\
\midrule
$u_{\infty}$ & 10 & \meter\per\second \\ $\nu_f$ & 1 & \meter\squared\per\second \\  $\rho_f$ & 1 & \kilogram\per\meter\cubed \\[.5em]

\multicolumn{3}{c}{Solid}\\
\midrule
$E$& 4e6 & \newton\per\meter\squared \\  $\nu_s$ & 3e-1 & \\  $\rho_s$ & 3e3 & \kilogram\per\meter\cubed \\  \end{tabular}
\end{minipage}
\end{center}
\caption{Flow in a channel with an elastic perpendicular flap FSI tutorial: 
The setup is depicted on the left and physical parameters are listed in the table on the right. 
The bottom of the flap is clamped, the solid participant reads forces at the interface, while the fluid participant reads displacement values. The inflow velocity at the channel inlet is
$\SI{10}{\meter / \second}$ and the outflow sets a zero velocity gradient. $\text{Ma}_\infty$, $p_\infty$, $T_\infty$, $u_{\infty}$: Mach number, pressure, temperature, and velocity at the inflow. $\nu_f$, $\rho_f$: kinematic viscosity and density of the fluid.
$E$: Young's modulus. $\nu_s$: Poisson's ratio of the solid. $\rho_s$: density of the solid. $z$ is the out-of-plane thickness: even if the coupled case is described as 2D, OpenFOAM and CalculiX are still 3D solvers.}
\label{fig:fsi_setup}
\end{figure}

For the fluid participant, the user can choose between:
\begin{enumerate}
  \item the incompressible OpenFOAM solver pimpleFoam,
  \item an incompressible CFD solver written in Nutils, or
  \item the compressible CFD solver of SU2.
\end{enumerate}

For the solid participant, the user can choose among:
\begin{enumerate}
  \item the linear structure solver of CalculiX with linear, rectangular finite elements,
  \item a linear structure solver provided with the deal.II adapter, using fourth order, rectangular finite elements, or
  \item a linear structure solver example in FEniCS, using quadratic, triangular finite elements.
\end{enumerate}

All possible combinations ($\{\text{OpenFOAM, Nutils, SU2}\} \times \{\text{CalculiX, deal.II, FEniCS}\}$)
use the same preCICE configuration file and the user can select any combination. The fluid solvers read absolute displacement values at the interface (Dirichlet boundary condition) and write forces, while all solid solvers read forces (Neumann boundary condition) and write absolute displacements.
Some of the adapters also support additional coupling fields, for example relative displacement or stresses, but the employed combination here works with all involved adapters. By default, the tutorial is configured with RBF data mappings and a parallel-implicit coupling scheme using IQN-ILS acceleration.

\begin{figure}[h!]
\begin{center}
\includegraphics[scale=1]{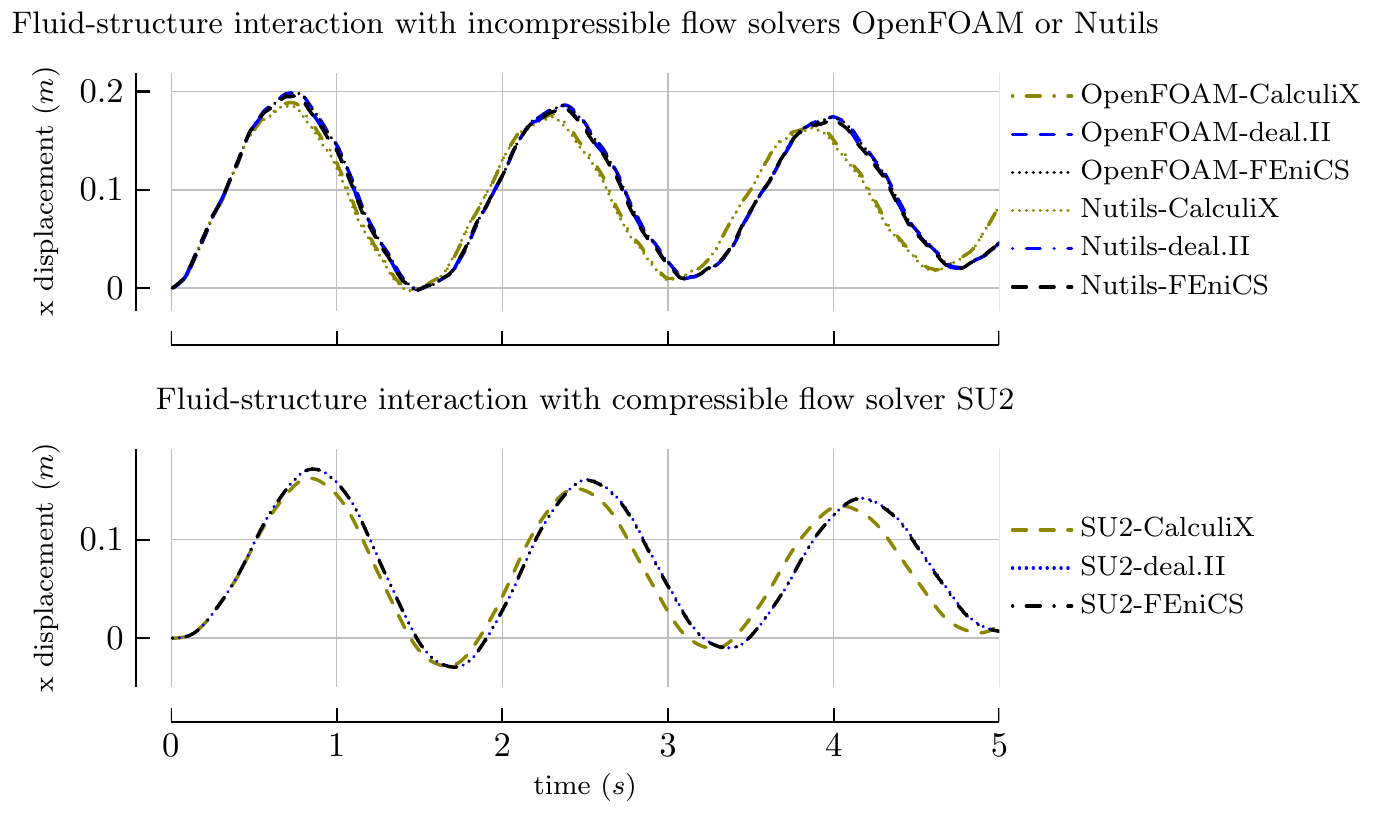}
\end{center}
\caption{Flow in a channel with an elastic perpendicular flap FSI tutorial: Comparison of the flap tip displacement for different combinations of solvers. The upper plot shows the results for an incompressible flow computed with Nutils or OpenFOAM. The lower plot shows the results for compressible flow computed with SU2. Incompressible and compressible flow give qualitatively different results, as expected. Good agreement within each class of flow simulation is achieved when using FEniCS or deal.II as solid solver. Using CalculiX as solid solver leads to different results, presumably due to the use of linear elements. The CalculiX adapter is only capable of handling quasi 2D-3D cases with out-of-plane thickness for linear elements.}
\label{fig:fsi_results}
\end{figure}

\begin{figure}[h!]
\begin{center}
\includegraphics[width=\textwidth]{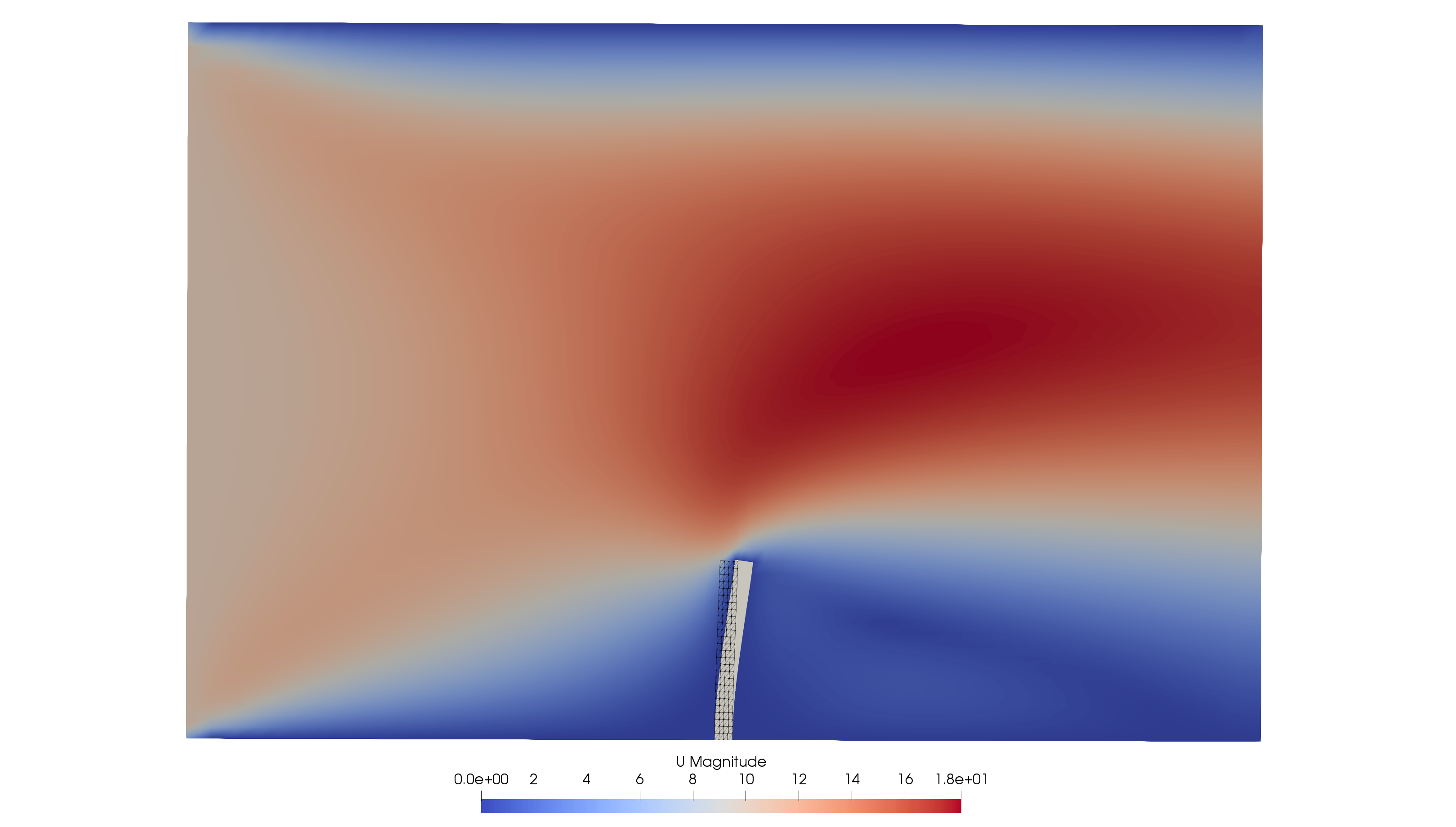}
\end{center}
\caption{Flow in a channel with an elastic perpendicular flap FSI tutorial: Visualization of flow field and deformed flap at time $t=2s$ using OpenFOAM as fluid and FEniCS as solid solver. For comparison, the deformed solid mesh of an SU2-FEniCS simulation is shown in black to make the difference between FSI with a compressible and an incompressible fluid simulation visible.}
\label{fig:fsi_paraview}
\end{figure}

A quantity that is commonly monitored in this scenario is the displacement of the tip of the flap. We track this quantity using a preCICE watchpoint and compare the results across different solver combinations in~\autoref{fig:fsi_results}. \autoref{fig:fsi_paraview} shows a direct comparison snapshot of an FSI simulation with an incompressible and a compressible fluid solver.

 \section{Testing and continuous integration} 
\label{sec:tests}

\begin{figure}[h!]
  \includegraphics[width=\textwidth]{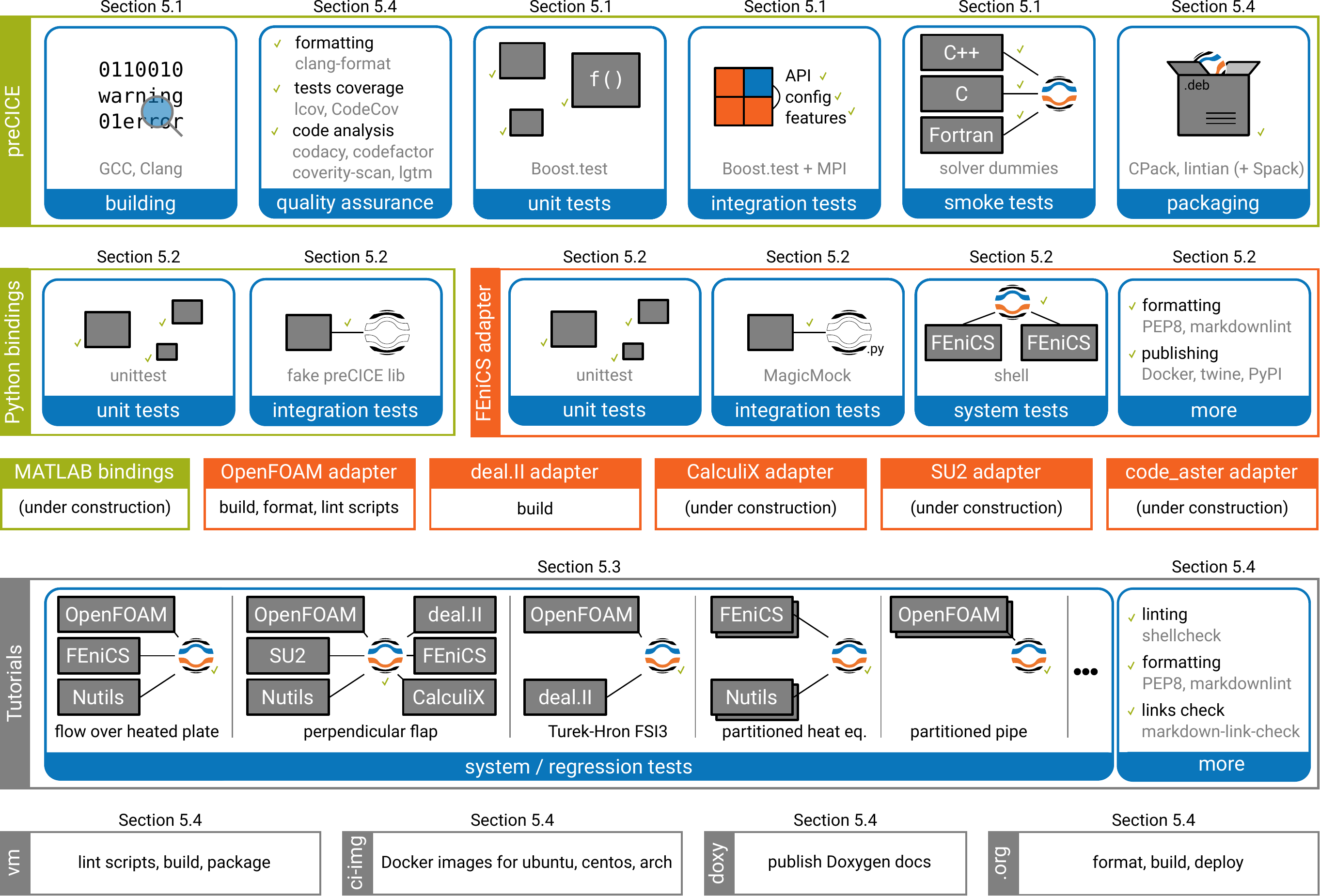}
  \caption{Overview of the testing and continuous integration workflows for different preCICE components.
  Each box represents a separate project repository. The non-native language bindings depend on preCICE.
  The adapters depend on language bindings or directly on preCICE.
  The tutorials and system tests depend on all the adapters.
  The components at the bottom-most row (virtual machine image, CI images, Doxygen documentation, website) depend on one or more other components.
  Read more details about each workflow in the section number listed above it.}
  \label{fig:tests-overview}
\end{figure}

The core preCICE library and the non-core components are developed in separate repositories,
each with a number of continuous integration (CI) workflows. These workflows are tests, quality assurance checks,
or operations to prepare and validate packages. We find such workflows to be indispensable for multi-component,
multi-developers projects, as they answer questions such as
``will the code still compile and behave in the same way if we integrate these changes?'',
``will a simulation still give the same results?'', and ``will these changes have side-effects in other (potentially not regularly updated) components?''.
In other words, these workflows facilitate further development by ensuring that everything still works.

Figure~\ref{fig:tests-overview} depicts the currently deployed workflows.
As one may observe, the granularity of testing and CI correlates to the
number of users and developers involved in each subproject.
In some cases, it may also be enabled or hindered by the respective programming language environment.
As the most important and actively developed component, the core library is rigorously tested in a wide range of levels.
The rich tooling collection of Python enables the CI of the Python bindings and the FEniCS adapter,
while we are gradually adding similar workflows to the rest of the adapters.
The tutorials provide a platform to test every component in complete simulations (system tests with results regression checks).
Finally, a few additional workflows keep non-critical systems up-to-date.

The number and diversity of components required to construct a complete coupled simulation
(at least two participants + multiple components per participant), as well as the challenges in
testing each component in isolation, makes testing a coupling library significantly more complex than testing a linear algebra solver library, for example.
Complex testing approaches of significant novelty are required.
We structure the rest of the section following the different complexity levels.
In \autoref{ssec:unit_and_integration_tests_preCICE}, we present the CI of the core library, for which
no interaction with other components is required and the respective runtime is relatively short.
In \autoref{ssec:unit_and_integration_tests_other}, we continue with the CI of non-native language bindings and adapters. This layer depends on
the core library, as well as on external components (the solvers), leading to a need for testing in isolation.
In \autoref{ssec:system_tests}, we construct system tests for the complete software stack.
Finally, in \autoref{ssec:additional_checks}, we give an overview of additional checks and workflows which we use across the whole project.

\subsection{Tests for the preCICE core library}
\label{ssec:unit_and_integration_tests_preCICE}

To test the complete functionality of the preCICE core library, heterogeneous test setups are needed.
Individual tests may require one or more logical participants running on one or more MPI ranks. 
To solve this intrinsic problem of testing a communication and orchestration library in a parallel environment, the core library tests are run on 4 MPI ranks.
Partitioning these 4 ranks allows to cover various scenarios, from testing math functions on a single rank, over testing parallel mappings on 3 ranks, up to testing scenarios with a serial participant on a single rank coupled to a parallel participant on 3 ranks.
As mature MPI-aware testing frameworks are not available, we developed our own testing framework extending \incode{Boost.test} to support the aforementioned criteria.

\begin{figure}[h!]
  \centering
  \includegraphics[width=0.75\textwidth]{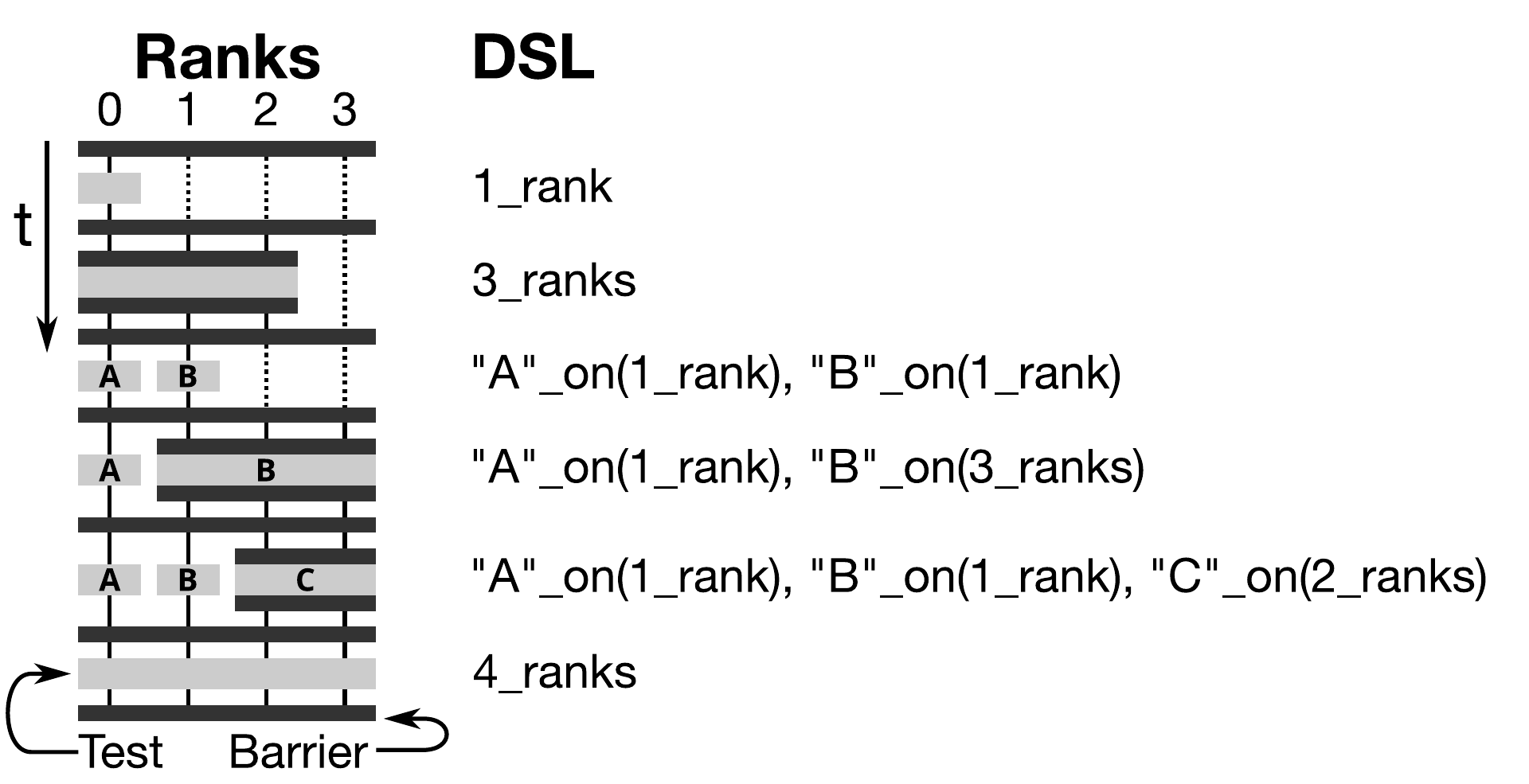}
  \caption{This figure depicts the MPI communicator setup on the left corresponding to a sequence of test setups on the right.
    The testing framework uses 4 MPI ranks, depicted by the vertical lines.
    Horizontal black bars are barriers and dotted lines are ranks which are unused during a test setup and hence idle.
Each test starts with \incode{PRECICE_TEST(...)}, containing an expression based on the DSL, which results in a complete test setup.
    The test DSL specifies the name of local participants followed by the amount of ranks.
    The name is optional for tests containing only a single participant.
  }
  \label{fig:testdsl}
\end{figure}

The extension of \incode{Boost.Test} provides a custom domain-specific language (DSL), which is used to set up a \incode{PRECICE_TEST()}.The DSL specifies the name of local participants used in the test, followed by the amount of ranks and optional requirements.
If a test contains only a single participant, then its name can be omitted.
The DSL is human-readable, examples are \incode{"A"_on(2_ranks), "B"_on(1_rank)} or simply \incode{1_rank}.
The implementation of the DSL firstly restricts the MPI communicator size to the required amount of total ranks, followed by grouping ranks by name, forming communicators for the logical participants.
Further specified requirements on logical participants, such as initialization of sub-components, are then handled inside the isolated state of each participant.
The result of each \incode{PRECICE_TEST()} is an immutable object which, for each test rank, provides access to the context including name and communicator information of the local participant.
The \incode{context} holds further information about the setup, information which allows to sanitize user input provided to utility functions.
At the end of each test, the context object firstly reverts all changes made by setup requirements, secondly ungroups the communicators, and finally synchronizes all ranks, including ranks not needed by the test.
See \autoref{fig:testdsl} for example configurations of this framework.

The core library tests can be categorized into unit tests, integration tests, and code-example tests.
Currently, preCICE is tested with a total of 438 unit and integration tests.

\paragraph{Unit tests} This type tests a component in isolation, using its public interface.
The test functions manually setup the majority of required components and partition the available ranks according to the needs of the test.
The needs for partitioning vary:
First, many unit tests handling geometric functions, VTK exports, and mesh internals require only a single logical participant and run on a single rank.
Furthermore, components such as VTK exports and radial-basis-function mappings have additional functionality when running in parallel, hence require multiple ranks on a single logical unit.
Finally, inter-code communication, the coupling schemes, and the mesh partitioning require multiple logical participants.
Each of these logical participants may run on one or multiple ranks.
See \autoref{lst:example-unit-test} for an example of a unit test.

\begin{listing}[h!]
  \centering
\begin{minipage}{0.85\linewidth}
\begin{minted}[linenos,numbersep=5pt,gobble=0,frame=none,framesep=0mm]{cpp}
BOOST_AUTO_TEST_CASE(ParallelMappingTest)
{
  PRECICE_TEST(""_on(4_ranks).setupMasterSlaves(), Require::PETSc);
  constexpr int dims = 2;
  // Setup InMesh
  mesh::PtrMesh inMesh(new mesh::Mesh("InMesh", dims));
  mesh::PtrData inData = inMesh->createData("InData", dims);
  getDistributedInMesh(context, inMesh, inData);
  // Setup OutMesh ...
  // Setup Mapping
  PetRadialBasisFctMapping<Gaussian> mapping{
      Mapping::CONSISTENT, dims, Gaussian{5.0}};
  mapping.setMeshes(inMesh, outMesh);
  // Test the Mapping preparation
  BOOST_TEST(not mapping.hasComputedMapping());
  mapping.computeMapping();
  BOOST_TEST(mapping.hasComputedMapping());
  // Test the Data Mapping
  BOOST_TEST(not mapping.hasComputedMapping());
  mapping.map(inData->getID(), outData->getID());
  BOOST_TEST(outData->values() == expectedData);
}
\end{minted}
\end{minipage}
\caption{Example unit test of a parallel RBF mapping running on 4 ranks.
Further requirements on the test are the setup of the master-slave communication and the initialization of PETSc.
First the test defines meshes and associated data followed by setting up and executing the mapping.
The result is the checked against the expected outcome.
The example showcases the preparation involved in testing individual components of preCICE in a parallel context.
}
\label{lst:example-unit-test}
\end{listing}

\paragraph{Integration tests} This type uses the API of preCICE itself to test specific scenarios, hence the test setup is handled using a preCICE configuration file.
Individual logical participants may run on a single rank each to test
coupling of serial solvers with various setups.
Another very common setup consists of two logical participants running on two ranks each.
This allows to thoroughly test the partitioning behavior given various mapping schemes.
Integration tests are also used to reproduce and fix bugs reported by users.
See \autoref{lst:example-integration-test} for an example of an integration test.

\begin{listing}[h!]
\begin{minted}[linenos,numbersep=5pt,gobble=0,frame=none,framesep=20mm]{cpp}
BOOST_AUTO_TEST_CASE(ParallelIntegrationTest2x2)
{
  PRECICE_TEST("SolverOne"_on(2_ranks), "SolverTwo"_on(2_ranks));
  std::string meshName, writeDataName, readDataName;
  if (context.isNamed("SolverOne")) {
    meshName      = "MeshOne";
    writeDataName = "Data1";
    readDataName  = "Data2";
  } else {
    // ...
  }
  SolverInterface interface(context.name, _pathToTests + "test-config.xml",
      context.rank, context.size);
  // set mesh vertices and initialize
  std::array<double, 4> inValues, outValues;
  while (interface.isCouplingOngoing()) {
    interface.readBlockScalarData(readDataID, 4, vertexIDs, inValues);
    if (context.isNamed("SolverOne")) {
      outValues = inValues;
    } else {
      outValues = solveSystem(inValues);
    }
    interface.writeBlockScalarData(writeDataID, 4, vertexIDs, outValues);
    interface.advance(1.0);
  }
  BOOST_TEST(outValues == expectedValues);
}
\end{minted}
\caption{Example integration test involving two parallel participants \incode{SolverOne} and \incode{SolverTwo} running on two ranks each.
  The \incode{context} object provides information about the identity of the current rank.
  This information is used to setup further local information such as mesh and data names.
  Integration tests then directly construct a \incode{SolverInterface} and use preCICE API calls to run the test.
}
\label{lst:example-integration-test}
\end{listing}

\paragraph{Code-example tests} This type smoke-tests native bindings using the provided examples.
Native-bindings are C and Fortran bindings, which are implemented using the C++ API of preCICE and linked directly into the library.
Non-native bindings such as Python are covered in \autoref{ssec:unit_and_integration_tests_other}.
Each language binding comes with an example program called a \emph{solverdummy}. All solverdummies implement the same functionality and provide a template for using the preCICE API in the respective programming language.
The code-example tests themselves consist of three steps:
First, the tests build each solverdummy and link it to the preCICE library. This tests a common subset of the interface of the bindings for completeness and assures that the build system is functional.
Second, they run a small coupled simulation coupling each solverdummy to itself. This ensures that the used language binding is working correctly.
Finally, they run a small coupled simulation coupling different solverdummies to each other. This ensures that the bindings (of different languages) are compatible.

\subsection{Tests for adapters and bindings}
\label{ssec:unit_and_integration_tests_other}

As explained in \autoref{ssec:building} and \autoref{sec:adapters}, language bindings and adapters are organized in independent repositories. The requirements for tests of such non-core components are identical to the ones for the core library: the non-core components must also comprise of valid code, their individual units should behave in the correct way and work together, while continuous integration tests should be performed on every commit to each component repository. We again distinguish unit tests and integration tests.

\paragraph{Unit tests} We do not need a special treatment for unit tests of non-core components. Note that non-core components are written in various languages. This requires the use of suitable testing frameworks for each language, such as the Python module \inpython{unittest} for the Python bindings and the FEniCS adapter.

\paragraph{Integration tests} Non-core components use preCICE through its API, treating it as a regular, black-box dependency. This safeguards low software coupling, but also leads to a technical complication as already mentioned above: Due to the very nature of coupled simulations, preCICE requires at least two participants for executing most steps of a simulation. This cannot be avoided easily (since the components are not able to modify preCICE itself) and, therefore, it is not trivial to test each component independently.

To solve this problem we use a strategy commonly known as \emph{mocking}. This is a well-known and widely established software engineering practice~\cite{fowler2007mocks}, but not as widespread in the scientific software community. Mocking is useful, if the system under test (non-core component) has another component (preCICE) as a dependency and interacts with this component through its API. Since we want to avoid starting a second participant in our integration tests, we use a mocked version of preCICE instead of the original one. This mocked preCICE returns fake output for testing and does not rely on any other components.
In the following, we give two examples for our implementation of this testing pattern: first, integration tests in the FEniCS adapter and, second, integration tests in the Python language bindings of preCICE. In both cases, API calls to the fake version of preCICE do not require any initialization of a second participant and hard-coded fake values are returned. This allows to write short and simple tests.

Mock testing for the FEniCS adapter heavily relies on the Python module \inpython{unittest.mock}, which allows to create a \inpython{MagicMock} object that is used to provide a fake implementation of functions or objects. Additionally, the module provides a \inpython{patch} function that allows to replace a module that a test imports with a fake version of the same module, at runtime. These two pieces allow us to replace the Python bindings of preCICE with a fake version. For a detailed example, please refer to~\cite{rodenberg2021fenicsprecice}.

Mock testing for the Python bindings is more involved, since the bindings rely on two different languages, C++ and Python, and (to the authors' knowledge) no mocking framework exists for this purpose. We, thus, test the Python bindings by building a specific executable, where we link against the mocked version of preCICE: a single \inbash{SolverInterface.cpp} with a fake implementation. We do include the original interface of preCICE (\inbash{SolverInterface.hpp}) to make sure that the API is consistent. This allows us to keep the application code of the Python bindings clean and to decide whether we want to use the real or fake implementation of the preCICE library at compile time.
The integration tests of the Python bindings then allow us to check for the correctness of type conversions done by the language bindings, such as converting a C++ \incode{double*} array to a numpy array~\cite{NumPy} using Cython~\cite{behnel2011cython}. An example is given in \autoref{lst:setMeshVertices}).
Our mocking approach leads to a non-standard \inpython{setup.py} build script, which allows us to choose whether we want to build the real executable or the one intended for testing through the standard interface of \inpython{setuptools}\footnote{\inbash{python3 setup.py install} or \inbash{python3 setup.py test}}. A nice side effect of this testing pattern is that preCICE itself is not even needed and does not have to be installed on the system running the tests.

\begin{listing}[h!]
\begin{minted}{cpp}
void setMeshVertices(int meshID, int size, const double *positions, int * ids){
  std::vector<int> fake_ids = get_hardcoded_ids(size);
  std::copy(fake_ids.begin(), fake_ids.end(), ids);
}
\end{minted}
\caption{Fake implementation of \incpp{setMeshVertices} used for integration tests of Python bindings. The fake version of \incpp{setMeshVertices} just returns fake vertex IDs. If an integration test of the Python bindings is calling this function, one can easily check whether the obtained vertex IDs are the expected ones using Python's unit testing framework.}
\label{lst:setMeshVertices}
\end{listing}

\paragraph{Outlook} Testing of other language bindings and adapters is under current development: Our prototype for integration tests for the OpenFOAM adapter uses the mock testing pattern and the C++ mocking framework \inbash{FakeIt}\footnote{FakeIt: \url{https://github.com/eranpeer/FakeIt}}. For testing the MATLAB bindings, the existing Python bindings testing approach may be used. Additional restrictions apply to each set of bindings and adapters, including language-specific challenges, level of interaction with the solver code, as well as licensing compatibility with our open testing infrastructure (e.g., in case of the MATLAB bindings).

\subsection{System and regression tests}
\label{ssec:system_tests} 

Fine-grained unit and integration tests can give us detailed insight into each component, but these tests only study each component or group of components in isolation.
System tests give us the user perspective of all components working together: ``does the coupled simulation black-box still behave in the same way?'' and ``if not, which change in which component introduced the regression?''.
While system tests can be quite straight-forward in their implementation, developing effective system tests for multi-component, multi-participant
simulations becomes a complex task, especially when considering different stakeholder perspectives.

Let us look at a few examples of such stakeholder perspectives. As a release \textit{manager}, \textit{\textbf{Ma}ria} wants to know that the latest state of all development branches to be released works flawlessly together on the release day, so that she can release a new version.
As a developer of the core \textit{library}, \textit{\textbf{Li}sa} wants to know that her proposed changes do not cause any unintended regression in results or behavior in the context of a complete simulation.
As a developer of an \textit{adapter}, \textit{\textbf{Ad}am} has even more questions. First, similarly to Lisa, he wants to know that his proposed changes do not cause any regressions downstream. Additionally, Adam wants to know if he needs to update his adapter to support breaking changes (of installation/configuration) in the development branches of upstream components or new solver and dependency versions.
As a developer of \textit{tutorial} cases, \textit{\textbf{Tu}dor} wants to know that configuration updates do not cause regressions and that the tutorials still work as expected with newer solver and dependency versions.
Finally, as a maintainer of the \textit{system tests}, \textit{\textbf{Sy}} wants to know that their proposed changes do not cause any downtime to the operations.

\begin{figure}[h!]
  \includegraphics[width=\textwidth]{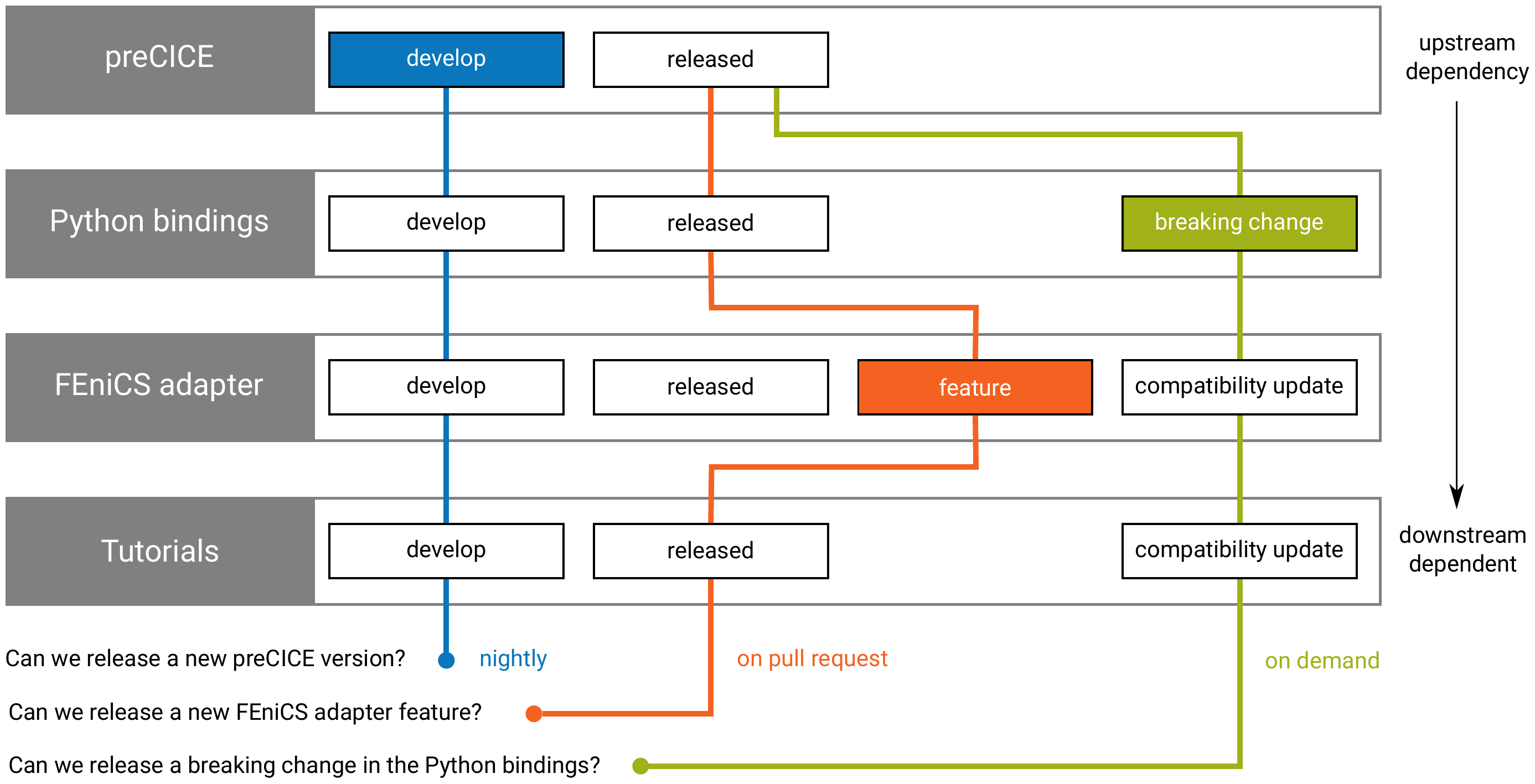}
  \caption{Example questions that the system (regression) tests of preCICE help us answer (planned workflow).
  The system tests are always executed at the bottom-most layer (tutorials repository), using Docker images prepared in the layers above.
  Every night, all develop branches are tested together, to ensure that a new version of preCICE can be released at any time (blue).
  Pull requests that introduce non-breaking changes are automatically tested against the latest released versions of the other repositories (orange).
  Pull requests that introduce breaking changes need additional coordination with downstream projects and the tests need to compose compatible branches (green).
  The Python bindings and the FEniCS adapter only serve here as examples; in practice, more repositories are involved.}
  \label{fig:tests-branches}
\end{figure}

The situation we just described becomes apparent looking at~\autoref{fig:tests-branches}.
The test matrix evolves into a cross product of:
\begin{equation*}
  \{\text{platform}\} \times \{\text{preCICE branches}\} \times \{\text{bindings b.}\} \times \{\text{adapter b.}\} \times \{\text{tutorial b.}\} \times \{\text{system tests b.}\}
\end{equation*}

As these tests take a long time to prepare and execute and as they are particularly challenging to log in a structured and effective way,
executing the complete test matrix is not realistic and we need to select representative configurations.

We restrict the test matrix to a set of strategically important combinations. In terms of platforms, we execute most tests on the platform most common among users, currently the latest Ubuntu LTS version. We also execute selected tests on the oldest supported platform (previous Ubuntu LTS) and on the latest state of the continuously-updated Arch Linux. In terms of branches, we test each proposed branch with the rest of the components in their latest released state: this helps Lisa, Adam, and Tudor develop their projects independently, without worrying about untested new features. In case a breaking change is introduced in one component, then this needs to be tested in combination with corresponding compatibility updates in the downstream components. We also test the development branches of all components together in nightly builds, so that Maria and every developer can confidently release new versions of each component.

Maintaining the system tests and reference data up-to-date requires significant effort and any failing tests need to be addressed quickly, so that they remain useful and trusted. We observed that deep understanding and documentation of known issues that trigger test failures is crucial for the infrastructure to facilitate instead of hinder the development. Similarly, easily accessible logging of different levels is very important to verify that all relevant tests have succeeded and to precisely identify any faults. Finally, even though the operations need to be automatic enough to get green lights at the right places, developers do want to be able to form a clear mental model of the system behind the automation in order to trust the system and try to debug it, if needed.

Since preCICE v1 and till v2.1, we maintained the system tests on a dedicated repository\footnote{preCICE system tests on GitHub: \url{https://github.com/precice/systemtests}} using Travis CI. This repository contained scripts to run tests, scripts to prepare the test cases, as well as reference data for each test case. Because of policy changes in Travis CI and increasing flexibility offered by newer alternatives, we phased-out this implementation and we have been migrating to a different system. We describe here the outdated architecture of the tests used for preCICE v2.1 and discuss issues and potential solutions.

With every new commit pushed to an adapter repository, the Travis CI instance of the adapter instructed the Travis CI instance of the \incode{systemtests} repository to run any tests (tutorials) it knew to involve this adapter. Travis CI then built Docker images of preCICE adapters and pushed these to Docker Hub so that they could be reused. It then started one Docker container per simulation participant, as well as one Docker container serving the tutorial configuration. We used Docker Compose to build connections between the containers and we set a commonly accessible directory to exchange necessary connection tokens for the inter-code communication, as described in \autoref{ssec:com}.

At the end of each test case, a script compared the results. We originally compared every available results file excluding lines unique to every run. To account for sporadic rounding errors, we filtered arithmetic data and applied a numerical comparison. As this approach was very tedious to maintain for every new solver, we switched to comparing only the exported VTK files of the preCICE interface meshes. With a common file format at hand, the comparison scripts became much easier to execute and maintain. We also found this simplification to be enough for identifying regressions in the coupling, which is our main interest.
After Travis CI compared the results, it archived key log files to a dedicated repository \incode{precice_st_output}.
This was a far-from-ideal logging solution, leading to cumbersome workflows to discover more details about the executed tests and potential failures.

We are currently redesigning our system tests. Key decisions so far have been to separate the machinery from the reference data, hosting the (reduced) reference data together with the configurations that produce them (tutorials), so that they can be updated at the same time. More recent tools, such as GitHub Actions and GitLab CI, offer multi-project pipelines and more possibilities for storing artifacts and archiving logs. With such additional options to avoid complex workarounds and with our experience from the aforementioned approaches, a redesigning was deemed reasonable and is expected to fruit in the near future. Until then, we rely on regular manual runs with every release.

\subsection{Additional checks} 
\label{ssec:additional_checks}

To maintain the quality and consistency of the codebase, the preCICE CI runs additional checks on the latest state of every pull request.
The CI uses a fixed version of clang-format\footnote{clang-format: \url{https://clang.llvm.org/docs/ClangFormat.html}} to check all C++ and C files, as well as a custom formatter based on the Python lxml\footnote{lxml: \url{https://lxml.de/}} package to check all XML configuration files for correct formatting.
As most CI environments provide multiple CPUs, the system uses GNU parallel~\cite{gnuparallel} to leverage the available compute resources.

When building and testing on Ubuntu, the CI additionally generates the Debian packages using CPack\footnote{CPack: \url{https://cmake.org/cmake/help/latest/module/CPack.html}} and test them using the Debian package checker Lintian\footnote{Lintial: \url{https://wiki.debian.org/Lintian}}.
Furthermore, the tests generate code testing coverage information using GCC.
The resulting coverage information is then gathered by LCOV\footnote{LCOV: \url{http://ltp.sourceforge.net/coverage/lcov.php}} and uploaded to the Codecov\footnote{Codecov: \url{https://about.codecov.io/}} service, which integrates the coverage report into the GitHub user interface.
This informs the reviewer about the coverage of the code change, as well as the resulting coverage change of the whole project.

Moreover, the external code quality services lgtm\footnote{lgtm: \url{https://lgtm.com/}}, CodeFactor\footnote{CodeFactor: \url{https://www.codefactor.io/}}, and Codacy\footnote{Codacy: \url{https://www.codacy.com/}} are integrated into the core library's GitHub project and automatically run to perform code analyses using webhooks.
A scheduled job runs the proprietary static code analysis tool Coverity Scan\footnote{Coverity Scan: \url{https://scan.coverity.com/}} on the codebase once per week and reports the results to the developer mailing list.
These tools are useful to find less obvious issues such as code complexity, code duplication, misspellings as well as technical issues such as unreachable code, code paths leading to using uninitialized variables, incorrect exception handling, and more.

We use publicly available GitHub Actions from the marketplace\footnote{GitHub Actions Marketplace: \url{https://github.com/marketplace?type=actions}} to apply such checks on less critical components:
we validate shell scripts using shellcheck\footnote{shellcheck: \url{https://github.com/koalaman/shellcheck}, via GitHub Action \incode{ludeeus/action-shellcheck}},
we validate and format Python scripts using autopep8\footnote{autopep8: \url{https://github.com/hhatto/autopep8}, via \incode{peter-evans/autopep8}},
we validate the syntax of our documentation files with markdownlint\footnote{markdown-lint: \url{https://github.com/DavidAnson/markdownlint}, via \incode{articulate/actions-markdownlint}},
and we check for broken hyperlinks using markdown-link-check\footnote{markdown-link-check: \url{https://github.com/tcort/markdown-link-check},\\ via \incode{gaurav-nelson/github-action-markdown-link-check}}.
Finally, we publish Python packages using twine\footnote{twine: \url{https://github.com/pypa/twine}}.

Code reviews provide an additional safety check, which can prevent issues that are otherwise difficult to check automatically. Pull request templates provide checklists for authors and reviewers. Most non-trivial code contributions to preCICE since 2018 are reviewed by at least one further core developer. The master branches are protected from pushing and from merging without reviews.

Apart from tests and quality checks, a few more operations contribute to maintaining the resources available to the user up-to-date. A GitHub Actions workflow builds and packages a Vagrant\footnote{Vagrant: \url{https://www.vagrantup.com/}, see also~\autoref{ssec:building}.} box for VirtualBox with the latest Ubuntu LTS and all common components and tutorials pre-installed. A similar workflow prepares and publishes Docker images with all the preCICE dependencies, images which we use for our CI\footnote{preCICE CI images: \url{https://github.com/precice/ci-images}, Docker: \url{https://hub.docker.com/u/precice}}. The website of preCICE is also automatically generated using GitHub Pages\footnote{preCICE Website sources: \url{https://github.com/precice/precice.github.io}}, integrating content from additional repositories (tutorials and adapters). Finally, a dedicated workflow periodically updates the Doxygen-based C++ source documentation.
 \section{Community} 
\label{sec:community}

As the purpose of preCICE is to connect different simulation software, preCICE naturally also helps connecting researchers -- imagine the fluid mechanics group and the solid mechanics group of a computational mechanics faculty with their individual in-house CFD and FEM codes.
In the last five years, a significant community of users has been formed around preCICE, with some of them also contributing back code or tutorials. The \textit{preDOM} project\footnote{More about preDOM on our blog: \url{https://precice.discourse.group/t/how-did-precice-get-popular/321}}, funded by the German Research Foundation, played an important role for this development. In fact, most of the improvements described in this paper were part of the project: building and packaging, adapters, tutorials, tests and continuous integration, but also user documentation and community building.   

\begin{figure}[h!]
\begin{minipage}{\textwidth}
\begin{minipage}{\dimexpr0.5\textwidth-10pt\relax}
\includegraphics[width=\textwidth]{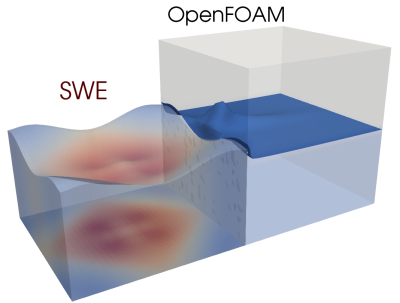}
\end{minipage}\hfill
\begin{minipage}{\dimexpr0.5\textwidth-10pt\relax}
\includegraphics[width=0.8\textwidth]{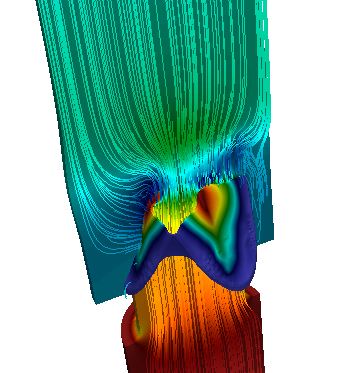}
\vspace{0.5cm}
\end{minipage}\end{minipage}
\begin{minipage}{\textwidth}
\begin{minipage}{\dimexpr0.45\textwidth-10pt\relax}
\includegraphics[width=\textwidth]{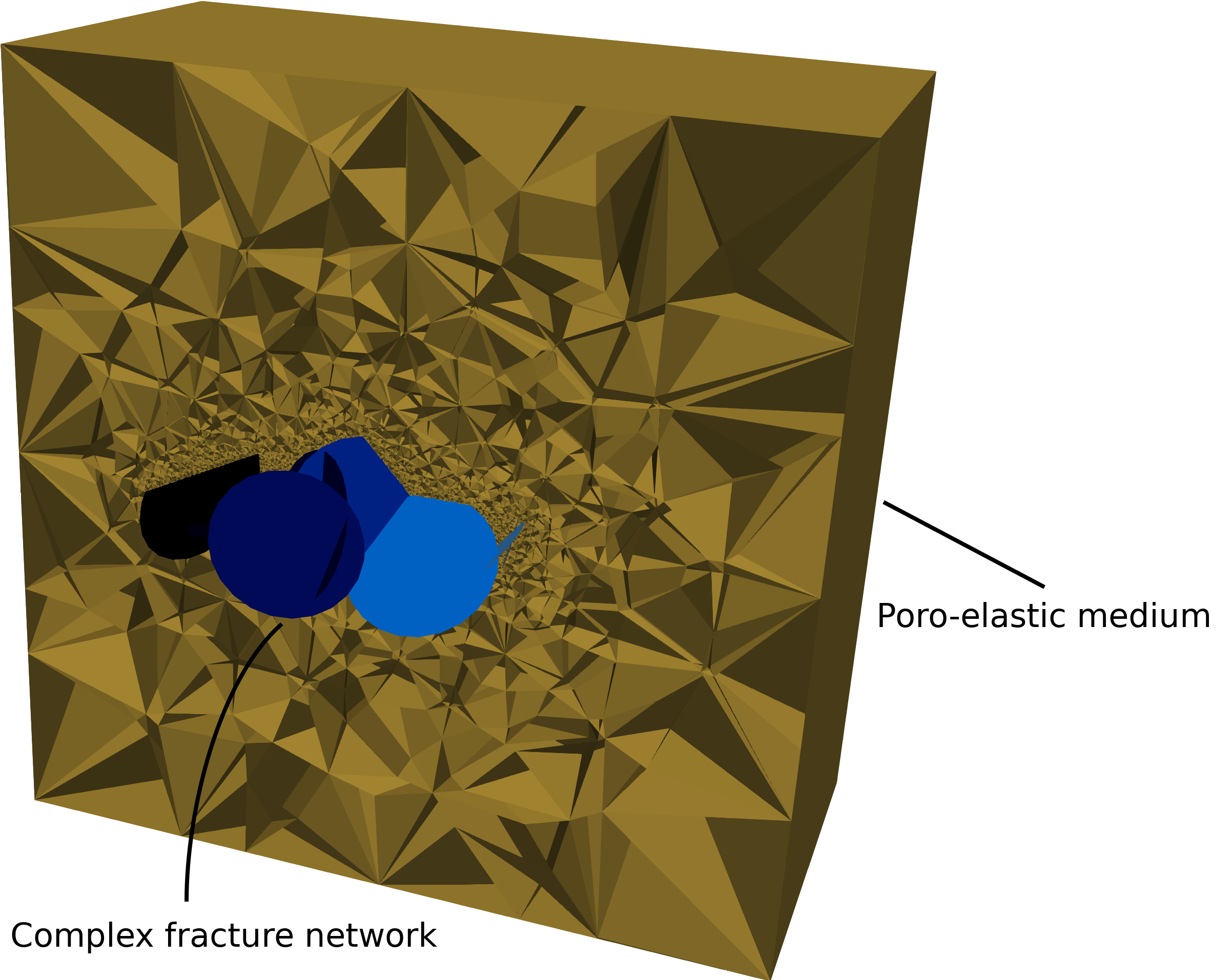}
\end{minipage}\hfill
\begin{minipage}{\dimexpr0.55\textwidth-10pt\relax}
\includegraphics[width=\textwidth]{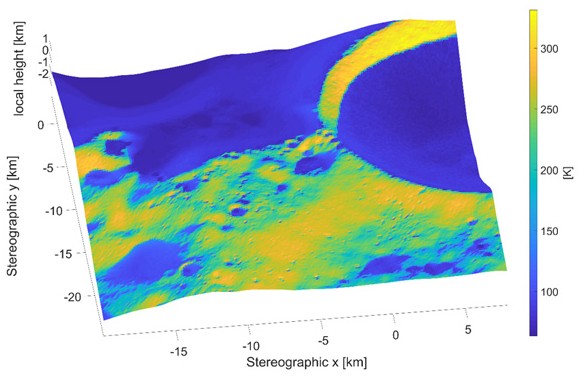}
\end{minipage}\end{minipage}
\caption{
\label{fig:testimonials}
Various simulations from the preCICE community. All pictures taken from the page \emph{Community stories} on \url{precice.org}. Top left: a shallow-water equations solver coupled to OpenFOAM~\cite{Espinosa2020}. Top right: an artificial heart valve simulated with OpenFOAM and CalculiX~\cite{Davis2019}. Bottom left: A 3D poro-mechanics model coupled to 2D fluid equations, both implemented in FEniCS~\cite{schmidt2021simulation}. Bottom right: a MATLAB heat equation solver coupled to a GPU ray-tracing software package to simulate heat conduction and radiation on the surface of the moon~\cite{Volland2019}.
}
\end{figure}

Today, we know through forum discussions, conferences, workshops, and publications of more than 100 research groups using preCICE. Roughly one half of them are from academia, while the other half comes from non-academic research centers (e.g., the German Max Planck Institute for Plasma Physics, the German Helmholtz-Zentrum Hereon, the Italian Aerospace Research Centre, or A*STAR in Singapore) or industry (e.g., MTU Aero Engines or Bitron).   
We collect some user stories on our website\footnote{preCICE community stories: \url{https://precice.org/community-projects.html}}
and depict some highlights in \autoref{fig:testimonials}.
Presumably half of the users apply preCICE for fluid-structure interaction or conjugate heat transfer applications. The other half uses preCICE for more uncommon setups, for example, coupling of different fluid models with each other (e.g.,~\cite{revell2020coupled}) or coupling of CFD to particle methods (e.g.,~\cite{besseron2021eulerian}). 

\paragraph{Applications and Software}
A non-exhaustive list of application fields includes mechanical and civil engineering
(astronautics~\cite{Volland2019}, manufacturing processes~\cite{Seufert2019_COUPLED, scheiblhofer2019coupling}, aerodynamics~\cite{marinoinvestigation, folkersma2020steady, Cocco2020, Risseeuw2019, CINQUEGRANA2021103264, Huang2021_Helicopters, srivastava2019computational}, urban wind modeling~\cite{revell2020coupled}, aeroacoustics~\cite{Lindner2020_ExaFSA, kersschot2020simulation}, explosions~\cite{nguyen2015fluid, zhang2020numerical}), marine engineering~\cite{andrun2020simulating}, bio engineering (heart valves~\cite{Davis2019}, aortic blood flow~\cite{naseri2020scalable}, fish locomotion~\cite{luo2020fluid}, muscle-tendon systems~\cite{Maier2021_Diss}), nuclear fission and fusion reactors~\cite{deSantis2019, fan2020study, Desai2020_Fusion}, and geophysics~\cite{Jaust2020, schmidt2021simulation, geoKW}.
Many users do not only use the official adapters (cf.~\autoref{sec:adapters}), but couple further community codes or their in-house codes. A non-exhaustive list of available coupled software (under a commercial or an open-source license) includes CAMRAD II and TAU~\cite{Huang2021_Helicopters}, DUST~\cite{Cocco2020}, DuMuX~\cite{koch2021dumux, Jaust2020}, DUNE~\cite{Firmbach2021}, Rhoxyz~\cite{andrun2020simulating}, Ateles~\cite{klimach2014end}, XDEM~\cite{besseron2021eulerian}, and FLEXI~\cite{krais2021flexi}.

\begin{figure}[h!]
\includegraphics[width=0.95\textwidth]{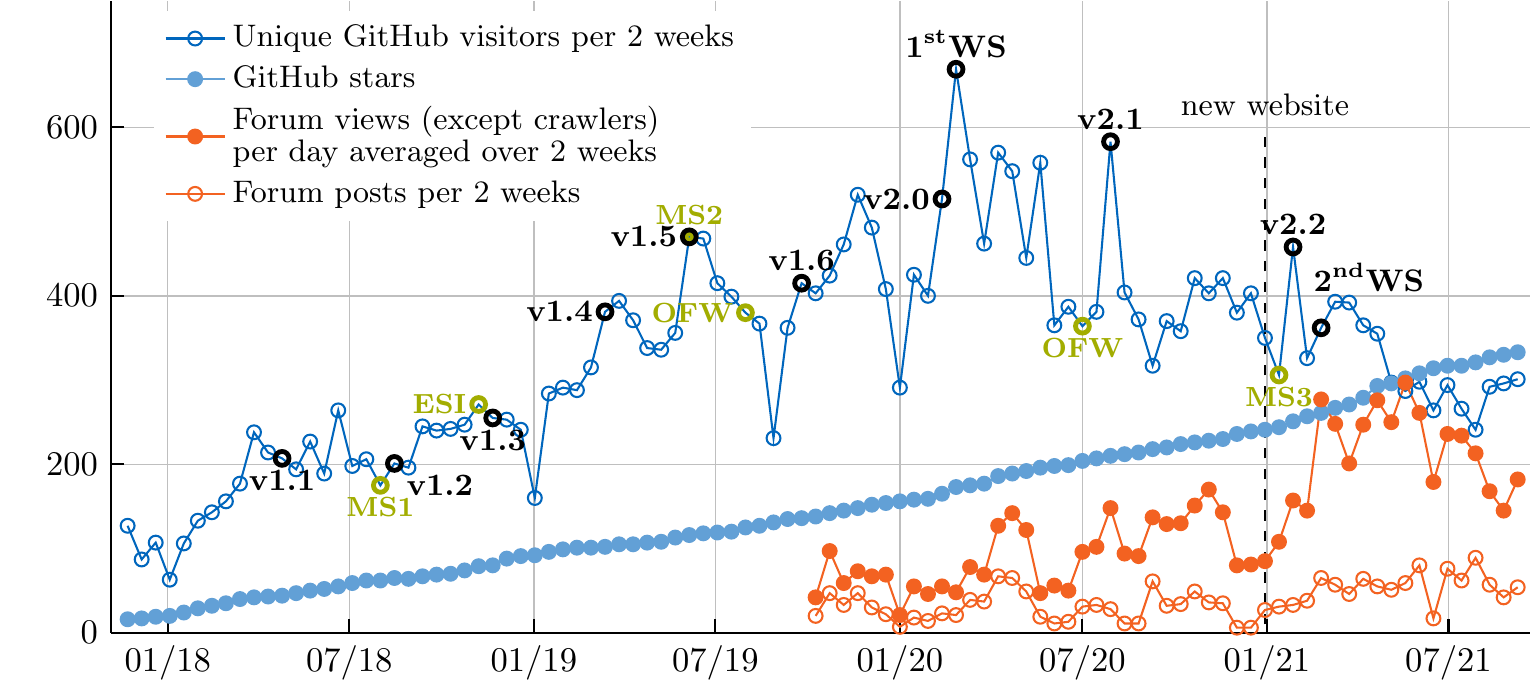}
\caption{
\label{fig:traffic}
Various traffic data showing community growth over time connected to key events. Whereas new releases have a clear impact on GitHub traffic, conferences, such as the ECCOMAS mini-symposia (MS1, MS2, MS3), different OpenFOAM workshops (ESI, OFW), and the preCICE workshops (WS) have not always a clear direct impact. With the start of the preCICE forum (fall 2019) and, in particular, with moving the user documentation to the new website at the end of 2020, traffic is shifted away from GitHub. At the online preCICE Workshop 2021, we used the forum to let attendees introduce themselves. This led to sustainable increase of forum traffic.
}
\end{figure}

\paragraph{Community Building}
preCICE users can interact with developers and with each other through various channels. 
We provide and moderate a Discourse forum\footnote{preCICE forum on Discourse: \url{https://precice.discourse.group/}} and a Gitter chat room\footnote{preCICE chat room on Gitter: \url{https://gitter.im/precice/Lobby}}. 
The forum replaced a previously used mailing list as discussions in the forum can be much better structured through categories, labels, and \textit{solution} posts. Moreover, Discourse can be customized to great extent, which allows us to hand over moderation responsibilities to the community at a suitable pace.   
For feature requests and bug reports, we use the issue trackers of the different repositories on GitHub\footnote{preCICE GitHub organization: \url{https://github.com/precice}}. Moreover, we organize yearly mini-symposia at ECCOMAS conferences (ECCM-ECFD 2018, COUPLED 2019, WCCM 2020, COUPLED 2021) and our own preCICE Workshops (preCICE Workshop 2020 in Munich\footnote{Aftermath of the 2020 workshop: \url{https://precice.discourse.group/t/precice-workshop-2020-updates/40}}, preCICE Workshop 2021 online). The workshops include an introduction course, which we plan to further extend in the next years. \autoref{fig:traffic} shows a static growth of the preCICE community over the previous 3 to 4 years.

\paragraph{Contributions}
To strengthen the sustainability of preCICE, we encourage users to also contribute back. Example contributions encompass code, tutorials, bug reports, or documentation. On our website, we provide detailed contributing guidelines\footnote{Contributing guidelines: \url{https://precice.org/community-contribute-to-precice.html}}. Our long-term goal is to hand over development of the official adapters (cf.~\autoref{sec:adapters}) to the community. In recent years, the OpenFOAM adapter has, in particular, seen various external contributions~\cite{nlrse19-chourdakis} and serves as an example of how the community may successfully contribute to isolated, smaller \emph{compartments} of a software project, as they can be easier to understand and contribute to. As of July 2021, 20\% (18 out of 89) pull requests and 26\% (26 out of 100) issues in the OpenFOAM adapter repository have been contributed by external contributors (not from the academic groups of the core team). While half of the external pull requests were ultimately not merged, they still serve as proof of concept for features that were at the time not aligned with the direction of the project. We have observed that several non-merged contributions were still useful for the community and we expect that tooling, automation, and clear guidelines will increase the ratio of successful external contributions in the long run.

  \section{Conclusions and outlook}
\label{sec:conclusion}

We have shown on the basis of various aspects that there is a tremendous gap between a working prototype software -- a software with state-of-the-art numerical and HPC methods (preCICE in 2016) and a sustainable and user-friendly software (preCICE in 2021). While the first one allows for scientific discoveries in scientific computing, only the latter allows for scientific discoveries in application areas as well. This can also be observed in the user numbers of preCICE. While the software today has a large and vivid community of users in a wide variety of application areas, it hardly had any users in 2016.
To bridge this gap, we presented necessary efforts in documentation, building, packaging, integration with external software, tutorials, tests, continuous integration, and community building.
Nearly all of these aspects are more complicated for a multi-component coupling software such as preCICE than for most other scientific computing software. This is not only due to the fact that preCICE is a library and, thus, needs another program that calls preCICE, but also that a coupled simulation needs by definition at least two different programs to be coupled. Therefore, often novel solutions are necessary for usually standard problems, such as the variety of testing concepts introduced in \autoref{sec:tests}.  

In forthcoming years, preCICE will undergo various extensions to make the software applicable beyond low-order, mesh-based, surface-coupled problems, such as fluid-structure interaction. Current work focuses on geometric multi-scale coupling, dynamic coupling meshes, waveform iteration~\cite{QNWI}, mesh-particle coupling, macro-micro coupling, and coupling to data-based approaches. An important topic will also be the efficient support of volume-coupled problems, which requires novel ideas in all main ingredients of preCICE: communication, coupling schemes, and data mapping.
To further increase the sustainability of preCICE, we will build on and extend the system test concept introduced in \autoref{ssec:system_tests}.

\section*{Acknowledgments}

Besides the authors of this paper, many more contributed to preCICE in the last five years. We want to thank 
Francisco Espinosa,
Carme Homs Pons,
Yakup Hoshaber,
Qunsheng Huang,
Alexander Jaust,
Gilberto Lem Carrillo,
Christopher Lorenz,
David Sommer,
Michel Takken,
Alexander Trujillo,
and everybody else who contributed to any repository of the preCICE organization on GitHub.

We thank the Deutsche Forschungsgemeinschaft (DFG, German Research Foundation) for supporting this work by funding -- EXC2075 -- 390740016 under Germany's Excellence Strategy. Furthermore, we acknowledge the support by the Stuttgart Center for Simulation Science (SimTech). 
This work was further funded by SPPEXA, DFG's Priority Program 1648 -- \emph{Software for Exascale Computing},  
the European Union's Horizon 2020 research and innovation program under the Marie Sklodowska-Curie grant agreement No 754462,
the International Graduate Research Group on \emph{Soft Tissue Robotics} (GRK 2198/1),
the DFG project \emph{preDOM}, project number 391150578,
the DFG SFB 1313, project number 327154368,
the Competence Network for Scientific High Performance Computing in Bavaria (KONWIHR) by the Bavarian State Ministry of Science and the Arts,
and the German Federal Ministry for Economic Affairs and Energy (BMWi) projects \emph{preCICE-ATHLET} and \emph{geoKW}. 
Moreover, we thank the Leibniz  Supercomputing  Centre of  the  Bavarian  Academy  of Sciences and Humanities for compute time on SuperMUC-NG.

Declarations of interest: none

\bibliographystyle{elsarticle-num}
\bibliography{literature}

\end{document}